\documentclass[12pt,lettersize,journal,onecolumn]{IEEEtran}

\usepackage{amsfonts,amssymb}
\usepackage{graphicx}
\usepackage{color}
\usepackage{todonotes}
\usepackage{algorithm,algpseudocode}
\usepackage{xspace}
\usepackage{etoolbox}
\usepackage{xcolor}
\usepackage{multirow}
\usepackage{enumitem}
\usepackage[normalem]{ulem}
\usepackage{bm}
\usepackage{subcaption}
\usepackage{placeins}
\usepackage{balance}

\usepackage[numbers]{natbib}
\usepackage{hyperref}

\usepackage{mathtools}

\newtheorem{definition}{Definition}
\newtheorem{lemma}{Lemma}
\newtheorem{theorem}{Theorem}

\newtheorem{corollary}[theorem]{Corollary}

\usepackage{tikz,pgfplots}
\usetikzlibrary{matrix, decorations, patterns, positioning, shapes, 3d, calc, intersections, arrows, fit, pgfplots.fillbetween}
\usepackage{tikz-3dplot}

\usepackage{cleveref}
\Crefname{definition}{Definition}{Definitions}
\crefname{definition}{Def.}{Defs.}
\Crefname{theorem}{Theorem}{Theorems}
\crefname{theorem}{Theorem}{Theorems}
\Crefname{lemma}{Lemma}{Lemmas}
\crefname{lemma}{Lemma}{Lemmas}
\Crefname{section}{Section}{Sections}
\crefname{section}{\S}{\S}
\crefname{paragraph}{\S}{\S}
\crefname{algorithm}{Alg.}{Algs.}
\Crefname{algorithm}{Algorithm}{Algorithms}
\crefname{figure}{Fig.}{Figs.}
\Crefname{figure}{Figure}{Figures}
\crefname{table}{Tab.}{Tabs.}
\Crefname{table}{Table}{Tables}

\newcommand{\tmp}[1]{q_{prev}}

\newcommand{\Real}{\mathbb{R}}

\colorlet{patternblue}{blue!60}

\newcommand{\A}{\mathbf{A}}
\newcommand{\B}{\mathbf{B}}
\newcommand{\CC}{\mathbf{C}}

\newcommand{\x}{x_1}
\newcommand{\y}{x_2}

\newcommand{\pluseq}{\mathrel{+}=}
\newcommand{\vc}[1]{\bm{#1}}

\title{Communication Lower Bounds and Optimal Algorithms for Symmetric Matrix Computations}

\author{
    \IEEEauthorblockN{Hussam Al~Daas}\IEEEauthorrefmark{1},
    \IEEEauthorblockN{Grey Ballard}\IEEEauthorrefmark{2},
    \IEEEauthorblockN{Laura Grigori}\IEEEauthorrefmark{3},
    \IEEEauthorblockN{Suraj Kumar}\IEEEauthorrefmark{4},
    \IEEEauthorblockN{Kathryn Rouse}\IEEEauthorrefmark{5},
    \IEEEauthorblockN{Mathieu V\'erit\'e}\IEEEauthorrefmark{6},

    \IEEEauthorblockA{%
        \IEEEauthorrefmark{1}STFC, Scientific Computing Department, Rutherford Appleton Laboratory,\\Didcot, Oxfordshire, UK, (hussam.al-daas@stfc.ac.uk)
    }

    \IEEEauthorblockA{%
        \IEEEauthorrefmark{2}Wake Forest University, Winston-Salem, NC, USA, (ballard@wfu.edu)
    }

    \IEEEauthorblockA{%
        \IEEEauthorrefmark{4}EPFL, Institute of Mathematics, Lausanne, Switzerland and\\PSI, Center for Scientific Computing, Theory and Data, Villigen, Switzerland (laura.grigori@epfl.ch)
    }

    \IEEEauthorblockA{%
        \IEEEauthorrefmark{5}Institut national de recherche en sciences et technologies du num\'erique,\\Lyon, France, (suraj.kumar@inria.fr)
    }

    \IEEEauthorblockA{
        \IEEEauthorrefmark{6}Inmar Intelligence, Winston-Salem, NC, USA, (kathryn.rouse@inmar.com)
    }

    \IEEEauthorblockA{
        \IEEEauthorrefmark{3}EPFL, Institute of Mathematics, Lausanne, Switzerland, (mathieu.verite@epfl.ch)
    }
}

\begin{document}

\maketitle

\begin{abstract}
In this article, we focus on the communication costs of three symmetric matrix computations: i) multiplying a matrix with its transpose, known as a symmetric rank-k update (SYRK) ii) adding the result of the multiplication of a matrix with the transpose of another matrix and the transpose of that result, known as a symmetric rank-2k update (SYR2K) iii) performing matrix multiplication with a symmetric input matrix (SYMM).
All three computations appear in the Level 3 Basic Linear Algebra Subroutines (BLAS) and have wide use in applications involving symmetric matrices.
We establish communication lower bounds for these kernels using sequential and distributed-memory parallel computational models, and we show that our bounds are tight by presenting communication-optimal algorithms for each setting.
Our lower bound proofs rely on applying a geometric inequality for symmetric computations and analytically solving constrained nonlinear optimization problems.
The symmetric matrix and its corresponding computations are accessed and performed according to a triangular block partitioning scheme in the optimal algorithms.
\end{abstract}

\section{Introduction}
\label{sec:intro}

Exploiting symmetry in matrix computations can reduce memory, computation, and data movement costs.
An $n \times n$ symmetric matrix requires storing only approximately $n^2/2$ unique entries in memory.
In the cases of computing the product of a matrix and its transpose, which is known as symmetric rank-$k$ update (SYRK), or computing the sum of a matrix product and its transpose, which is known as symmetric rank-$2k$ update (SYR2K), the output matrix is symmetric.
Thus, only the unique entries need to be computed, saving approximately half the computation.
In the case of multiplying a symmetric matrix by a nonsymmetric matrix, a computation denoted SYMM, no computation is saved compared to a general matrix multiplication, but fewer entries of the symmetric input need to be accessed.
Our goal in this paper is to establish upper and lower bounds on the communication costs, or the amount of data movement, required to perform computations with symmetric matrices like SYRK, SYR2K, and SYMM.
We show that exploiting symmetry indeed reduces the communication by constant factors when compared to nonsymmetric counterparts, but establishing the lower bounds and structuring the algorithms require very different techniques in the symmetric case.

We address two types of memory models in this work, namely, the two-level sequential memory model and the distributed-memory parallel model.
In the sequential model, computations are performed on data in a fast memory or cache with limited size, which is too small to store all the data, and we wish to minimize data movement between fast memory and an unbounded slow memory.
In the parallel model, each processor has its own local memory, and the data must be distributed across the memories of the processors.
For both lower bounds and algorithms, we consider the cases where the size of the local memories is not restrictive as well as where the size of the memories limits the possible algorithms.

Following the terminology of \cite{Ballard:3NL}, we classify SYRK, SYR2K, and SYMM as symmetric three nested loop (symmetric 3NL) computations.
In \cref{sec:memdeplb,sec:memindeplb}, we derive sequential and parallel communication lower bounds for this class of computations.
The lower bound proofs rely on a geometric inequality, which we call the symmetric Loomis-Whitney inequality \cite{ABGKR23}, that relates a computation embedded in a 3D iteration space to the required data access within 2D matrices.
In each of the sequential and parallel settings, we use this inequality along with other constraints to express communication lower bounds as solutions to constrained optimization problems.
We solve these problems analytically and derive bounds that include constant factors of leading order terms.
In the parallel case, we establish memory-dependent bounds, which depend on the size of the local memory of each processor and are derived from sequential lower bounds, and memory-independent bounds, which depend only on the matrix dimensions and the number of processors.
Either type of bound may be tightest depending on the matrix dimensions, number of processors, and memory size.

Our communication-optimal algorithms for symmetric 3NL computations, which attain the lower bounds with matching constants, utilize triangle block partitioning schemes \cite{BELV22}.
As we explain in \cref{sec:triangularDistrib}, triangle block partitioning corresponds to partitioning the lower triangle of a symmetric matrix into smaller lower triangles of symmetric submatrices of fixed dimension.
This problem is equivalent to the balanced clique partitioning problem, where the edges of a complete graph are partitioned into sets of edges that form cliques of the same size.
We generalize the partitioning scheme introduced by Beaumont et al.~\cite{BELV22} and used by Al Daas et al.~\cite{ABGKR23} by constructing new schemes using tools from affine and projective geometry and deriving necessary and sufficient conditions for the existence of schemes using established combinatorial results.

We present our sequential algorithms in \cref{sec:seqAlgorithms}.
They use triangle block partitioning schemes to define triangular blocks that fit into fast memory so that the computation associated with each triangle block can benefit from the symmetry of the computation.
In \cref{sec:memIndepAlgorithms,sec:limitedMemory}, we use the triangle block partitions to distribute the symmetric matrix across processors so that each processor can benefit from the reduced data access required of its symmetric computation.
Following the terminology for nonsymmetric matrix multiplication \cite{DE+13}, we classify our parallel algorithms as 1D, 2D, or 3D, depending on how many dimensions of the iteration space are partitioned across processors, and we adapt our algorithms in the case of limited memory to optimally navigate the memory-communication tradeoff.

\Cref{tab:lbs,tab:algs} highlight the contributed results of this paper, as they generalize existing lower bounds and algorithms.
We discuss more related work in \cref{sec:related}.
Beaumont et al.~\cite{BELV22} proves the sequential SYRK lower bound with a different argument relying on dominating sets and give an optimal sequential SYRK algorithm using their triangle block partitioning scheme.
Al~Daas et al.~\cite{ABGKR23} derive parallel memory-independent communication lower bounds for SYRK and present 1D, 2D, and 3D parallel algorithms that are communication optimal and use a specific instance of the triangle block partitioning of Beaumont et al.~\cite{BELV22}.
Agullo et al.~\cite{ABC+23} implemented triangle block distributions for 2D and 3D parallel algorithms for SYMM and reported practical speedups over nonsymmetric computations.
Our results generalize to all three computations and both sequential and parallel settings, and we show that each type and case of the lower bounds are matched by a different optimal algorithm, covering all configurations of matrix dimension, memory size, and number of processors.

\begin{table}
    \begin{center}
        \begin{tabular}{|c|c|c|}
            \hline
            & Sequential or memory-dependent &  \multicolumn{1}{|c|}{Memory-independent}\\
            \multirow{1}{*}{}&  \multirow{1}{*}{parallel bound} & \multicolumn{1}{|c|}{parallel bound}\\ \hline
            \cite{BELV22} & SYRK & \multicolumn{1}{|c|}{-}\\ \hline
            \cite{ABGKR23} & \multicolumn{1}{|c|}{-} & \multicolumn{1}{|c|}{SYRK}\\ \hline
            This paper & \multicolumn{2}{|c|}{SYRK, SYR2K and SYMM}\\ \hline
        \end{tabular}
        \caption{Existing communication lower bounds for symmetric three nested loop computations. Memory-dependent parallel bound can be derived straightforwardly from a sequential bound.}
        \label{tab:lbs}	
    \end{center}
\end{table}

\begin{table}
    \begin{center}
        \begin{tabular}{|c|c|c|c|c|}
            \hline
            & \multirow{2}{*}{Sequential} & \multicolumn{3}{|c|}{Parallel}\\ \cline{3-5}
            & &  1D & 2D & 3D \\ \hline
            \cite{BELV22} & SYRK &  \multicolumn{3}{|c|}{-}\\ \hline
            \cite{ABGKR23} & - & \multicolumn{3}{|c|}{SYRK}\\ \hline
            \cite{ABC+23} & - & - & \multicolumn{2}{|c|}{SYMM}\\ \hline
            This paper & \multicolumn{4}{|c|}{SYRK, SYR2K and SYMM}\\ \hline
        \end{tabular}
        \caption{Triangle block cyclic algorithms for symmetric three nested loop computations. 1/2/3D denotes that processors are arranged in a 1/2/3-dimensional grid.}
        \label{tab:algs}
    \end{center}
\end{table}
\FloatBarrier

\section{Related Work}
\label{sec:related}
The first communication lower bounds were derived by Hong and Kung~\cite{Hong:STOC81} for many sequential computations including matrix multiplication.
They modeled the data transfers between two levels of memory as a pebbling game with rules determined by a Computational Directed Acyclic Graph (CDAG) constrained by the size of the memory.
The first memory-independent parallel communication lower bounds were given by Aggarwal et al.~\cite{ACS90} after they extended the matrix multiplication lower bounds of Hong and Kung to the LPRAM parallel computation model.
By using the Loomis-Whitney inequality~\cite{LW49}, Irony et al.~\cite{ITT04} were able to reproduce the existing matrix multiplication lower bounds with a geometric argument that relates the volume of computation that can be performed to the number of elements accessed.
They also noted that parallel algorithms with access to more memory can perform less communication, known as the memory-communication trade off.

Despite their different proofs, all of the previously mentioned lower bounds indicate that minimizing the communication is achieved by partitioning the computation into cubical subsets.
This is because a cube maximizes the volume, which corresponds to the amount of computation performed, given the size of the projection onto each of its faces, which corresponds to the number of elements accessed.
When the iteration space is not cubical, as in the case of multiplying rectangular matrices, it is possible for the memory-independent lower bounds to degenerate.
To give lower bounds even in situations where the original memory-independent lower bounds degenerate, Demmel et al.~\cite{DE+13} derived a family of three parallel memory-independent matrix multiplication lower bounds.

Ballard et al.~\cite{Ballard:3NL} demonstrated the broad applicability of the Loomis-Whitney inequality for finding asymptotic communication lower bounds by applying it to all three nested loop computations.
McColl and Tiskin~\citep{MT99} and Solomonik and Demmel~\citep{SD11} generalized existing 2D and 3D parallel matrix multiplication algorithms to introduce the class of 3D parallel algorithms with limited memory (also referred as 2.5D algorithms in the literature) that are able to take advantage of all available memory, to minimize communication in light of this trade off.

While the previous work gave tight asymptotic lower bounds, the leading constants were not optimal.
Smith et al.~\cite{SLLvdG19} gave a tight leading constant to the sequential matrix multiplication lower bound by applying the Loomis-Whitney inequality then using Lagrange multipliers to further optimize their arguments.
Kwasniewski et al.~\cite{KK+19} matched their constant for a restricted class of parallelization using CDAG analysis.
Al~Daas et al.~\cite{ABGKR22} gave tight constants for all three memory-independent matrix multiplication lower bounds by applying the Loomis-Whitney inequality and using the Karush-Kuhn-Tucker (KKT) conditions to handle the additional optimizations.

Given the importance of the leading constant when comparing algorithms for the same computation, recent work has focused on methods that derive not only asymptotic lower bounds but also pursue tight constants for families of computations.
Both Olivry et al. and Kwasniewski et al.~\cite{OLPSR20,KK+21} applied CDAG analysis to automatically derive communication lower bounds, including leading constants, for large classes of computations.
However, the automated analysis methods were not able to derive tight leading order constants for all computations.
In particular, symmetric computations that have an iteration space that is only half a cube, can be performed with less communication than non-symmetric computations as Beaumont et al.~\cite{BELV22} demonstrated when they gave a tight leading constant for the sequential SYRK computation.
In order to show that their constant was tight, they provided a new algorithm with a novel data access pattern based on triangle blocks.

Inspired by the new sequential communication lower bounds for SYRK, Al Daas et al.~\cite{ABGKR23} proved a symmetric version of the Loomis-Whitney inequality, which allowed them to develop a family of three memory-independent communication lower bounds with new leading order constants.
To prove that the leading order constants are tight, they also derived 1D, 2D and 3D algorithms by extending the triangle blocking given by Beaumont et al.~\cite{BELV22}.
Agullo et al.~\cite{ABC+23} applied triangle block distributions to the parallel SYMM computation, deriving 2D and limited memory 3D parallel algorithms, and were able to improve communication by the expected factor.
As shown in \cref{tab:lbs,tab:algs}, our results build on and generalize this previous work by covering SYRK, SYR2K, and SYMM in both sequential and parallel settings.

\section{Preliminaries}
\label{sec:prelim}

\subsection{Symmetric Atomic Three Nested Loop Algorithms}
\label{subsec:symmAtomic3NL}
A symmetric three nested loop algorithm implements a computation over the iteration space $S=\{(i,j,k)\in [1:n_1]\times[1:n_1]\times[1:n_2] \; | \; i\geq j\}$ with one symmetric matrix of dimensions $n_1\times n_1$, and $m$ non-symmetric matrices of dimensions $n_1\times n_2$.
The algorithm is atomic if, at each point $(i,j,k)$ in the iteration space where $i\neq j$, both elements indexed by $(i,k)$ and $(j,k)$ of every non-symmetric matrix and the element indexed by $(i,j)$ of the symmetric matrix are accessed and all operations on these elements are performed atomically.

Three examples of computations that can be performed with symmetric atomic three nested loop algorithms are SYRK (\cref{alg:syrk}), SYR2K (\cref{alg:syr2k}), and SYMM (\cref{alg:symm}).

\begin{algorithm}
	\caption{\label{alg:syrk}Pseudo code of SYRK}
	\begin{algorithmic}
		\Require matrix $\A$ is $n_1\times n_2$, symmetric matrix $\CC$ is $n_1\times n_1$
		\Ensure $\CC\pluseq\A\cdot \A^T$
		\For{$i=1 \text{ to } n_1$}
		\For{$j=1 \text{ to } i$}
		\For{$k=1 \text{ to } n_2$}
		\State $\CC_{ij} \pluseq \A_{ik}\cdot \A_{jk}$
		\EndFor
		\EndFor
		\EndFor
	\end{algorithmic}
\end{algorithm}

\begin{algorithm}
	\caption{\label{alg:syr2k}Pseudo code of SYR2K}
	\begin{algorithmic}
		\Require matrix $\A$ is $n_1\times n_2$, matrix $\B$ is $n_1\times n_2$, symmetric matrix $\CC$ is $n_1\times n_1$
		\Ensure $\CC\pluseq\A\cdot \B^T+\B\cdot \A^T$
		\For{$i=1 \text{ to } n_1$}
		\For{$j=1 \text{ to } i$}
		\For{$k=1 \text{ to } n_2$}
		\State $\CC_{ij} \pluseq \A_{ik}\cdot \B_{jk} + \B_{ik}\cdot \A_{jk}$
		\EndFor
		\EndFor
		\EndFor
	\end{algorithmic}
\end{algorithm}

\begin{algorithm}
	\caption{\label{alg:symm}Pseudo code of SYMM}
	\begin{algorithmic}
		\Require matrix $\A$ is symmetric $n_1\times n_1$, $\B$ and $\CC$ are $n_1\times n_2$
		\Ensure $\CC\pluseq\A\cdot \B$
		\For{$i=1 \text{ to } n_1$}
		\For{$j=1 \text{ to } i-1$}
		\For{$k=1 \text{ to } n_2$}
		\State $\CC_{ik} \pluseq \A_{ij}\cdot \B_{jk}$
                \State $\CC_{jk} \pluseq \A_{ij}\cdot \B_{ik}$
		\EndFor
		\EndFor
		\EndFor
                \For{$i=1 \text{ to } n_1$}
		\For{$k=1 \text{ to } n_2$}
		\State $\CC_{ik} \pluseq \A_{ii}\cdot \B_{ik}$
		\EndFor
		\EndFor
	\end{algorithmic}
\end{algorithm}

\subsection{Computation Models}
\label{sec:prelim:compModel}

\subsubsection{Sequential Computation Model}

We consider the sequential execution of all three symmetric computations under a machine model with two levels of memory: one distant/slow unlimited memory and one local/fast memory of limited size $M$.
In this model, computations can only occur using input and output operands that are present in the fast memory.
During the execution of an algorithm, elements are communicated between the two levels of memory depending on the computation needs.
Additionally, communications and computations are constrained by the size of the local memory and the dependencies between the operations.
Communication cost, also referred to as bandwidth cost, is evaluated as the sum of the number of words \emph{read} from the slow to the fast memory and the number of words \emph{written} from the fast to the the slow memory, to complete the whole computation.
Such a model allows to assess the theoretical communication performance of any sequential algorithm and has been extensively used~\cite{CDKSY13,Ballard:3NL,KK+21}.
Bandwidth communication lower bounds derived using this model are directly transferable to a parallel computation setting: for each processor we can indeed consider that all other processors' local memories (see \cref{sec:par-comp-model}) play the role of the slow memory.
The two levels memory model is relevant from a practical point of view because it closely approximates the actual sequential execution of an algorithm on a single computing unit communicating data with a distant memory, for example the last level cache of a CPU with the RAM.

\subsubsection{Parallel Computation Model}
\label{sec:par-comp-model}

We consider the parallel execution of symmetric three nested loop algorithms under the MPI or $\alpha-\beta-\gamma$ model \cite{Thakur:CollectiveCommunications:2005,Chan:CollectiveCommunications:2007} of parallel computation in which a computation is distributed over $P$ processors.
Each processor has its own local memory of size $M$, and can only perform computations with data which is stored in its local memory.
We consider both the memory constrained and memory unconstrained settings when deriving lower bounds and algorithms.
In the former, the size of each processor's local memory, $M$, may be smaller than the amount of memory required for the optimal algorithms that do not take into account the size of a processor's memory.
In the latter, $M$ is sufficiently large that each processor can store any necessary data to perform the algorithms of interest and the memory size is not explicitly considered.
We call the lower bounds derived in the first setting memory-dependent and the lower bounds derived in the second setting memory-independent following~\cite{BDHLS12-SS}.

Each processor can share data from its local memory with any other processor by sending a message over a fully connected network with bidirectional links.
Any input data required for the computations performed by a processor which is not in the processor's local memory must be received from another processor.
While a processor can only send or receive one message at a time, disjoint pairs of processors may communicate at the same time.
We refer to the sending of messages between processors over the network as communication.
An algorithm's total communication cost is determined by the number of messages, the latency cost of $\alpha$ per message, and the size of the messages, the bandwidth cost of $\beta$ per word.
When messages are large, bandwidth cost dominates the overall communication, hence we seek to minimize the bandwidth cost of algorithms that perform the computation of interest.
The final parameter of the model, $\gamma$, is the per-operation arithmetic cost.

\paragraph{Collective Communication Costs}
\label{sec:prelim:collectiveCommunicationCosts}
Our parallel algorithms in \cref{sec:memIndepAlgorithms} use \Call{All-to-All}{} and \Call{Reduce-Scatter}{} collectives to perform communication between processors.
As pairwise exchange algorithms have optimal bandwidth costs, we assume that they are used for both collectives.
The latency cost is $P-1$ when either collective is being performed on $P$ processors.
The bandwidth cost of either collective is $\left(1-\frac{1}{P}\right)n$, where $n$ is the number of words on each processor both before and after an \Call{All-to-All}{} collective or before a \Call{Reduce-Scatter}{} collective.
Each processor also performs $\left(1-\frac{1}{P}\right)n$ computations for \Call{Reduce-Scatter}{}.
We refer the reader to \cite{Thakur:CollectiveCommunications:2005,Chan:CollectiveCommunications:2007,BCKUW97} for the derivation of the costs of each collective.

\subsection{Fundamental Results}
\label{sec:pastResults}

The Loomis-Whitney inequality~\cite{LW49} is a geometric result that allows us to relate the volume of a computation to the area of the elements it requires from the input and output arrays.
The relationship determines a constraint on the amount of data reuse an algorithm can take advantage of, which has been used as a constraint in optimization problems to derive lower bounds for many linear algebra computations~\cite{ITT04, Ballard:3NL, DE+13, SLLvdG19, ABGKR22}.

\begin{lemma}[Loomis-Whitney \cite{LW49}]
\label{lem:LW}
Let $V$ be a finite set of points in $\mathbb{Z}^3$.
Let $\phi_i(V)$ be the projection of $V$ in the $i$-direction, i.e. all points $(j,k)$ such that there exists an $i$ so that $(i,j,k) \in V$.
Define $\phi_j(V)$ and $\phi_k(V)$ similarly.
Then
$$|V| \leq |\phi_i(V)|^{1/2} \cdot |\phi_j(V)|^{1/2} \cdot |\phi_k(V)|^{1/2}\text,$$
where $|\cdot|$ denotes the cardinality of a set.
\end{lemma}

The goal when deriving communication lower bounds is to maximize the amount of computation given the number of elements accessed, or alternatively minimize the number of elements accessed given the amount of computation performed.
In both cases, the amount of computation corresponds to the volume, while the number of elements accessed corresponds to the total size of all the projections.
As the Loomis-Whitney inequality is tight when the set $V$ is a cube, and the projections are squares, the intuition for minimizing communication is to make each segment of computation correspond to a cubical set of indices in the iteration space.
While applying the constraint derived from the Loomis-Whitney inequality to problems with cubical iteration spaces leads to tight lower bounds, additional constraints are necessary in the case of non-cubical iteration spaces as was previously shown for rectangular matrix multiplication~\cite{DE+13,ABGKR22}.
The iteration space of symmetric three nested loop algorithms is a triangular prism.
Thus the Loomis-Whitney inequality does not yield a tight bound on the volume of computation that can be performed given a fixed number of elements.
While the first tight communication lower bounds for a symmetric three nested loop algorithm, given by Beaumont et al.~\cite{BELV22}, used a pebbling argument with CDAG analysis, we will follow Al~Daas et al.~\cite{ABGKR23} and use their symmetric Loomis-Whitney inequality as a foundation for our lower bound arguments.

\begin{lemma}[{\cite[Lemma 3]{ABGKR23}}]
	\label{lem:symLW}
	Let $V$ be a finite set of points contained in $\{(i, j, k)\in\mathbb{Z}^3 \; | \; j<i\}$.
	Let $\phi_i(V)$ be the projection of $V$ in the $i$-direction, $\phi_i(V) = \{(j,k)\in\mathbb{Z}^2|\text{ there exists } i\in\mathbb{Z}\text{ such that } (i, j, k)\in V\}$.
	Define $\phi_j(V)$ and $\phi_k(V)$ similarly.
	Then
	$$2|V| \leq |\phi_i(V)\cup\phi_j(V)| \cdot (2|\phi_k(V)|)^{1/2}\text,$$
	where $|\cdot|$ denotes the cardinality of a set.
\end{lemma}

\section{Memory Dependent Lower Bound Results}
\label{sec:memdeplb}
\subsection{Key Optimization Problem}
The following optimization problem is the key to our memory dependent lower bound argument where we want to maximize the number of atomic symmetric inner loop iterations that can be performed given the amount of memory available.
The following lemma presents an abstract nonlinear optimization problem and its corresponding solution that will be applied later on to derive a lower bound for memory-dependent sequential computations.

\begin{lemma}
  \label{lem:memdepopt}
  Consider the optimization problem
  $$\max_{\vc{x} > 0}\frac{\sqrt2}{2}x_1 x_2^{1/2}$$
  such that
  $$mx_1 + x_2 \leq X$$
  for positive constants $m,X$.
  Then the optimal solution
  $$\frac{\sqrt2}{3\sqrt{3}m}X^{3/2}$$
  occurs when $x_1^* = \frac{2X}{3m}$, and $x_2^* = \frac{X}{3}$.
\end{lemma}
\begin{IEEEproof}
  Consider the constraint given in the problem
  \[x_2 \leq X - m x_1 \text.\]
  Since $x_2\geq 0$ we can take the square root of both sides of the inequality and multiply them by $\frac{\sqrt{2}}{2} x_1$. This yields the following inequality
  \[\frac{\sqrt{2}}{2} x_1 \sqrt{x_2} \leq \frac{\sqrt{2}}{2} x_1\sqrt{X - mx_1}\text.\]
Now, the right hand side of this inequality can be seen as a function in one variable, $x_1$. We denote this function by $w$, that is, $w(x) = \frac{\sqrt{2}}{2} x\sqrt{X - mx}$ defined on the interval of interest $\left[0,\frac{X}{m}\right]$. This function is continuous on $\left[0,\frac{X}{m}\right]$ and twice continuously differentiable on $\left[0,\frac{X}{m}\right)$, with
  \begin{align*}
    w^\prime(x) &= \frac{ \sqrt{2} }{ 4 } \frac{ 2X - 3 mx }{ \sqrt{X-mx} }\text,\\
    w^{\prime\prime}(x) &= \frac{\sqrt{2}}{8} \frac{m (3mx-4X)}{(X-mx)^{\frac{3}{2}}}\text.
  \end{align*}
  The first derivative of $w$ vanishes at $x = \frac{2X}{3m}$ where the second derivative takes the value $-\frac{3\sqrt{6}m}{4\sqrt{X}} < 0$. Therefore, the function $w$, which is positive, on the interval of interest $\left[0,\frac{X}{m}\right]$, and vanishes on its domain boundary, attains its maximum value at $x=\frac{2X}{3m}$, that is
  \[
  w(x) \leq w\left(\frac{2X}{3m}\right) =  \frac{\sqrt{2}}{3\sqrt{3}} \frac{X\sqrt{X}}{m}\text,\quad \text{ for } x \in \left[0,\frac{X}{m}\right]\text.
  \]
  Using this upper bound on $w$ in the inequality derived from the constraint above, we have
  \[
  \frac{\sqrt{2}}{2} x_1 \sqrt{x_2} \leq  \frac{\sqrt{2}}{3\sqrt{3}} \frac{X\sqrt{X}}{m}\text.
  \]
  By setting $x_1=\frac{2X}{3m}$ and $x_2 = X-mx_1 = \frac{X}{3}$, the objective function attains its maximum.
\end{IEEEproof}

While we state the optimization problem abstractly, each term relates to the size of an entity in the lower bound.
One should think of $x_1$ as the size of the projections onto each of the non-symmetric matrices, and $x_2$ as the size of the projection onto the symmetric matrix.
The constants $m$ and $X$ correspond to the number of non-symmetric matrices and the amount of data available, respectively.
Thus the constraint corresponds to the requirement that the projections of the iteration space onto the matrices must not exceed the data available, while the objective function is the largest number of atomic symmetric inner loop iterations that can be performed with those projections from \cref{lem:symLW}.

\subsection{Sequential Lower Bounds}
\label{subsec:seqLowerBound}
Following \cite{BELV22}, we begin our lower bound proofs by finding an upper bound on the largest subcomputation $F$ that can be performed while accessing at most $X$ elements by a symmetric atomic three nested loop computation.
Our notation and the crux of our proof for the main theorem will follow \cite{ABGKR23}.
The first result, our extension of \cite[Proposition 3.4]{BELV22}, relates the size of the set of elements accessed to projections of the iteration space.
\begin{lemma}
  \label{lem:dataaccess}
  Let $F\subseteq \{(i,j,k) \in \mathbb{Z}^3 \; | \; i > j\}$ be a subset of atomic inner loop iterations performed by a symmetric atomic three nested loop computation with $m$ non-symmetric matrices.
  Define $\phi_i(F), \phi_j(F), \phi_k(F)$ as in \cref{lem:symLW}.
  Then the number of elements required to perform $F$ must be
  $$m|\phi_i(F)\cup\phi_j(F)|+|\phi_k(F)|\text.$$
\end{lemma}
\begin{IEEEproof}
  To begin, note that the elements of the symmetric matrix are indexed by $\{(i,j) \; | \; (i,j,k) \in F\} = \phi_k(F)$, thus the second term corresponds to elements of the symmetric matrix that must be accessed by the atomic inner loop iterations in $F$.
  For the first term, recall that a symmetric three nested loop computation is atomic if any inner loop iteration where $i \neq j$ accesses both elements of each non-symmetric matrix.
  Consider an inner loop iteration indexed by $(i,j,k)\in F$.
  By the atomicity assumption, both the elements indexed by $(i,k)$ and $(j,k)$ of every non-symmetric matrix must be accessed by the inner loop iteration.
  Thus, we can see that for each non-symmetric matrix, the set of elements accessed must equal those indexed by $\phi_i(F)\cup\phi_j(F)$, and the result follows.
\end{IEEEproof}

Now we restate \cite[Theorem 4.1]{BELV22} in our more general situation.
While the result is an extension, our proof technique is different as we rely on the Symmetric Loomis Whitney inequality instead of CDAG analysis.

\begin{theorem}
  \label{thm:oi}
Let
$\mathcal{P}(X)$
  be the largest number of atomic inner loop iterations of a symmetric atomic three nested loop computation with $m$ non-symmetric matrices that can be performed accessing at most $X$ matrix elements.
  Then $\mathcal{P}(X) \leq \frac{\sqrt{2}}{3\sqrt{3}m}X^{3/2}.$
\end{theorem}
\begin{IEEEproof}
Let $F$ denote a subset associated with atomic inner loop iterations that can be performed by accessing at most $X$ matrix elements.
  By \cref{lem:dataaccess}, the number of elements accessed to perform the atomic inner loop iterations in $F$ is $m|\phi_i(F)\cup\phi_j(F)|+|\phi_k(F)|\leq X$.
  Additionally, by \cref{lem:symLW}, we know that $|F|\leq\frac{1}{2}|\phi_i(F)\cup\phi_j(F)|(2|\phi_k(F)|)^{1/2}$.
  Substituting $x_1=|\phi_i(F)\cup\phi_j(F)|$, and $x_2=|\phi_k(F)|$ in \cref{lem:memdepopt}, we obtain $|F|\leq \frac{\sqrt{2}}{3\sqrt{3}m}X^{3/2}$. Taking the maximum over the possible subsets $F$, we see that $\mathcal{P}(X) \leq \frac{\sqrt{2}}{3\sqrt{3}m}X^{3/2}$ bounds the number of atomic inner loop iterations that may be performed accessing at most $X$ elements of the matrices.
\end{IEEEproof}

From the above, we are able to derive sequential lower bounds for any symmetric atomic three nested loop algorithm.
Let $M$ be the size of the local fast memory.
To prove the lower bound, we break the stream of instructions into segments each of which performs exactly $R$ element reads (a complete segment), except the last segment which may contain less than $R$ element reads.
Then we derive a general upper bound on the number of matrix elements that can be accessed which we use with \cref{thm:oi} to bound the number of atomic inner loop iterations that may be performed in a segment.
From this upper bound on the number of atomic inner loop iterations that may be performed in a segment, we find a lower bound for the number of complete segments that must be performed.
The lower bound on the number of segments allows us to determine a lower bound on the number of read operations in terms of the free variable $R$.
Finally, we determine the $R$ that maximizes the lower bound.

\begin{theorem}
  The number of element reads required to perform a symmetric atomic three nested loop computation with $m$ non-symmetric matrices of dimension $n_1\times n_2$, one symmetric matrix of size $n_1\times n_1$, and the fast memory size $M$ is at least
  $$\frac{m}{\sqrt{2}}\frac{n_1(n_1-1)n_2}{M^{1/2}}-2M\text.$$
\end{theorem}
\begin{IEEEproof}
  We break down the atomic inner loop iterations involving the strict lower triangle of the symmetric matrix into complete segments where at most $R$ elements are read from the slow memory.
  As there are at most $M$ elements of the matrices in memory at the start of each segment, at most $R$ additional elements are read during the segment, and any element must either be in memory or be read into the memory, at most $M+R$ elements of the matrices are accessed during the segment.
  Let $F$ correspond to the indices $(i, j, k)$ associated with the atomic inner loop iteration during a complete segment.
  Then $|F|\leq \frac{\sqrt{2}}{3m\sqrt{3}}(M+R)^{3/2}$ bounds the number of atomic inner loop iterations that may be performed in a complete segment.
  As there are $n_1(n_1-1)n_2/2$ atomic inner loop iterations that access elements of the strict lower triangle of the symmetric matrix, any algorithm must be broken into at least $\left\lfloor\frac{3m\sqrt{3}}{2\sqrt{2}}\frac{n_1(n_1-1)n_2}{(M+R)^{3/2}}\right\rfloor$ segments each of which performs $R$ element reads.
  Thus, any such algorithm must perform at least $\frac{3m\sqrt{3}}{2\sqrt{2}}\frac{Rn_1(n_1-1)n_2}{(M+R)^{3/2}}-R$ element reads.
  In order to maximize this lower bound, we select $R$ that maximizes $R\frac{3m\sqrt{3}}{2\sqrt{2}}\frac{n_1(n_1-1)n_2}{(M+R)^{3/2}}$, which occurs when $R=2M$.
  Hence, setting $R=2M$, we see that any algorithm that computes a symmetric atomic three nested loop computation with $m$ non-symmetric matrices of dimension $n_1\times n_2$, and one symmetric matrix of size $n_1\times n_1$ must perform at least $\frac{m}{\sqrt{2}}\frac{n_1(n_1-1)n_2}{M^{1/2}}-2M$ element reads.
\end{IEEEproof}

We are able to derive the sequential communication lower bounds for each of our symmetric atomic three nested loop algorithms by substituting the number of non-symmetric matrices for $m$.

\begin{corollary}
  {\cite[Corollary 4.7]{BELV22}}
  The number of element reads required to perform a SYRK computation where $\A$ has dimensions $n_1\times n_2$ and the fast memory size is $M$ is at least
  $$\frac{1}{\sqrt{2}}\frac{n_1(n_1-1)n_2}{M^{1/2}}-2M\text.$$
\end{corollary}
\begin{corollary}
  The number of element reads required to perform a SYR2K computation with atomic inner loop iterations where matrices $\A$ and $\B$ have dimensions $n_1\times n_2$ and the fast memory size is $M$ is at least
  $$\frac{2}{\sqrt{2}}\frac{n_1(n_1-1)n_2}{M^{1/2}}-2M\text.$$
\end{corollary}
\begin{corollary}
  The number of element reads required to perform a SYMM computation with atomic loop iterations where matrices $\B$ and $\CC$ have dimensions $n_1\times n_2$, symmetric matrix $\A$ has dimensions $n_1\times n_1$, and the fast memory size is $M$ is at least
  $$\frac{2}{\sqrt{2}}\frac{n_1(n_1-1)n_2}{M^{1/2}}-2M\text.$$
\end{corollary}

We note that the number of element reads for a computation is also a valid lower bound on the number of elements communicated. We will see later in \Cref{sec:seqAlgorithms} that the number of element reads of our algorithms match the leading terms in the lower bounds exactly and they are often the dominating terms of overall communication cost.

\subsection{Parallel Lower Bounds}
It is straightforward to extend a sequential lower bound to a memory-dependent parallel lower bound.
Note that any algorithm performing $N$ atomic inner loop iterations using $P$ processors must have at least one processor that performs $N/P$ atomic inner loop iterations.
We can then apply the relevant sequential lower bound argument to this processor to determine a parallel memory dependent lower bound on the number of words that processor must receive.
\begin{corollary}
  Any parallel SYRK computation using $P$ processors where $\A$ has dimensions $n_1\times n_2$ and the each processor has local memory of size $M$ involves at least
  $$\frac{1}{\sqrt{2}}\frac{n_1(n_1-1)n_2}{PM^{1/2}}-2M$$
  receives.
\end{corollary}
\begin{corollary}
  Any parallel SYR2K computation with atomic inner loop iterations using $P$ processors where matrices $\A$ and $\B$ have dimensions $n_1\times n_2$ and each processor has local memory of size $M$ involves at least
  $$\frac{2}{\sqrt{2}}\frac{n_1(n_1-1)n_2}{PM^{1/2}}-2M$$
  recieves.
\end{corollary}
\begin{corollary}
  Any parallel SYMM computation with atomic loop iterations using $P$ processors where matrices $\B$ and $\CC$ have dimensions $n_1\times n_2$, symmetric matrix $\A$ has dimensions $n_1\times n_1$, and each processor has local memory of size $M$ involves at least
  $$\frac{2}{\sqrt{2}}\frac{n_1(n_1-1)n_2}{PM^{1/2}}-2M$$
  recieves.
\end{corollary}

\section{Memory Independent Parallel Lower Bound Results}
\label{sec:memindeplb}

We now derive communication lower bounds in the parallel case that do not depend on the size of the local memory.
Our proof for memory independent communication lower bounds also relies on optimizing the relationship between the computation performed and the data available, but it takes a different approach.
For memory dependent lower bounds derived in \cref{sec:memdeplb}, the proof maximizes the amount of computation given the amount of data available.
For memory independent lower bounds, the proof minimizes the amount of data given the computation that a processor performs, and this optimization problem requires different techniques to solve.

We now present definitions and results that allow us to solve the optimization problem used in the derivation of the memory-independent lower bounds in \cref{sec:memindeplb:lbexpression}.
\begin{definition}[{\cite[eq. (3.2)]{BV04}}]
	\label{def:convex}
	A differentiable function $f:\Real^d\rightarrow \Real$ is \emph{convex} if its domain is a convex set and for all $\vc{x},\vc{y} \in \textbf{dom} \; f$,
	$$f(\vc{y}) \geq f(\vc{x}) + \langle \nabla f(\vc{x}), \vc{y} - \vc{x} \rangle\text.$$
\end{definition}

\begin{definition}[{\cite[eq. (3.20)]{BV04}}]
	\label{def:quasiconvex}
	A differentiable function $g:\Real^d\rightarrow \Real$ is \emph{quasiconvex} if its domain is a convex set and for all $\vc{x},\vc{y} \in \textbf{dom} \; g$,
	$$g(\vc{y}) \leq g(\vc{x}) \text{ implies that } \langle \nabla g(\vc{x}), \vc{y} - \vc{x} \rangle \leq 0\text.$$
\end{definition}

\begin{definition}[{\cite[eq. (5.49)]{BV04}}]
	\label{def:KKT}
	Consider an optimization problem of the form
	\begin{equation}
	\label{eq:optprob}
	\min_{\vc{x}} f(\vc{x}) \quad \text{ subject to } \quad \vc{g}(\vc{x}) \leq \vc{0}\text,
	\end{equation}
	where $f:\Real^d \rightarrow \Real$ and $\vc{g}:\Real^d\rightarrow \Real^c$ are both differentiable.
	Define the dual variables $\vc{\mu}\in\mathbb{R}^c$, and let $\vc{J}_{\vc{g}}$ be the Jacobian of $\vc{g}$.
	The \emph{Karush-Kuhn-Tucker (KKT)} conditions of $(\vc{x},\vc{\mu})$ are as follows:
	\begin{itemize}
		\item \emph{Primal feasibility}: $\vc{g}(\vc{x}) \leq \vc{0}$;
		\item \emph{Dual feasibility}: $\vc{\mu} \geq 0$;
		\item \emph{Stationarity}: $\nabla f(\vc{x}) + \vc{\mu} \cdot \vc{J}_{\vc{g}}(\vc{x}) = \vc{0}$;
		\item \emph{Complementary slackness}: $\mu_i g_i(\vc{x})=0$ for all $i\in \{1,\dots,c\}$.
	\end{itemize}
\end{definition}

\begin{lemma}[{\cite[Lemma 3]{ABGKR22}}]
	\label{lem:KKT}
	Consider an optimization problem of the form given in \cref{eq:optprob}.
	If $f$ is a convex function and each $g_i$ is a quasiconvex function, then the KKT conditions are sufficient for optimality.	
\end{lemma}

\begin{lemma}[{\cite[Lemma 4]{ABGKR23}}]
	\label{lem:quasiconvex}
	The function $g_0(\vc{x}) = L - x_1^2x_2$, for some constant $L$, is quasiconvex in the positive quadrant.
\end{lemma}

\subsection{Key Optimization Problem}
As in the previous section about memory-dependent lower bounds, we begin the argument by solving an abstract optimization problem that is the key to proving our lower bounds.
For the memory independent bounds, we want to minimize the number of elements required given a number of symmetric atomic inner loop iterations.

\begin{lemma}
\label{lem:memindepopt}
Consider the optimization problem:
$$\min_{\vc{x} \in \mathbb{R}^2} mx_1+x_2$$
such that
$$\left(\frac{n_1(n_1-1)n_2}{\sqrt{2}P}\right)^2 \leq x_1^2x_2\text,$$
$$ 0 \leq x_1,$$
$$\frac{n_1(n_1-1)}{2P} \leq x_2 \leq \frac{n_1(n_1-1)}{2}\text,$$
where $P \geq 1 $, and $n_1, n_2$ are positive integers.
The optimal solution $\vc{x}^*$ depends on the relative values of the constraints, yielding three cases:
\begin{enumerate}
	\item if $n_1 \leq mn_2$ and $P \leq \frac{mn_2}{\sqrt{n_1(n_1-1)}}$, then $x_1^*=\frac{n_2\sqrt{n_1(n_1-1)}}{P}$, $x_2^*=\frac{n_1(n_1-1)}{2}$;
	\item if $mn_2 < n_1$ and $P \leq  \frac{n_1(n_1-1)}{m^2n_2^2}$, then $x_1^*=n_2\sqrt{\frac{n_1(n_1-1)}{P}}$, $x_2^*=\frac{n_1(n_1-1)}{2P}$;
	\item if ($n_1 \leq mn_2$ and $P > \frac{mn_2}{\sqrt{n_1(n_1-1)}}$) or ($mn_2 < n_1$ and $P >  \frac{n_1(n_1-1)}{m^2n_2^2}$) then $x_1^*=\frac{1}{m^{1/3}}\left(\frac{n_1(n_1-1)n_2}{P}\right)^{2/3}$, and $x_2^*=\frac{m^{2/3}}{2}\left(\frac{n_1(n_1-1)n_2}{P}\right)^{2/3}$.
\end{enumerate}
\end{lemma}
\begin{IEEEproof}
  To begin, we rewrite our objective functions and constraints in the notation of \cref{def:KKT}.
  Let $f(\vc{x}) = mx_1+x_2 $ and
  \begin{equation*}
    \vc{g}(\vc{x}) =
    \begin{bmatrix}
      \left(\frac{n_1(n_1-1)n_2}{\sqrt{2}P}\right)^2-x_1^2x_2 \\
      - x_1 \\
      \frac{n_1(n_1-1)}{2P} - x_2 \\
      x_2 - \frac{n_1(n_1-1)}{2}
    \end{bmatrix}\text.
  \end{equation*}
  We note that the first constraint is quasiconvex by \cref{lem:quasiconvex}, and that the remaining constraints and the objective function are all affine functions hence convex functions.
  Thus by \cref{lem:KKT} it is sufficient to prove optimality of the solutions by finding dual variables $\vc{\mu}^*$ such that the KKT conditions in \cref{def:KKT} hold.

  Before considering the cases, we note that  $\nabla f(\vc{x}) = \begin{bmatrix} m & 1\end{bmatrix}$, and
  \begin{equation*}
    \vc{J}_{\vc{g}}(\vc{x}) =
    \begin{bmatrix}
      -2x_1x_2 & -x_1^2 \\
      -1 & 0 \\
      0 & -1 \\
      0 & 1
    \end{bmatrix}\text.
  \end{equation*}

  \paragraph{Case 1 ($n_1\leq mn_2$ and $P \leq \frac{mn_2}{\sqrt{n_1(n_1-1)}}$)}
  Let
  $$\vc{x}^* = \begin{bmatrix} \frac{n_2\sqrt{n_1(n_1-1)}}{P} & \frac{n_1(n_1-1)}{2}\end{bmatrix}$$
  and
  $$\vc{\mu}^* = \begin{bmatrix}\frac{mP}{(n_1(n_1-1))^{3/2}n_2} & 0 & 0 & \frac{mn_2}{(n_1(n_1-1))^{1/2}P}-1\end{bmatrix}\text.$$
    To begin, we note that the first constraint is tight, $x_2^*$ is set to the boundary of one of its constraints, and $x_1^* >0$.
    Thus it is clear that primal feasibility holds.
    As $\mu_1^* >0$ and $\mu_4^*\geq0$ from the conditions of the case, dual feasibility follows.
    Stationarity can be directly verified.
    Complementary slackness is satisfied as the first and fourth constraints are tight and the dual variables $\mu_2^*$ and $\mu_3^*$ are 0.

    \paragraph{Case 2 ($mn_2 < n_1$ and $P \leq \frac{n_1(n_1-1)}{m^2n_2^2}$)}
    Let
    $$\vc{x}^* = \begin{bmatrix} n_2\sqrt{\frac{n_1(n_1-1)}{P}} & \frac{n_1(n_1-1)}{2P}\end{bmatrix}$$
    and
    $$\vc{\mu}^* = \begin{bmatrix} \frac{mP^{3/2}}{(n_1(n_1-1))^{3/2}n_2} & 0 & 1-mn_2\sqrt{\frac{P}{n_1(n_1-1)}} & 0\end{bmatrix}\text.$$
    As in Case 1, primal feasibility is verified by noting the first constraint is tight, $x_2^*$ is set to the boundary of one of its constraints, and $x_1^* >0$.
    As $\mu_1^* >0$ and $\mu_3^*\geq0$ from the conditions of the case, dual feasibility follows.
    Stationarity can be directly verified.
    Complementary slackness is satisfied as the first and third constraints are tight and the dual variables $\mu_2^*$ and $\mu_4^*$ are 0.

      \paragraph{Case 3 ($n_1\leq mn_2$ and $P > \frac{mn_2}{\sqrt{n_1(n_1-1)}}$) or ($mn_2 < n_1$ and $P > \frac{n_1(n_1-1)}{m^2n_2^2}$)}
      Let
      $$\vc{x}^* = \begin{bmatrix}\frac{1}{m^{1/3}}\left(\frac{n_1(n_1-1)n_2}{P}\right)^{2/3} &\frac{m^{2/3}}{2}\left(\frac{n_1(n_1-1)n_2}{P}\right)^{2/3}\end{bmatrix}$$
      and
      $$\vc{\mu}^* = \begin{bmatrix} m^{2/3}\left(\frac{P}{n_1(n_1-1)n_2}\right)^{4/3} & 0 & 0 & 0\end{bmatrix}\text.$$
        Primal feasibility can be verified by noting that the first constraint is tight, that $x_2^*$ falls between its two constraints and $x_1^* > 0$.
        Dual feasibility is clear as $\mu_1^* >0$.
        Stationarity can be directly verified.
        Complementary slackness is satisfied as the first constraint is tight and dual variables $\mu_2^*, \mu_3^*, \mu_4^*$ are all 0.
\end{IEEEproof}

Each term in the abstract optimization problem maps to the same quantities as in~\Cref{sec:memdeplb}.
Specifically $x_1$ is the size of the projections onto each of the $m$ non-symmetric matrices, and $x_2$ is the size of the projection onto the symmetric matrix.
Here the objective function gives the total number of elements accessed, while the first constraint comes from \cref{lem:symLW}.
The final constraints are derived from the sizes of the matrices, and the minimum number of elements that a single processor must contribute to.

\subsection{Communication Lower Bounds}
\label{sec:memindeplb:lbexpression}

To derive the parallel memory independent lower bounds, we consider the atomic inner loop iterations and the number of elements required by one processor.
We begin by proving that for any load balanced parallel algorithm that handles the inner loop iterations atomically there exists a processor that performs at least its share of the atomic inner loop iteration and owns at most its share of the matrices' elements.
Then we derive lower bounds on the number of elements such a processor must access from each matrix.
Finally we apply these along with the optimization problem in \cref{lem:memindepopt} to derive the lower bound.

\begin{lemma}
  \label{lem:loadbalance}
  Consider any parallel algorithm using $P$ processors that performs $N$ atomic inner loop iterations and starts the computation with exactly one copy of $W$ words.
  If the algorithm either load balances the computation of atomic inner loop iterations or load balances the distribution of elements, there must exist a processor that performs at least $N/P$ atomic inner loop iterations, and owns at most $W/P$ words.
\end{lemma}
\begin{IEEEproof}
  To begin, we note that there must exist a processor that owns at most $W/P$ words at the start of the computation.
  If no such processor existed, then the algorithm would need to start the computation with more than one copy of some elements.
  We also note that there must be at least one processor that performs at least $N/P$ atomic inner loop iterations.
  If this were not the case, then it would be impossible to compute all $N$ atomic inner loop iterations.

  Suppose that the algorithm load balances the computation.
  Then each processor performs $N/P$ atomic inner loop iterations, and by our first observation there must exist a processor that owns at most $W/P$ words at the start of the computation.

  Alternatively, suppose that the algorithm load balances the distribution of elements.
  Then each processor owns exactly $W/P$ words and there must be a processor that performs at least $N/P$ atomic inner loop iterations by our second observation.
\end{IEEEproof}

In the following result, a generalization of \cite[Lemma 5]{ABGKR23}, we determine the minimum number of elements from each matrix that must be accessed by a processor performing a symmetric atomic three nested loop computation based on the number of atomic inner loop iterations that this processor is performing.
These lower bounds provide additional constraints to our optimization problem which allows us to determine all applicable lower bounds with a unified argument.

\begin{lemma}
	\label{lem:projlb}
	Given a parallel algorithm that computes a symmetric atomic three nested loop algorithm involving $m$ non-symmetric matrices of size $n_1\times n_2$ and a symmetric matrix of size $n_1\times n_1$ using $P$ processors, any processor that performs at least $1/P$th of the inner loop iterations associated with elements in the strict lower triangle of the symmetric matrix must access at least $n_1n_2/2P$ elements of each non-symmetric matrix and also $n_1(n_1-1)/2P$ elements of the strict lower triangle of the symmetric matrix.
\end{lemma}
\begin{IEEEproof}
  The total number of atomic inner loop iterations that must be performed involving the elements in the strict lower triangle of the symmetric matrix is $n_1(n_1-1)n_2/2$.
  Consider a processor that computes at least $1/P$th of these inner loop iterations.
  Every element of each of the non-symmetric matrices is involved in $n_1-1$ scalar multiplications.
  If the processor accesses fewer than $n_1n_2/2P$ elements of any non-symmetric matrix, then it would perform fewer than $(n_1-1) \cdot n_1n_2/2P$ atomic inner loop iterations, which is a contradiction.
  Finally, each element of the symmetric matrix is involved in $n_2$ atomic inner loop iterations.
  If the atomic inner loop iterations assigned to the processor involve fewer than $n_1(n_1-1)/2P$ elements of of the symmetric matrix, then it would perform fewer than $n_2\cdot n_1(n_1-1)/2P$ atomic inner loop iterations, which is again a contradiction.
\end{IEEEproof}

\begin{theorem}
  \label{thm:memindeplb}
  Consider a parallel symmetric atomic three nested loop algorithm using $P$ processors that starts and ends the computation with exactly one copy of $m$ non-symmetric matrices of dimensions $n_1\times n_2$ and one symmetric matrix of dimensions $n_1\times n_1$ and load balances either the inner loop iterations involving computations with elements below the diagonal of the symmetric matrix or the elements required in those inner loop iterations; it must communicate at least
  \begin{equation*}
    W-\frac{n_1(n_1-1)/2+mn_1n_2}{P}
  \end{equation*}
  words where
  \begin{equation*}
    W=\begin{cases}
    m\frac{n_2\sqrt{n_1(n_1-1)}}{P} + \frac{n_1(n_1-1)}{2} &\text{if}\qquad n_1\leq mn_2\text{ and }P \leq \frac{mn_2}{\sqrt{n_1(n_1-1)}}\\
    m n_2\sqrt{\frac{n_1(n_1-1)}{P}} + \frac{n_1(n_1-1)}{2P} &\text{if}\qquad mn_2< n_1\text{ and }P \leq \frac{n_1(n_1-1)}{m^2n_2^2}\\
    \frac{3m}{2}\left(\frac{n_1(n_1-1)n_2}{\sqrt{m}P}\right)^{2/3} &\text{if} \qquad (mn_2<n_1\text{ and }\frac{n_1(n_1-1)}{m^2n_2^2} < P)\\
    &\qquad\text{ or }(n_1\leq mn_2\text{ and }\frac{mn_2}{\sqrt{n_1(n_1-1)}} < P)\text.\\
    \end{cases}
  \end{equation*}
\end{theorem}
\begin{IEEEproof}
  To establish the lower bounds, we consider the inner loop iterations performed by a single processor, and the elements required to perform those inner loop iterations.
  We restrict our focus to the inner loop iterations and computations involving elements of the strict lower triangle of the symmetric matrix.
  Our focus will be on a processor that performs at least $1/P$th of the inner loop iterations and owns at most $1/P$th of the elements required for those inner loop iterations.
  Such a processor exists by \cref{lem:loadbalance}.

  We consider the number of elements required to perform the inner loop iterations assigned to this processor.
  Let $F$ be the set of indices $(i,j,k)$ assigned to the inner loop iterations performed by the processor.
  Then
  $$ |F| \geq \frac{n_1(n_1-1)n_2}{2P} $$
  as this processor performs at least $1/P$th of the inner loop iterations involving elements of the strict lower triangle of the symmetric matrix.
  Thus by \cref{lem:symLW}, we must have
  $$|\phi_i(F)\cup\phi_j(F)|\cdot\left(2|\phi_k(F)|\right)^{1/2} \geq 2|F| \geq\frac{n_1(n_1-1)n_2}{P}$$
  which we simplify to
  $$|\phi_i(F)\cup\phi_j(F)|^2|\phi_k(F)|\geq\left(\frac{n_1(n_1-1)n_2}{\sqrt{2}P}\right)^2\text.$$

  Considering the projections of $F$, we note that $\phi_i(F) \cup \phi_j(F)$ give the indices of the elements of the $m$ non-symmetric matrices required to perform the inner loop iterations assigned to the processor.
  Meanwhile, $\phi_k(F)$ yields the indices of the elements of the lower triangle of the symmetric matrix involved in the inner loop iterations assigned to the processor.
  We note that the number of elements from the non-symmetric matrices required to perform the inner loop iterations must be non-negative, and thus $0 \leq |\phi_i(F)\cup\phi_j(F)|$.
  Additionally, from \cref{lem:projlb}, the processor must access at least $n_1(n_1-1)/(2P)$ elements of the lower triangle of the symmetric matrix.
  Finally, the processor can not perform inner loop iterations involving more elements of the strict lower triangle of the symmetric matrix than exist.
  Thus
  $$ \frac{n_1(n_1-1)}{2P}\leq |\phi_k(F)|\leq \frac{n_1(n_1-1)}{2}\text.$$

  To minimize the communication, we want to minimize $m|\phi_i(F)\cup\phi_j(F)|+|\phi_k(F)|$ subject to the constraints above, and the result follows by \cref{lem:memindepopt}.
\end{IEEEproof}

\begin{corollary}
  {\cite[Theorem 1]{ABGKR23}}
  \label{cor:syrkmemindeplb}
  Consider the SYRK computation, $\CC \pluseq \A\cdot\A^T$, where $\A$ has dimensions $n_1\times n_2$.
  Any parallel algorithm using $P$ processors that begins with one copy of the input matrix $\A$ and ends with one copy of the strict lower triangle of $\CC$ and load balances either the computation of the elements below the diagonal of $\CC$ or the distribution of elements required to compute the elements below the diagonal of $\CC$; it must communicate at least
\begin{equation*}
  W-\frac{n_1(n_1-1)/2 + n_1n_2}{P}
\end{equation*}
words where
\begin{equation*}
  W = \begin{cases}
    \frac{n_2\sqrt{n_1(n_1-1)}}{P} + \frac{n_1(n_1-1)}{2} &\text{if}\qquad n_1\leq n_2\text{ and }P \leq \frac{n_2}{\sqrt{n_1(n_1-1)}}\\
    n_2\sqrt{\frac{n_1(n_1-1)}{P}} + \frac{n_1(n_1-1)}{2P} &\text{if}\qquad n_2< n_1\text{ and }P \leq \frac{n_1(n_1-1)}{n_2^2}\\
    \frac{3}{2}\left(\frac{n_1(n_1-1)n_2}{P}\right)^{2/3} &\text{if} \qquad \left(n_1\leq n_2\text{ and } P > \frac{n_2}{\sqrt{n_1(n_1-1)}}\right)\\
    &\qquad\text{ or } \left(n_2<n_1\text{ and } P > \frac{n_1(n_1-1)}{n_2^2} \right)\text.\\
  \end{cases}
\end{equation*}
\end{corollary}

\begin{corollary}
\label{cor:syr2kmemindeplb}
Consider the SYR2K computation, $\CC\pluseq\A^T\cdot\B + \B^T\cdot \A$ where $\A$ and $\B$ have dimensions $n_1\times n_2$.
Any atomic parallel algorithm using $P$ processors that begins with one copy of the input matrices $\A, \B$ and ends with one copy of the strict lower triangle of $\CC$ and load balances either the computation of the elements below the diagonal of $\CC$ or the distribution of elements required to compute the elements below the diagonal of $\CC$; it must communicate at least
\begin{equation*}
  W-\frac{n_1(n_1-1)/2 + 2n_1n_2}{P}
\end{equation*}
words where
\begin{equation*}
  W = \begin{cases}
    2\frac{n_2\sqrt{n_1(n_1-1)}}{P} + \frac{n_1(n_1-1)}{2} &\text{if}\qquad n_1\leq 2n_2\text{ and }P \leq \frac{2n_2}{\sqrt{n_1(n_1-1)}}\\
    2 n_2\sqrt{\frac{n_1(n_1-1)}{P}} + \frac{n_1(n_1-1)}{2P} &\text{if}\qquad 2n_2< n_1\text{ and }P \leq \frac{n_1(n_1-1)}{4n_2^2}\\
    3\left(\frac{n_1(n_1-1)n_2}{\sqrt{2}P}\right)^{2/3} &\text{if} \qquad (2n_2<n_1\text{ and }\frac{n_1(n_1-1)}{4n_2^2} < P)\\
    &\qquad\text{ or }(n_1\leq 2n_2\text{ and }\frac{2n_2}{\sqrt{n_1(n_1-1)}} < P)\text.\\
  \end{cases}
\end{equation*}
\end{corollary}

\begin{corollary}
  \label{cor:symmmemindeplb}
  Consider the SYMM computation $\CC \pluseq \A\cdot\B$ where $\A$ is a symmetric matrix of dimensions $n_1\times n_1$ and $\B$ has dimensions $n_1\times n_2$.
  Any atomic parallel algorithm using $P$ processors that begins with one copy of $\B, \CC$ and the strict lower triangle of $\A$, ends with one copy of $\CC$ and load balances either the inner loop iterations involving elements below the diagonal of $\A$ or the distribution of elements required to compute the elements of $\CC$ associated with elements below the diagonal of $\A$; it must communicate at least
  \begin{equation*}
    W-\frac{n_1(n_1-1)/2 + 2n_1n_2}{P}
  \end{equation*}
  words where
  \begin{equation*}
    W = \begin{cases}
      2\frac{n_2\sqrt{n_1(n_1-1)}}{P} + \frac{n_1(n_1-1)}{2} &\text{if}\qquad n_1\leq 2n_2\text{ and }P \leq \frac{2n_2}{\sqrt{n_1(n_1-1)}}\\
      2 n_2\sqrt{\frac{n_1(n_1-1)}{P}} + \frac{n_1(n_1-1)}{2P} &\text{if}\qquad 2n_2< n_1\text{ and }P \leq \frac{n_1(n_1-1)}{4n_2^2}\\
      3\left(\frac{n_1(n_1-1)n_2}{\sqrt{2}P}\right)^{2/3} &\text{if} \qquad (2n_2<n_1\text{ and }\frac{n_1(n_1-1)}{4n_2^2} < P)\\
        &\qquad\text{ or }(n_1\leq 2n_2\text{ and }\frac{2n_2}{\sqrt{n_1(n_1-1)}} < P).\\
    \end{cases}
  \end{equation*}
\end{corollary}

\section{Triangle Block Partitions}
\label{sec:triangularDistrib}

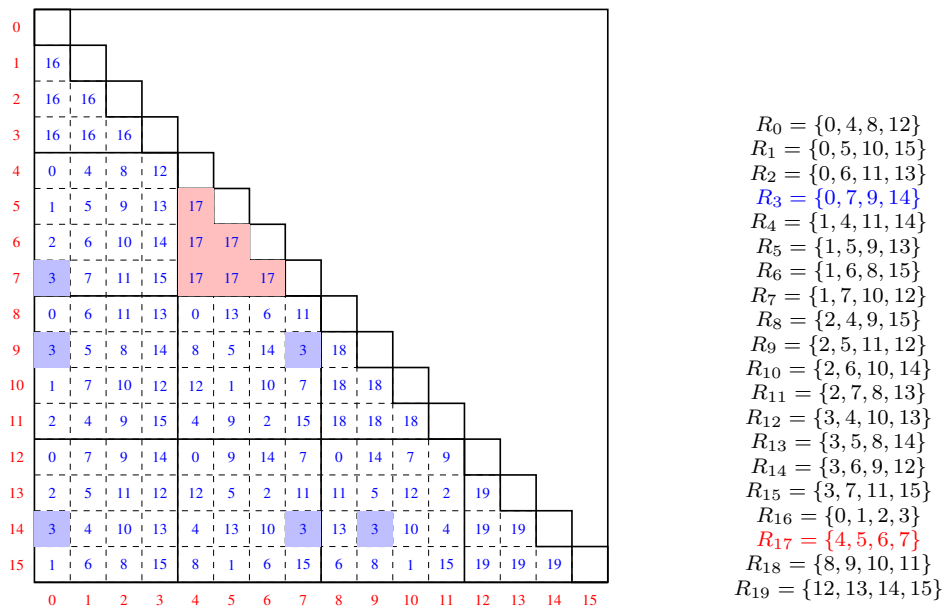
\begin{figure}
\pgfmathsetmacro{\c}{4}
\def\R{%
0/{0,4,8,12}/0,%
1/{0,5,10,15}/0,%
2/{0,6,11,13}/0,%
3/{0,7,9,14}/0,%
4/{1,4,11,14}/0,%
5/{1,5,9,13}/0,%
6/{1,6,8,15}/0,%
7/{1,7,10,12}/0,%
8/{2,4,9,15}/0,%
9/{2,5,11,12}/0,%
10/{2,6,10,14}/0,%
11/{2,7,8,13}/0,%
12/{3,4,10,13}/0,%
13/{3,5,8,14}/0,%
14/{3,6,9,12}/0,%
15/{3,7,11,15}/0,%
16/{0,1,2,3}/0,%
17/{4,5,6,7}/0,%
18/{8,9,10,11}/0,%
19/{12,13,14,15}/0%
}
\pgfmathsetmacro{\sp}{3}
\pgfmathsetmacro{\ssp}{17}

\centering
\resizebox{.5\textwidth}{!}{
%!TEX root = ../paper.tex

% set offset and matrix dim
\pgfmathsetmacro{\offset}{.33}
\pgfmathsetmacro{\m}{\c*\c}
\pgfmathsetmacro{\n}{\m-1}

% set styles
\tikzset{proclabel/.style={yscale=-1,blue}}
\tikzset{rowlabel/.style={yscale=-1,red}}

% for drawing help	
\newcommand{\gridhelp}{
	\coordinate (ll) at (-1,-1);
	\coordinate (ur) at (10,10);
	\draw[help lines,dashed] (ll) grid (ur);
	\node at (0,0) {0};
}	

%\begin{tikzpicture}[every node/.append style={transform shape},yscale=-1,scale=.675]
\begin{tikzpicture}[every node/.append style={transform shape},yscale=-1,scale=.615]
%\gridhelp
	% draw little boxes
	\foreach \x in {1,...,\m} {\draw[dashed] (0,\x) -- (\x,\x) -- (\x,\m);}
	% draw big boxes
	\foreach \x in {1,...,\c} {\draw[thick] (0,\c*\x) -- (\c*\x,\c*\x) -- (\c*\x,\m);}
	% draw diagonal
	\foreach \x in {1,...,\m} {
		\draw[thick] (\x-1,\x-1) -- (\x,\x-1) -- (\x,\x);
		\draw[thick] (\x-1,\x-1) -- (\x-1,\x) -- (\x,\x);
	}
	% draw outer box
	\draw[thick] (0,0) rectangle +(\m,\m);
	% draw processor labels
	\foreach \p/\list/\d in \R {
		% off-diagonal blocks
		\foreach \x in \list {
			\foreach \y in \list {
				\ifnum \x<\y 
					\ifnum \p=\sp
						\fill [blue!25] (\x,\y) rectangle (\x+1,\y+1);
					\fi
					\ifnum \p=\ssp
						\fill [red!25] (\x,\y) rectangle (\x+1,\y+1);
					\fi
					\node[proclabel] at (\x+.5,\y+.5) {\p};
				\fi
			}
		}
		% diagonal block
		\ifnum \d>0
			\node[proclabel] at (\d-.5,\d-.5) {\p};
		\fi
	}
	% draw row/col labels
	\foreach \x in {0,...,\n} {
		\node[rowlabel] at (-.5,\x+.5) {\x};
		\node[rowlabel] at (\x+.5,\m+.5) {\x};
	}
\end{tikzpicture}
} \hspace{1cm}
\begin{tikzpicture}
\pgfmathsetmacro{\vscl}{.325}
	\foreach \p/\list/\d in \R {
		\ifnum \p=\sp
			\node[blue] at (0,\vscl*19-\vscl*\p) {\scriptsize $R_{\p} = \{\list\}$};
		\else
			\ifnum \p=\ssp
				\node[red] at (0,\vscl*19-\vscl*\p) {\scriptsize $R_{\p} = \{\list\}$};
			\else
				\node at (0,\vscl*19-\vscl*\p) {\scriptsize $R_{\p} = \{\list\}$};
			\fi
		\fi
	}
\end{tikzpicture}
\caption{Triangle block partition for $n_1=16$ and $|R_k|=4$.  Triangle blocks for $R_{\sp}$ and $R_{\ssp}$ are highlighted to illustrate both non-contiguous and contiguous triangle blocks.}
\label{fig:TBD-ex}
\pgfmathsetmacro{\sp}{-1}
\pgfmathsetmacro{\ssp}{-1}
\end{figure}

The key to achieving the symmetric communication lower bounds is the observation that the lower triangle of a symmetric matrix can be partitioned into triangle blocks \cite{BELV22}.
Let $R$ be a subset of the indices of the rows of the symmetric matrix, $R\subseteq \{0,\ldots,n_1-1\}$, then the triangle block defined by $R$ is denoted by $TB(R)=\{(i,j) \; | \; i,j\in R, i > j\}$.
For example, given a set of indices $R=\{1,2,4\}$, the corresponding triangle block is the set of pairs $TB(R) = \{(2,1),(4,1),(4,2)\}$.
If $R= \{0,\ldots,n_1-1\}$, then $TB(R)$ is the entire strict lower triangle.
We will see that for many values of $n_1$, we can achieve balanced partitions of the strict lower triangle into triangle blocks.
An example triangle block partition for $n_1=16$ using 20 triangle blocks corresponding to sets of 4 elements each is given in \cref{fig:TBD-ex}.

The cyclic $(c,k)$-indexing family introduced by \cite{BELV22}, of which a special case is used by \cite{ABGKR23,ABC+23}, is one method of partitioning the lower triangle of the symmetric matrix into triangle blocks.
A $(c,k)$-indexing family decomposes the symmetric matrix into $k(k-1)/2$ square zones of dimensions $c\times c$ and $k$ triangular zones on the diagonal.
The triangle blocks specified by a valid $(c,k)$-indexing family \cite[Definition 5.2]{BELV22} contain exactly one element from each off diagonal zone.
One valid $(c,k)$-indexing family is the cyclic $(c,k)$-indexing family given in \cite[Definition 5.4]{BELV22} when $c$ is co-prime with all positive integers less than $k-1$.

Additional families of triangle block partitions come from solutions to the clique partitioning problem for the complete graph with $n_1$ vertices.
To see the relationship between generating triangle blocks and the clique partitioning problem, we begin by noting that each element in the strict lower triangle of the symmetric matrix corresponds to the edge in the complete graph connecting the vertex associated with its row index and column index.
Additionally, note that the triangle block over a set of indices corresponds to the edges of the complete graph connecting the nodes associated with those vertices.
Thus, partitioning the strict lower triangle of the symmetric matrix of dimensions $n_1\times n_1$ into triangle blocks is equivalent to the problem of partitioning the edges of a complete graph on $n_1$ vertices into cliques.

\subsection{Clique Partitions of $K_{n}$}
\label{sec:cliquePartitions}
Clique partitions of graphs have long been studied by mathematicians, who have been particularly interested in the smallest number of cliques required to non-trivially partition the edges of a graph.
De Bruijn and Erd{\"o}s~\cite[Theorem 1]{dBE48} first determined that the minimum number of cliques required to partition the complete graph $K_{n}$ on $n$ vertices was $n$.
When $n=c^2+c+1$ for a prime power $c$, Wallis~\cite{Wallis82} gave a novel construction of a minimal clique partition of $K_n$ where all cliques are the same size using projective geometry.

The construction given by Wallis is of particular interest as clique partitions where all cliques are the same size coincide with partitions of the lower triangle of the symmetric matrix where all subsets in the partition have equal size .
This yields balanced sub-computations in the sequential setting and load balanced algorithms in the parallel setting.
From Wallis' construction of a clique partition using projective geometry, one can derive a balanced clique partition using the relationship between projective geometry for $n=c^2+c+1$ and affine geometry for $n=c^2$.
Note that the derived clique partition for affine geometry is not minimal as it contains $c^2+c$ cliques.

In this section, we remind the readers of the properties of affine and projective geometry over finite fields that are required to understand these constructions.
In particular, we will define affine and projective planes over finite fields, define lines in these planes, count the number of points and lines in each type of plane, and count the number of points on each line.
For a more complete treatment, we refer the reader to the introductions to projective geometry in either Gibson \cite{Gibson98} or Silverman and Tate \cite[Appendix A]{ST92}.

To begin, we recall that a finite field is a finite set with two operations, addition and multiplication, that satisfy the field axioms of associativity, commutativity, distributivity, identity and inverses.
We denote a finite field with $c$ elements by $\mathbb{F}_c$.
Example finite fields include the integers modulo $c$ for any prime number $c$, under the standard operations of modular addition and multiplication.
In particular, there exists a finite field of order $c$ if and only if $c$ is a prime power.

Now we will define projective and affine planes over finite fields.
\begin{definition}[{\cite[Example 2.6]{Gibson98}}]
  \label{def:affplane}
  Let $c$ be a prime power and $\mathbb{F}_c$ a finite field with $c$ elements.
  The affine plane over $\mathbb{F}_c$ is $\mathbb{A}^2(\mathbb{F}_c)=\{(a_1,a_2) \; | \; a_1,a_2\in \mathbb{F}_c\}$.
\end{definition}
Note that $|\mathbb{A}^2(\mathbb{F}_c)| = c^2$.
\begin{definition}[{\cite[Example 2.6]{Gibson98}}]
  \label{def:affline}
  A line in $\mathbb{A}^2(\mathbb{F}_c)$ is any subset of points of the form $\{(x_1,x_2)\in\mathbb{A}^2(\mathbb{F}_c) \; | \; ax_1+bx_2+d=0\}$ for some $a,b,d\in\mathbb{F}_c$ where either $a\neq 0$ or $b\neq 0$.
\end{definition}
We deduce that a line contains exactly $c$ points by noting that if $a\neq 0$, for each $x_2\in\mathbb{F}_c$ there exists a unique $x_1$ that will satisfy $ax_1+bx_2+d=0$.
Similarly, if $b\neq 0$, there is a unique solution $x_2$ given $x_1$.
As noted in \cite[Example 2.6]{Gibson98}, one can prove that any pair of distinct points in $\mathbb{A}^2(\mathbb{F}_c)$ are contained in a unique line.
To determine how many lines are in the affine plane we note that there are $\binom{c^2}{2}$ ways of choosing two distinct points in $\mathbb{A}^2(\mathbb{F}_c)$, and there are $\binom{c}{2}$ ways of selecting two distinct points from the $c$ points on a line.
Thus there are $\binom{c^2}{2}/\binom{c}{2}=c^2+c$ lines in $\mathbb{A}^2(\mathbb{F}_c)$.

\begin{definition}[{\cite[Page 109]{Gibson98}}]
  \label{def:projplane}
  Let $c$ be a prime power and $\mathbb{F}_c$ a finite field with $c$ elements.
  The projective plane over $\mathbb{F}_c$, $\mathbb{P}^2(\mathbb{F}_c)$, is the set $\{(a_1,a_2,a_3) \; | \; a_1,a_2,a_3\in\mathbb{F}_c, (a_1,a_2,a_3)\neq (0,0,0)\}$ under the equivalence relation $(a_1,a_2,a_3)\sim\lambda(a_1,a_2,a_3)$ for any $\lambda\in\mathbb{F}_c$, $\lambda\neq 0$.
  We use the homogeneous coordinates $(a_1:a_2:a_3)$ to denote the equivalence class of elements containing $(a_1,a_2,a_3)$.
\end{definition}

Given the equivalence relationship used to define $\mathbb{P}^2(\mathbb{F}_c)$, we can write each point using a normalized form where the last non-zero element is 1.
Then one can deduce that $|\mathbb{P}^2(\mathbb{F}_c)| = c^2+c+1$ as there are $c^2$ elements with normalized form $(a_1:a_2:1)$, $c$ elements with normalized form $(a_1:1:0)$ and one element with normalized form $(1:0:0)$.
We note that there is an injective map from $\mathbb{A}^2(\mathbb{F}_c)$ to $\mathbb{P}^2(\mathbb{F}_c)$ where a point $(a_1, a_2)$ is mapped to the equivalence class $(a_1:a_2:1)$.
The points in $\mathbb{P}^2(\mathbb{F}_c)$ that are not in the image of this map are precisely those points where the final coordinate is 0.

\begin{definition}[{\cite[Page 110]{Gibson98}}]
  \label{def:projline}
  A line in $\mathbb{P}^2(\mathbb{F}_c)$ is any subset of the form $\{(x_1:x_2:x_3)\in\mathbb{P}^2(\mathbb{F}_c) \; | \; ax_1+bx_2+dx_3=0\}$ for some $a,b,d\in\mathbb{F}_c$ where at least one of $a,b,d$ is not 0.
\end{definition}
We can count the number of points on each line by considering how many equivalence classes it contains.
If $a=b=0$ and the line is of the form $dx_3=0$, then none of the points on the line may fall in the embedded affine plane as all points must take the form $(x_1:x_2:0)$.
Thus that line contains the $c$ points with normalized form $(x_1:1:0)$ and the single point with normalized form $(1:0:0)$ for $c+1$ points.

If $a\neq 0$ or $b\neq 0$, the equivalence classes $(x_1,x_2,1)$ correspond to the $c$ elements on the line $ax_1+bx_2+d=0$ in the embedded affine plane.
Thus a line with $(a,b)\neq(0,0)$, contains $c$ equivalence classes of the form $(x_1,x_2,1)$.
Additionally, such a line contains the single equivalence $(x_1:x_2:0)$ that satisfies $ax_1+bx_2=0$, and thus the line contains $c+1$ points.
Therefore, any line in $\mathbb{P}^2(\mathbb{F}_c)$ must contain $c+1$ points.

As with the points, there in an injective map from the set of lines in $\mathbb{A}^2(\mathbb{F}_c)$ to the set of lines in $\mathbb{P}^2(\mathbb{F}_c)$ where the line $ax_1+bx_2+d = 0$ is mapped to $ax_1+bx_2+dx_3=0$.
To determine the number of lines in the projective plane one can note that there are $c^2+c$ lines in the affine plane which correspond to lines where $a\neq 0$ or $b\neq 0$, and one line which does not intersect the affine plane which corresponds to the line with $a=b=0$ for a total of $c^2+c+1$ lines.
We also note that, as in affine planes, given any pair of distinct points in $\mathbb{P}^2(\mathbb{F}_c)$, there is a unique line that contains that pair of points \cite[Example 9.3]{Gibson98}.

We now return to the results on clique partitions discussed at the beginning of this section.
The following theorem of de Bruijn and Erd{\"o}s \cite[Theorem 1]{dBE48} bounds the number of cliques in a partition of the complete graph on $n$ vertices.
We use the statement given by Wallis \cite{Wallis82} in the discussion in the final paragraph of page 99 as it most clearly states the theorem in terms of the projective plane construction which we will use.
\begin{theorem}[\cite{dBE48}{\cite[Page 99]{Wallis82}}]
\label{thm:clique-partition}
  Any clique partition of $K_n$ other than the trivial one-clique case, must contain at least $n$ cliques, and $n$ is attained in precisely two ways:
  \begin{enumerate}
  \item the cliques are of size $n-1$ (once) and 2 ($(n-1)$ times).
  \item $n=c^2+c+1$, there is a finite projective plane with $n$ points, and the cliques are all of size $c+1$ (corresponding to lines).
  \end{enumerate}
\end{theorem}
To understand the second construction, think of the graph $G$ whose vertices are the points in the projective plane and whose edges are given by the set of all line segments that connect two points in the projective plane.
As we just counted, there are $c^2+c+1$ points in $\mathbb{P}^2(\mathbb{F}_c)$ for a prime power $c$, and, as there is a unique line that contains any pair of distinct points in $\mathbb{P}^2(\mathbb{F}_c)$ there is a unique edge between every pair of points in this graph.
There are $c+1$ points on each line, every pair of points on the line has an edge between them which is defined by that line, thus the lines define cliques of size $c+1$.
Finally, as the edges are constructed from the lines, the lines must partition the edges.
Thus the lines define a clique partition of the graph $G$.
These lines and the points on them can be enumerated algorithmically using a computer algebra system; we use Magma \cite{BCP97}.

To construct the clique partition for the affine plane, we can follow the same logic as the clique partition for the projective plane or inherit the clique partition from the projective construction.
We construct the graph $G$ whose vertices are the points in the affine plane and whose edges are given by the set of all line segments that connect two points in the affine plane.
This graph has $c^2$ vertices as there are $c^2$ points in the affine plane.
As there are $c^2+c$ lines each containing $c$ points in the affine plane, there are $c^2+c$ cliques each containing $c$ vertices.
To derive this construction from the projective construction recall our injective map from the lines in the affine plane over a finite field to the lines in the projective plane over the same finite field.
There is one line in the projective plane that is not in the image of this map.
By removing this line and all the points on it, we are left with the lines in the projective plane that correspond to lines in the affine plane and the points in the projective plane that correspond to points in the affine plane.
While it is not necessary to use this derived construction for the affine solution to the clique partition problem, it is convenient when comparing projective and affine constructions over the same finite field.

\subsection{Triangle Block Partitions of the Strict Lower Triangle of a Matrix}

Recall that by associating element $(i,j)$ in the strict lower triangle of a matrix of dimensions $n_1\times n_1$ to the edge between vertices $i$ and $j$ in $K_{n_1}$, we can use a clique partition of a complete graph to create a triangle block partition of the strict lower triangle of a square matrix.
In particular, we can use the solutions to the clique partition problem discussed in the previous section to generate triangle block partitions of the strict lower triangle of a square matrix.

\begin{figure}[htb]
	\centering
	%!TEX root = ../paper.tex

% set offset
\pgfmathsetmacro{\offset}{.33}
% set styles
\tikzset{proclabel/.style={yscale=-1,blue}}
\tikzset{rowlabel/.style={yscale=-1,red}}

% for drawing help	
\newcommand{\gridhelp}{
	\coordinate (ll) at (-1,-1);
	\coordinate (ur) at (10,10);
	\draw[help lines,dashed] (ll) grid (ur);
	\node at (0,0) {0};
}

%\begin{tikzpicture}[every node/.append style={transform shape},yscale=-1,scale=.675]
\begin{tikzpicture}[every node/.append style={transform shape},yscale=-1,scale=.55]
%\gridhelp
        \begin{scope}[shift={(-5,0)}]
	% draw little boxes
	\foreach \x in {1,...,9} {\draw[dashed] (0,\x) -- (\x,\x) -- (\x,9);}
	% draw big boxes
	\foreach \x in {3,6} {\draw[thick] (0,\x) -- (\x,\x) -- (\x,9);}
	% draw diagonal
	\foreach \x in {1,...,9} {\draw[thick] (\x-1,\x-1) -- (\x,\x-1) -- (\x,\x);	}
	% draw outer box
	\draw[thick] (0,0) rectangle +(9,9);
	% draw processor labels
	\foreach \x/\y in {0/3,0/6,3/6,3/3} {\node[proclabel] at (\x+.5,\y+.5) {0};}
	\foreach \x/\y in {0/4,0/8,4/8} {\node[proclabel] at (\x+.5,\y+.5) {1};}
	\foreach \x/\y in {0/5,0/7,5/7} {\node[proclabel] at (\x+.5,\y+.5) {2};}
	\foreach \x/\y in {1/3,1/8,3/8,1/1} {\node[proclabel] at (\x+.5,\y+.5) {3};}
	\foreach \x/\y in {1/4,1/7,4/7,7/7} {\node[proclabel] at (\x+.5,\y+.5) {4};}
	\foreach \x/\y in {1/5,1/6,5/6,6/6} {\node[proclabel] at (\x+.5,\y+.5) {5};}
	\foreach \x/\y in {2/3,2/7,3/7,2/2} {\node[proclabel] at (\x+.5,\y+.5) {6};}
	\foreach \x/\y in {2/4,2/6,4/6,4/4} {\node[proclabel] at (\x+.5,\y+.5) {7};}
	\foreach \x/\y in {2/5,2/8,5/8} {\node[proclabel] at (\x+.5,\y+.5) {8};}
	\foreach \x/\y in {0/1,0/2,1/2,0/0} {\node[proclabel] at (\x+.5,\y+.5) {9};}
	\foreach \x/\y in {3/4,3/5,4/5,5/5} {\node[proclabel] at (\x+.5,\y+.5) {10};}
	\foreach \x/\y in {6/7,6/8,7/8,8/8} {\node[proclabel] at (\x+.5,\y+.5) {11};}
	% draw row/col labels
	\foreach \x in {0,...,8} {
		\node[rowlabel] at (-.5,\x+.5) {\x};
		\node[rowlabel] at (\x+.5,9+.5) {\x};
	}
        \end{scope}
        \begin{scope}[shift={(5,0)}]
	% draw little boxes
	\foreach \x in {1,...,13} {\draw[dashed] (0,\x) -- (\x,\x) -- (\x,13);}
	% draw big boxes
	\foreach \x in {3,6,9,12} {\draw[thick] (0,\x) -- (\x,\x) -- (\x,13);}
	% draw diagonal
	\foreach \x in {1,...,13} {\draw[thick] (\x-1,\x-1) -- (\x,\x-1) -- (\x,\x);	}
	% draw outer box
	\draw[thick] (0,0) rectangle +(13,13);
	% draw processor labels
	\foreach \x/\y in {0/3,0/6,0/9,3/6,3/9,6/9,3/3} {\node[proclabel] at (\x+.5,\y+.5) {0};}
	\foreach \x/\y in {0/4,0/8,0/10,4/8,4/10,8/10,10/10} {\node[proclabel] at (\x+.5,\y+.5) {1};}
	\foreach \x/\y in {0/5,0/7,0/11,5/7,5/11,7/11,11/11} {\node[proclabel] at (\x+.5,\y+.5) {2};}
	\foreach \x/\y in {1/3,1/8,1/11,3/8,3/11,8/11,1/1} {\node[proclabel] at (\x+.5,\y+.5) {3};}
	\foreach \x/\y in {1/4,1/7,1/9,4/7,4/9,7/9,7/7} {\node[proclabel] at (\x+.5,\y+.5) {4};}
	\foreach \x/\y in {1/5,1/6,1/10,5/6,5/10,6/10,6/6} {\node[proclabel] at (\x+.5,\y+.5) {5};}
	\foreach \x/\y in {2/3,2/7,2/10,3/7,3/10,7/10,2/2} {\node[proclabel] at (\x+.5,\y+.5) {6};}
	\foreach \x/\y in {2/4,2/6,2/11,4/6,4/11,6/11,4/4} {\node[proclabel] at (\x+.5,\y+.5) {7};}
	\foreach \x/\y in {2/5,2/8,2/9,5/8,5/9,8/9,9/9} {\node[proclabel] at (\x+.5,\y+.5) {8};}
	\foreach \x/\y in {0/1,0/2,0/12,1/2,1/12,2/12,0/0} {\node[proclabel] at (\x+.5,\y+.5) {9};}
	\foreach \x/\y in {3/4,3/5,3/12,4/5,4/12,5/12,5/5} {\node[proclabel] at (\x+.5,\y+.5) {10};}
	\foreach \x/\y in {6/7,6/8,6/12,7/8,7/12,8/12,8/8} {\node[proclabel] at (\x+.5,\y+.5) {11};}
	\foreach \x/\y in {9/10,9/11,9/12,10/11,10/12,11/12,12/12} {\node[proclabel] at (\x+.5,\y+.5) {12};}
	% draw row/col labels
	\foreach \x in {0,...,12} {
		\node[rowlabel] at (-.5,\x+.5) {\x};
		\node[rowlabel] at (\x+.5,13+.5) {\x};
	}
        \end{scope}

\end{tikzpicture}
	\caption{The triangle block partitions of the lower triangle defined by the affine and projective constructions for $n_1=9$, $c=3$ and $n_1=13$, $c=3$, respectively. The affine and projective constructions have 12 and 13 triangle blocks, respectively. Each entry of the lower triangle is marked with a triangle block from which it belongs. For example, $(5,0),(7,0)$ and $(7,5)$ entries belong to the $2$nd triangle block in the affine construction. Diagonal elements are assigned in a compatible way with the triangle blocks.}
	\label{fig:TBD-3}
\end{figure}
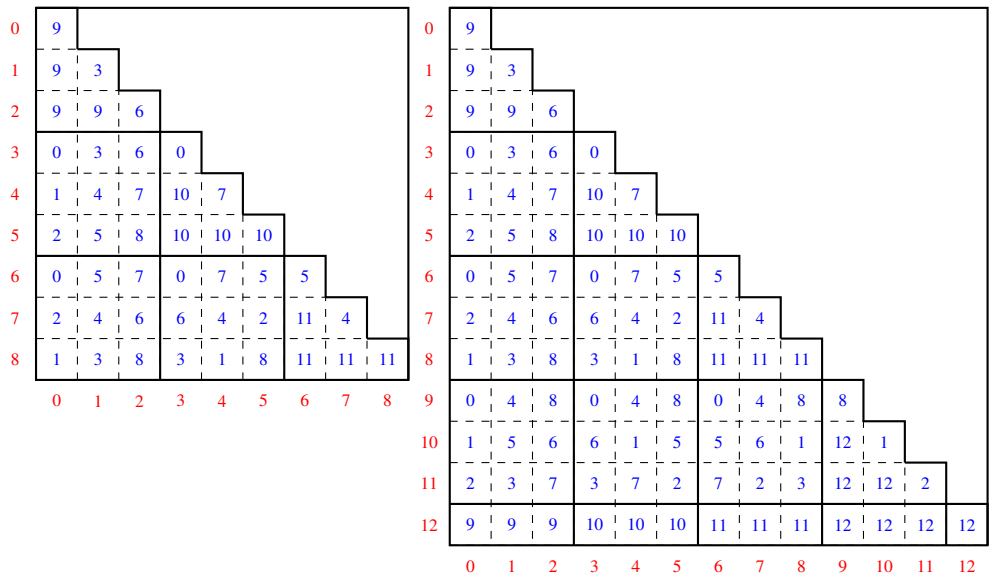

When $n_1 = c^2+c+1$ for a prime power $c$, the projective construction gives a partition of the strict lower triangle of the matrix into $c^2+c+1$ triangle blocks each defined by a set $R_k$ with $|R_k|=c+1$.
When $n_1 = c^2$ for a prime power $c$, the affine construction gives a partition of the strict lower triangle of the matrix into $c^2+c$ triangle blocks each defined by a set $R_k$ with $|R_k|=c$.
Unlike with the $(c,k)$-indexing families, there is not a clear pattern to the indices of the triangle block constructed using either the projective or affine plane constructions.
Thus we do not specify the triangle blocks in general using functions like was possible with the $(c,k)$-indexing families. \Cref{fig:TBD-3} shows triangle block partitions by the affine and projective constructions for $c=3$.

We can extend these constructions of triangle block partitions from lines in finite projective and affine planes to higher dimensional finite projective and affine spaces.
The projective space of dimension $\alpha$ over the finite field of order $c$, $\mathbb{P}^\alpha(\mathbb{F}_c)$, contains $(c^{\alpha+1}-1)/(c-1)$ points, and, as in the case of projective planes where $\alpha = 2$, lines have $c+1$ points on them.
Thus, it is possible to partition the strict lower triangle of a matrix of dimensions $n_1\times n_1$ where $n_1=(c^{\alpha+1}-1)/(c-1)$ for any prime power $c$, $\alpha\geq 2$ into triangle blocks where the sets $\{R_k\}$ which define the blocks all have size $c+1$.
Similarly, the affine space of dimension $\alpha$ over the finite field of order $c$, $\mathbb{A}^\alpha(\mathbb{F}_c)$ contains $c^\alpha$ points, and lines all have $c$ points on them.
Hence, it is possible to partition the strict lower triangle of a matrix of dimensions $n_1\times n_1$ where $n_1=c^\alpha$ for any prime power $c$, $\alpha\geq 2$ into triangle blocks where the sets $\{R_k\}$ that define the blocks all have size $c$.
As before, we do this by associating the elements $\{0,1,\ldots,n_1-1\}$ with points in the affine or projective space and define the sets $\{R_k\}$ to be the sets of collinear points.
As with the planar triangle block constructions (where $\alpha=2$), the sets can be enumerated efficiently with a computer algebra system.

Both the cyclic $(c,k)$-indexing families of \cite{BELV22} and the triangle block partitions based upon the sets of lines in affine and projective spaces are special cases of Steiner systems.
\begin{definition}{\cite[Section 3]{Gowers2017}}
A Steiner $(n, r, s)$-system is a collection $\Sigma$ of subsets of size $r$ from the set $S=\{0,\ldots,n-1\}$ such that every subset of $S$ of size $s$ is contained in exactly one set from $\Sigma$.
\end{definition}
Steiner systems where $s=2$, also called balanced incomplete block designs, correspond to triangle block partitions.

Our triangle block partitions constructed from projective spaces correspond to Steiner $(c^{\alpha+1}-1)/(c-1), c+1, 2)$ systems, while the triangle block partitions constructed from affine space constructions correspond to Steiner $(c^\alpha, c, 2)$ systems.
The triangle block partitions constructed with cyclic $(c,k)$-indexing families are Steiner $(ck,c,2)$ systems.

While general results on the existence of Steiner systems are due to Keevash~\cite{Keevash2014}, existence results for $s=2$ are due to Wilson~\cite{Wilson1970,Wilson1972,Wilson1975}.

\begin{theorem}{\cite[Corollary A]{Wilson1975}}
\label{thm:divConditions}
Given a positive integer $r$, there exists an $n_0\in\mathbb{Z}$ such that Steiner $(n,r,2)$ systems exist for all $n \geq n_0$ for which $r-1$ divides $n-1$ and $r(r-1)$ divides $n(n-1)$.
\end{theorem}

The existence of the more general Steiner $(n,r,2)$ systems allows us to decouple the matrix dimension from the size of the triangle blocks under the requirement that the divisibility conditions be satisfied.
As any Steiner $(n,r,2)$ is a solution to the clique partition problem for $K_n$, $|\Sigma| \geq n$ if $n>r$ by \cref{thm:clique-partition}.
Note that the $|R_k|=r$ for all $k$, and each element appears in $(n-1)/(r-1)$ sets.
An example of triangle block partition using the Steiner $(15, 3, 2)$ system is provided in \cref{sec:appendix:Steiner-example}.

\subsection{Assignment of Diagonal Elements}

In the previous subsections, we discussed families of triangle block partitions of the strict lower triangle of the symmetric matrix.
We also need to assign diagonal elements in such a way that they are consistent with the triangle blocks, and ideally each triangle block has almost the same number of diagonal elements assigned to it.
Let $x$ be the row and column index of a diagonal element.
Assigning a diagonal element in a compatible way with triangle blocks requires assigning the diagonal element to a triangle block where the set $R_k$ that defines the triangle block contains $x$.
As we do not specify the sets algebraically, it is not feasible to write a function to assign the diagonal blocks as it was with the cyclic $(c,k)$-indexing families.
We instead prove that an assignment must exist where each diagonal element is assigned to a unique triangle block and that there are efficient algorithms for generating such an assignment.

The next theorem is used to prove that we are able to assign diagonal elements in a consistent manner with any triangle block partition derived from a Steiner $(n,r,2)$ system.

\begin{theorem}[Hall's Marriage Theorem, {\cite[Theorem 2.1.2]{Diestel2017}}]
	\label{thm:hall:graphformulation}
	Let $G(X,Y,E)$ be a finite bipartite graph with bipartite sets $X$ and $Y$ and edge set $E$. There exists a matching with a set of disjoint edges that covers every vertex in $X$ iff $|W| \leq |N_G(W)|$ for every subset $W\subseteq X$. Here $N_G(W)$ denotes the set of all vertices in $Y$ that are adjacent to at least one vertex of $W$.
\end{theorem}

Recall the sets $\{R_k\}$ that define the triangle blocks, and let $K$ be the number of triangle blocks.
Consider the bipartite graph  $G(X, Y, E)$ with $X =\{0,\ldots,n_1-1\}$ corresponding to diagonal elements, $Y = \{0,\ldots,K-1\}$ corresponding to triangle blocks, and $E = \{(i,k)\in X\times Y: i \in R_k\}$.
One can easily see that finding an assignment of diagonal elements to triangle blocks is equivalent to finding a matching with a set of disjoint edges in $G$ that covers all the vertices of $X$.

\begin{theorem}
	\label{thm:bipartitegraph:matching}
	Let $G$ be the bipartite graph constructed from the triangle blocks derived from a Steiner $(n,r,2)$ system where $n>r$.
	There exists a matching with a set of disjoint edges in $G$ that covers all the vertices of $X$.
\end{theorem}
\begin{IEEEproof}
	It is sufficient to show that $|W| \leq |N_{G}(W)|$ for every subset $W\subseteq X$ by \Cref{thm:hall:graphformulation}.
	From the construction, we know that each set $R_k$ which defines a triangle block contains $r$ elements in $X$, and each element in $X$ is contained in the set $R_k$ for $(n-1)/(r-1)$ triangle blocks.
	Thus for any $W\subseteq X$, $|N_G(W)| \geq \frac{n-1}{r(r-1)}|W|$.
	Hence we need only to show that $\frac{n-1}{r(r-1)}\geq 1$.
	Recall that there are $\frac{n(n-1)}{r(r-1)}$ triangle blocks, and that there must be at least $n$ triangle blocks.
	Thus $\frac{n(n-1)}{r(r-1)}\geq n$ and so $\frac{n-1}{r(r-1)}\geq 1$.
\end{IEEEproof}

We can use maximum cardinality matching algorithms, such as the Ford–Fulkerson algorithm~\cite{Ford87} and the Hopcroft-Karp algorithm~\cite{HK73}, on $G$ to obtain a set of disjoint edges. Each of these edges represents the assignment of a diagonal element to a unique triangular block.

We note that the only family of triangle block constructions where all triangle blocks are assigned exactly one diagonal element are precisely those that come from the projective plane construction.
For all other triangle block distributions, there are fewer diagonal elements than triangle blocks which will prevent perfect load balance of the diagonal element computation among the blocks.

\subsection{Triangle Block Partitions for Sequential and Parallel Algorithms}
\label{sec:TBD}

In this section we demonstrate how to apply the theoretical results presented above to design triangle block partitions that are used by algorithms for the SYRK, SYR2K and SYMM computations. We illustrate the underlying principles using SYMM but they can be used similarly for the other computations by applying the same triangle block partitions to the symmetric and non-symmetric matrices. For the simplicity of the explanations, we detail the elaboration of triangle block partitions based on the affine construction associated to a prime power $c$ and assuming $n_{1} = c^{2}$. The scheme associating elements to triangle blocks and the definitions presented below are however fully applicable to any triangle block partition.

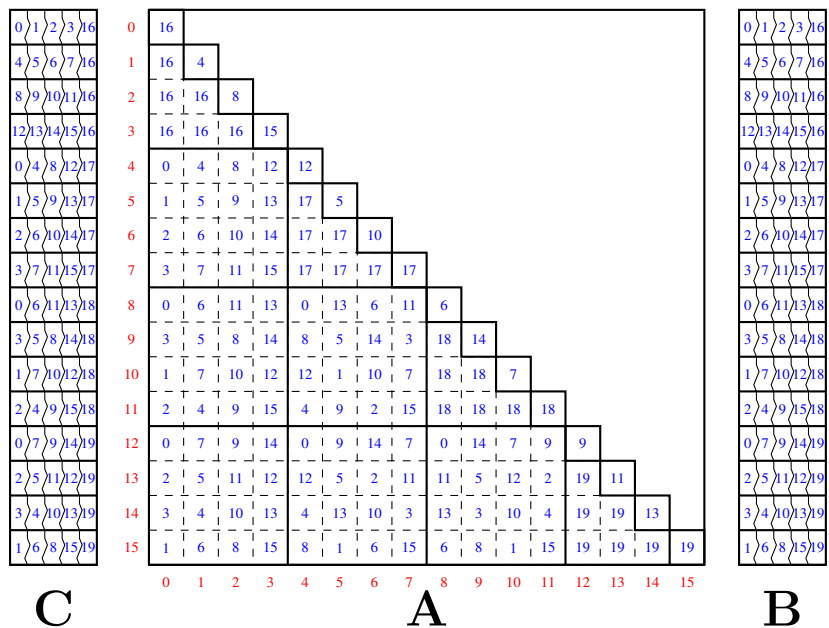
\begin{figure}[h]
	\pgfmathsetmacro{\c}{4}
	\def\R{%
		0/{1,5,9,13}/0,%
		1/{1,6,11,16}/0,%
		2/{1,7,12,14}/0,%
		3/{1,8,10,15}/0,%
		4/{2,5,12,15}/2,%
		5/{2,6,10,14}/6,%
		6/{2,7,9,16}/9,%
		7/{2,8,11,13}/11,%
		8/{3,5,10,16}/3,%
		9/{3,6,12,13}/13,%
		10/{3,7,11,15}/7,%
		11/{3,8,9,14}/14,%
		12/{4,5,11,14}/5,%
		13/{4,6,9,15}/15,%
		14/{4,7,10,13}/10,%
		15/{4,8,12,16}/4,%
		16/{1,2,3,4}/1,%
		17/{5,6,7,8}/8,%
		18/{9,10,11,12}/12,%
		19/{13,14,15,16}/16%
	}
	\centering
	%!TEX root = ../paper.tex

% set offset and matrix dim
\pgfmathsetmacro{\offset}{.33}
\pgfmathsetmacro{\m}{\c*\c}
\pgfmathsetmacro{\n}{\m-1}

% set styles
\tikzset{proclabel/.style={yscale=-1,blue}}
\tikzset{rowlabel/.style={yscale=-1,red}}

% for drawing help	
\newcommand{\gridhelp}{
	\coordinate (ll) at (-1,-1);
	\coordinate (ur) at (10,10);
	\draw[help lines,dashed] (ll) grid (ur);
	\node at (0,0) {0};
}	

%\begin{tikzpicture}[every node/.append style={transform shape},yscale=-1,scale=.675]
\begin{tikzpicture}[every node/.append style={transform shape},yscale=-1,scale=.4615]
%\gridhelp
	% draw little boxes
	\foreach \x in {1,...,\m} {\draw[dashed] (0,\x) -- (\x,\x) -- (\x,\m);}
	% draw big boxes
	\foreach \x in {1,...,\c} {\draw[thick] (0,\c*\x) -- (\c*\x,\c*\x) -- (\c*\x,\m);}
	% draw diagonal
	\foreach \x in {1,...,\m} {
		\draw[thick] (\x-1,\x-1) -- (\x,\x-1) -- (\x,\x);
		\draw[thick] (\x-1,\x-1) -- (\x-1,\x) -- (\x,\x);
	}
	% draw outer box
	\draw[thick] (0,0) rectangle +(\m,\m);
	% draw processor labels
	\foreach \p/\list/\d in \R {
		% off-diagonal blocks
		\foreach \x in \list {
			\foreach \y in \list {
				\ifnum \x<\y 
					\node[proclabel] at (\x-.5,\y-.5) {\p};
				\fi
			}
		}
		% diagonal block
		\ifnum \d>0
			\node[proclabel] at (\d-.5,\d-.5) {\p};
		\fi
	}
	% draw row/col labels
	\foreach \x in {0,...,\n} {
		\node[rowlabel] at (-.5,\x+.5) {\x};
		\node[rowlabel] at (\x+.5,\m+.5) {\x};
	}

	% label A
	\node[yscale=-1,scale=2] at (8,17.25) {\LARGE $\A$};
	
	\begin{scope}[shift={(-4,0)}]
		% draw row blocks
		\foreach \x in {1,...,16} {
			\foreach \y in {1,...,4} { \draw[decorate,decoration={zigzag,amplitude=1pt}] (\y/2,\x) -- (\y/2,\x-1);}
			\draw[thick] (0,\x) -- (2.5,\x);
		}
		% draw outer box
		\draw[thick] (0,0) rectangle +(2.5,16);
		% draw processor labels
		\foreach \x/\y in {1/0,2/1,3/2,4/3,5/16} {\node[proclabel] at (\x*.5-.25,.5) {\y};}
		\foreach \x/\y in {1/4,2/5,3/6,4/7,5/16} {\node[proclabel] at (\x*.5-.25,1.5) {\y};}
		\foreach \x/\y in {1/8,2/9,3/10,4/11,5/16} {\node[proclabel] at (\x*.5-.25,2.5) {\y};}
		\foreach \x/\y in {1/12,2/13,3/14,4/15,5/16} {\node[proclabel] at (\x*.5-.25,3.5) {\y};}
		\foreach \x/\y in {1/0,2/4,3/8,4/12,5/17} {\node[proclabel] at (\x*.5-.25,4.5) {\y};}
		\foreach \x/\y in {1/1,2/5,3/9,4/13,5/17} {\node[proclabel] at (\x*.5-.25,5.5) {\y};}
		\foreach \x/\y in {1/2,2/6,3/10,4/14,5/17} {\node[proclabel] at (\x*.5-.25,6.5) {\y};}
		\foreach \x/\y in {1/3,2/7,3/11,4/15,5/17} {\node[proclabel] at (\x*.5-.25,7.5) {\y};}
		\foreach \x/\y in {1/0,2/6,3/11,4/13,5/18} {\node[proclabel] at (\x*.5-.25,8.5) {\y};}
		\foreach \x/\y in {1/3,2/5,3/8,4/14,5/18} {\node[proclabel] at (\x*.5-.25,9.5) {\y};}
		\foreach \x/\y in {1/1,2/7,3/10,4/12,5/18} {\node[proclabel] at (\x*.5-.25,10.5) {\y};}
		\foreach \x/\y in {1/2,2/4,3/9,4/15,5/18} {\node[proclabel] at (\x*.5-.25,11.5) {\y};}
		\foreach \x/\y in {1/0,2/7,3/9,4/14,5/19} {\node[proclabel] at (\x*.5-.25,12.5) {\y};}
		\foreach \x/\y in {1/2,2/5,3/11,4/12,5/19} {\node[proclabel] at (\x*.5-.25,13.5) {\y};}
		\foreach \x/\y in {1/3,2/4,3/10,4/13,5/19} {\node[proclabel] at (\x*.5-.25,14.5) {\y};}
		\foreach \x/\y in {1/1,2/6,3/8,4/15,5/19} {\node[proclabel] at (\x*.5-.25,15.5) {\y};}
		% label C
		\node[yscale=-1,scale=2] at (1.25,17.25) {\LARGE $\CC$};
	\end{scope}

	\begin{scope}[shift={(17,0)}]
		% draw row blocks
		\foreach \x in {1,...,16} {
			\foreach \y in {1,...,4} { \draw[decorate,decoration={zigzag,amplitude=1pt}] (\y/2,\x) -- (\y/2,\x-1);}
			\draw[thick] (0,\x) -- (2.5,\x);
		}
		% draw outer box
		\draw[thick] (0,0) rectangle +(2.5,16);
		% draw processor labels
		\foreach \x/\y in {1/0,2/1,3/2,4/3,5/16} {\node[proclabel] at (\x*.5-.25,.5) {\y};}
		\foreach \x/\y in {1/4,2/5,3/6,4/7,5/16} {\node[proclabel] at (\x*.5-.25,1.5) {\y};}
		\foreach \x/\y in {1/8,2/9,3/10,4/11,5/16} {\node[proclabel] at (\x*.5-.25,2.5) {\y};}
		\foreach \x/\y in {1/12,2/13,3/14,4/15,5/16} {\node[proclabel] at (\x*.5-.25,3.5) {\y};}
		\foreach \x/\y in {1/0,2/4,3/8,4/12,5/17} {\node[proclabel] at (\x*.5-.25,4.5) {\y};}
		\foreach \x/\y in {1/1,2/5,3/9,4/13,5/17} {\node[proclabel] at (\x*.5-.25,5.5) {\y};}
		\foreach \x/\y in {1/2,2/6,3/10,4/14,5/17} {\node[proclabel] at (\x*.5-.25,6.5) {\y};}
		\foreach \x/\y in {1/3,2/7,3/11,4/15,5/17} {\node[proclabel] at (\x*.5-.25,7.5) {\y};}
		\foreach \x/\y in {1/0,2/6,3/11,4/13,5/18} {\node[proclabel] at (\x*.5-.25,8.5) {\y};}
		\foreach \x/\y in {1/3,2/5,3/8,4/14,5/18} {\node[proclabel] at (\x*.5-.25,9.5) {\y};}
		\foreach \x/\y in {1/1,2/7,3/10,4/12,5/18} {\node[proclabel] at (\x*.5-.25,10.5) {\y};}
		\foreach \x/\y in {1/2,2/4,3/9,4/15,5/18} {\node[proclabel] at (\x*.5-.25,11.5) {\y};}
		\foreach \x/\y in {1/0,2/7,3/9,4/14,5/19} {\node[proclabel] at (\x*.5-.25,12.5) {\y};}
		\foreach \x/\y in {1/2,2/5,3/11,4/12,5/19} {\node[proclabel] at (\x*.5-.25,13.5) {\y};}
		\foreach \x/\y in {1/3,2/4,3/10,4/13,5/19} {\node[proclabel] at (\x*.5-.25,14.5) {\y};}
		\foreach \x/\y in {1/1,2/6,3/8,4/15,5/19} {\node[proclabel] at (\x*.5-.25,15.5) {\y};}
		% label C
		\node[yscale=-1,scale=2] at (1.25,17.25) {\LARGE $\B$};
	\end{scope}

\end{tikzpicture}
	\vspace*{-0.1cm}\caption{Triangle block partition using the affine construction for SYMM ($\CC\pluseq\A\B$) with $c=4$. Segments $0\leq k < c^2+c$ are shown in blue to indicate ownership of an element and element indices $0\leq i < c^2$ are shown in red. Each row of $\B$ and each row of $\CC$ are required for all the $c+1$ segments listed in the row.}\vspace*{-0.215cm}
	\label{fig:TBD-SYMM-4}
\end{figure}

To specify a triangle block partition, we define three types of sets:
$R_k$ is a $c$-element set of row indices that defines the $k$th triangle block of the symmetric matrix;
$D_k\subset R_k$ is a one-element or empty set containing the index of the diagonal element of the symmetric matrix that is assigned to the $k$th triangle block;
and for each row or index $i$, $Q_i$ is a set of $(c+1)$ triangle blocks such that $i\in R_k$, meaning row $i$ of the non-symmetric matrices is needed for the computations performed by the $k$th triangle block.
An example of a triangle block partition for SYMM is detailed in \cref{fig:TBD-SYMM-4}.
The sets used to define the partition for these examples are given in \cref{tab:TBD-aff-4}.
This partition is an affine plane construction for $c=4$; a projective plane construction for $c=4$ is given in \cref{sec:appendix:projexample}.

\begin{table}[t]
    \begin{center}
        \begin{tabular}{|c|c|c||c|c|}
            \hline
            $k$ & $R_k$ & $D_k$ & $i$ & $Q_i$ \\
            \hline
            0 & \{0,4,8,12\} & \{\} & 0 & \{0,1,2,3,16\}\\
            1 & \{0,5,10,15\} & \{\} & 1 & \{4,5,6,7,16\} \\
            2 & \{0,6,11,13\} & \{\} & 2 & \{8,9,10,11,16\} \\
            3 & \{0,7,9,14\} & \{\} & 3 & \{12,13,14,15,16\} \\
            4 & \{1,4,11,14\} & \{1\} & 4 & \{0,4,8,12,17\} \\
            5 & \{1,5,9,13\} & \{5\} & 5 & \{1,5,9,13,17\} \\
            6 & \{1,6,8,15\} & \{8\} & 6 & \{2,6,10,14,17\} \\
            7 & \{1,7,10,12\} & \{10\} & 7 & \{3,7,11,15,17\}\\
            8 & \{2,4,9,15\} & \{2\} & 8 & \{0,6,11,13,18\} \\
            9 & \{2,5,11,12\} & \{12\} & 9 & \{3,5,8,14,18\} \\
            10 & \{2,6,10,14\} & \{6\} & 10 & \{1,7,10,12,18\} \\
            11 & \{2,7,8,13\} & \{13\} & 11 & \{2,4,9,15,18\} \\
            12 & \{3,4,10,13\} & \{4\} & 12 & \{0,7,9,14,19\} \\
            13 & \{3,5,8,14\} & \{14\} & 13 & \{2,5,11,12,19\} \\
            14 & \{3,6,9,12\} & \{9\} & 14 & \{3,4,10,13,19\} \\
            15 & \{3,7,11,15\} & \{3\} & 15 & \{1,6,8,15,19\} \\
            16 & \{0,1,2,3\} & \{0\} & & \\
            17 & \{4,5,6,7\} & \{7\} & & \\
            18 & \{8,9,10,10\} & \{11\} & & \\
            19 & \{12,13,14,15\} & \{15\} & & \\
            \hline
        \end{tabular}
        \caption{Row sets and segment/processor sets of triangle block partition for $c=4$, $c^2+c=20$ segments using the affine construction.} \label{tab:TBD-aff-4}
    \end{center}
\end{table}

To define the sets $\{R_k\}$, we enumerate the sets of points on distinct lines in $\mathbb{A}^2(\mathbb{F}_c)$ using a computer algebra system.
To do this, we take any two points $(a_1, a_2), (b_1,b_2)$, then find all other points that fall on the line between those points by taking $t(a_1-b_1,a_2-b_2)+(b_1,b_2)$ for all $t\in\mathbb{F_c}$ and repeating this process with pairs of points that are not yet in a set $R_k$ until all sets have been defined.
We have chosen an assignment of diagonal elements that ensures $D_k\subset R_k$, and $|D_k|=\{0,1\}$, as this implies that no extra elements of the non-symmetric matrices are required by the $k$th triangle block and each triangle block has to compute at most one diagonal element.
We proved in the section above that it is always possible to do this.

Given $R_k$ and $D_k$, we can specify the triangle block partition of the symmetric matrix: the $k$th triangle block owns element $ij$ of the symmetric matrix if $i,j \in R_k$ and $i>j$, and it owns element $ii$ of the symmetric matrix if $i\in D_k$.
Given an element $ij$ of the symmetric matrix with $i\neq j$, the indices $i$ and $j$ appear together in exactly one set $R_k$ (this is because the triangle block scheme partitioned the lower triangle), and the element is owned by the corresponding triangle block.
To specify the rows of the non-symmetric matrices that are needed by a block in the computations involving the relevant triangle block of the symmetric matrix, we define the sets $\{Q_i\}$ by looping over the $R_k$ and checking for the inclusion of element $i$ in $R_k$.

The triangle block partitions so defined are used by our sequential and parallel algorithms for SYRK, SYR2K and SYMM computations, detailed in the rest of the paper (\cref{sec:seqAlgorithms,sec:limitedMemory,sec:memIndepAlgorithms}). In the sequential algorithms, during each segment all the elements of a single triangle block are loaded in the fast memory. For parallel algorithms, each processor owns all the elements belonging to a triangle block.

A practical limitation of using such partitioning scheme is the divisibility conditions between the size of the symmetric matrix $n_{1}$ and the size of the set $R_{k}$ that must be satisfied in order for a triangle block partition to exist. On the one hand $n_{1}$ is an input of the problem considered. On the other hand, the size of $R_{k}$ is derived from the size of the fast memory $M$, for the sequential algorithms, or from the number of processors involved, for the parallel algorithms. To reconcile both constraints and design a feasible triangle block partition in all cases, we use the following technique: we consider extended versions of the symmetric matrix padded with zero elements, not involved in actual computations or communications, such that its dimensions and the size of the set $R_{k}$ satisfy the divisibility conditions.
In the sequential case we consider generic Steiner systems.

A different blocking technique is used for parallel algorithms, in which the partitioning is applied to chunks of contiguous elements rather than elementwise.
The main idea of our parallel algorithms is to assign a triangular block to a unique processor.
As working with the affine construction simplifies the algebra while analyzing communication costs, we will focus only on the affine construction for our parallel algorithms.

The following sections are dedicated to detailing our algorithms. The sequential algorithms are presented in \cref{sec:seqAlgorithms} based on generic triangle block partitions. The parallel algorithms are presented in \cref{sec:memIndepAlgorithms}, for the memory independent version, and in \cref{sec:limitedMemory}, for the memory dependent version.

\section{Sequential Algorithms}
\label{sec:seqAlgorithms}

In this section we present sequential algorithms for the SYRK, SYR2K and SYMM computations. The core idea of our sequential algorithms is to partition the symmetric matrix according to triangle blocks as described in \cref{sec:triangularDistrib} in order to maximize the usage of associated elements from the non symmetric matrix or matrices. We will see that the main part of bandwidth costs of our algorithms comes from the necessity of moving non symmetric matrix or matrices, either to read input elements, for SYRK and SYR2K, or read and write elements, for SYMM. We also show that the number of words read by our algorithms exactly matches the sequential lower bounds of \cref{subsec:seqLowerBound} in leading order terms. The number of words written by our SYRK and SYR2K algorithms are lower order terms, while it is of the same order as the number of words read in the case of SYMM.
We analyze algorithms for all three computations at once in a generic way, using the number of non-symmetric matrices $m$ involved.

The number of words read by our algorithms match the lower bounds provided that the symmetric matrix is sequentially loaded in the fast memory during each segment according to a triangle block partition, as described in \cref{sec:triangularDistrib}. Whether it derives from projective or affine plane construction or it is a general Steiner system, the divisibility conditions stated in \cref{thm:divConditions} are satisfied by any such triangle block partition. In the entire section, we assume that $n_1$ is large enough to be in the configuration described by this theorem, which means that the existence of a triangle block partition is equivalent to the divisibility conditions. Now for sequential algorithms, since the size of the fast memory $M$ is limited, the size $r$ of the set $R_{k}$ defining a triangle block partition of the symmetric matrix is constrained because each triangle block must fit in the memory. Hence the divisibility conditions for such $r$ and the size of the symmetric matrix $n_1$ do not hold in general.

In the following, we first present and analyze our algorithms in a specific case, assuming that the value of the parameter $r$ derived from the constraint on the memory size $M$ and the matrix size $n_1$ are such that the divisibility conditions hold, thus there exists a triangle block partition of the symmetric matrix with $|R_k| = r$ for each triangle block. Then in the general case, we define $\hat{n}_1 \geq n_1$ such that the divisibility conditions are satisfied for $\hat{n}_1$ and $r$ and there exists a triangle block partition applicable to a matrix of size $\hat{n}_1$. We consider in such cases $m$ non-symmetric matrices of dimensions $\hat{n}_1 \times n_2$ and a symmetric matrix of dimensions $\hat{n}_1 \times \hat{n}_1$ that are extended versions of the original matrices padded with zeros. We prove that our sequential algorithms applied to such padded matrices achieve the same bandwidth cost as in the special case, ie. the leading order term of the number of words read is identical.

The remainder of the section is divided as follows: \cref{subsec:seqAlgoDescription} describes the algorithms for each of the three computations, \cref{subsec:seqAlgoAnalysis} analyzes the constraint on the fast memory size and bandwidth costs for all our algorithms in the specific case where $r$ and $n_1$ satisfy the divisibility conditions, \cref{subsec:seqAlgoGeneralCase} extends the analysis to the general case and shows that the bandwidth cost remains optimal, finally \cref{subsec:seqAlgoLimitExtend} discusses a limitation of the SYMM algorithm in terms of number of words written.

\subsection{Description of the Algorithms}
\label{subsec:seqAlgoDescription}

We present our SYRK, SYR2K and SYMM algorithms in \cref{alg:seqSYRK,alg:seqSYR2K,alg:seqSYMM}, respectively. All the three algorithms proceed similarly using two embedded loops. During each iteration of the outer loop, one triangle block from the symmetric matrix is read in the fast memory. The behavior for the iterations of the inner loop depends on whether the symmetric matrix is an input or an output of the computation.

If the symmetric matrix is an output, as for SYRK, respectively SYR2K, the algorithm iteratively reads and uses the elements from the non-symmetric input matrix $\A$, respectively matrices $\A$ and $\B$, to perform all the operations inside the inner loop of the computation. Thus it computes the values of all elements inside the triangle block.

If the symmetric matrix is an input, as for SYMM, the algorithm simultaneously reads the elements from the non-symmetric input matrix $\B$ and output matrix $\CC$. The elements of $\B$ are the ones that perform all possible operations inside the inner loop with the elements of the triangle block already present in the fast memory. The elements of $\CC$ that can use the result of those operations are updated and written back to the slow memory. For example, let us consider $TB(R_{5})$ triangle block of \cref{fig:TBD-SYMM-4} is already present in the fast memory. The algorithm reads the rows $\{1,5,9,13\}$ of $\B$ and $\CC$ to the fast memory. All possible operations of the rows of $\B$ with $TB(R_{5})$ are performed and the rows of $\CC$ are modified. After that, the updated rows of $\CC$ are written back to the slow memory.

Because of the property of triangle blocks, each time a set of elements from the non symmetric matrix or matrices is loaded, the maximum number of operations involving them are performed. Hence, the algorithms achieve the maximum possible reuse. Furthermore, the use of triangle block partition ensures that the symmetric matrix is loaded only once in the local memory.

\begin{algorithm}
	\caption{Sequential algorithm for SYRK}
	\label{alg:seqSYRK}
	\begin{algorithmic}[1]
		\Require $\A$ an $n_1 \times n_2$ matrix, $\CC$ an $n_1 \times n_1$ symmetric matrix, $M$ the size of the fast memory
		\Ensure $\CC \pluseq \A\A^{\text{T}}$
		\Function{$\CC = $ Seq-SYRK}{$\CC,\A, M$}
                \State Let $r = \left\lfloor \sqrt{2M+1}-1 \right\rfloor$ and $K = \frac{n_1(n_1-1)}{r(r-1)}$ \label{line:seqSYRK:cParam}
                \State Let $R_k$ and $D_k$, $k \in \{0, \dots, K-1\}$, subsets of $\{0, \dots, n_1-1\}$ that define a triangle block partition of $\A$

		\ForAll{$k \in \{0, \dots, K-1 \}$}
		\State Load $TB(R_{k})$ of matrix $\CC$ in the fast memory \label{line:seqSYRK:TBLoad}
		\ForAll{$j \in \{0, \dots, n_2-1\}$}
		\State Load $(\A_{i, j})_{i \in R_{k}}$ in the fast memory \label{line:seqSYRK:colLoad}
		\ForAll{$i_1, i_2 \in R_{k}$}
		\If{$\lbrack i_2 < i_1 \rbrack$ or $\lbrack (i_2 = i_1)$ and $(i_2 \in D_{k}) \rbrack$}
		\State $\CC_{i_1, i_2} \pluseq \A_{i_1, j}\A_{i_2, j}$
		\EndIf
		\EndFor
		\State Flush $(\A_{i, j})_{i \in R_{k}}$ from the fast memory
		\EndFor
		\State Write all elements of $TB(R_{k})$ to the slow memory
		\EndFor
		\EndFunction
	\end{algorithmic}
\end{algorithm}

\begin{algorithm}
    \caption{Sequential algorithm for SYR2K}
    \label{alg:seqSYR2K}
    \begin{algorithmic}[1]
        \Require $\A$ and $\B$ two $n_1 \times n_2$ matrices, $\CC$ an $n_1 \times n_1$ symmetric matrix, $M$ the size of the fast memory
        \Ensure $\CC \pluseq \A\B^{\text{T}}+\B\A^{\text{T}}$
        \Function {$\CC = $ Seq-SYR2K}{$\CC, \A, \B, M$}
        \State Let $r = \left\lfloor \sqrt{2M+4}-2 \right\rfloor$ and $K = \frac{n_1(n_1-1)}{r(r-1)}$ \label{line:seqSYR2K:cParam}
        \State Let $R_k$ and $D_k$, $k \in \{0, \dots, K-1\}$, subsets of $\{0, \dots, n_1-1\}$ that define a triangle block partition of $\A$
        \ForAll{$k \in \{0, \dots, K-1\}$}
            \State Load $TB(R_{k})$ of matrix $\CC$ in the fast memory \label{line:seqSYR2K:TBLoad}
            \ForAll{$j \in \{0, \dots, n_2-1\}$}
                \State Load $(\A_{i, j})_{i \in R_{k}}$ and $(\B_{i, j})_{i \in R_{k}}$ in the fast memory \label{line:seqSYR2K:colLoad}
                \ForAll{$i_1, i_2 \in R_{k}$}
                    \If{$\lbrack i_2 < i_1 \rbrack$ or $\lbrack (i_2 = i_1)$ and $(i_2 \in D_{k}) \rbrack$}
                        \State $\CC_{i_1, i_2} \pluseq \A_{i_1, j}\B_{i_2, j} + \A_{i_2, j}\B_{i_1, j}$
                    \EndIf
                \EndFor
                \State Flush $(\A_{i, j})_{i \in R_{k}}$ and $(\B_{i, j})_{i \in R_{k}}$ from the fast memory
            \EndFor
            \State Write all elements of $TB(R_{k})$ to the slow memory
        \EndFor
        \EndFunction
    \end{algorithmic}
\end{algorithm}

\begin{algorithm}
    \caption{Sequential algorithm for SYMM}
    \label{alg:seqSYMM}
    \begin{algorithmic}[1]
        \Require $\A$ an $n_1 \times n_1$ symmetric matrix, $\B$  and $\CC$ two $n_1 \times n_2$ matrices, $M$ the size of the fast memory
        \Ensure $\CC \pluseq \A\B$
        \Function {$\CC = $ Seq-SYMM}{$\CC, \A, \B, M$}
        \State Let $r = \left\lfloor \sqrt{2M+4}-2 \right\rfloor$ and $K = \frac{n_1(n_1-1)}{r(r-1)}$ \label{line:seqSYMM:cParam}
        \State Let $R_k$ and $D_k$, $k \in \{0, \dots, K-1\}$, subsets of $\{0, \dots, n_1-1\}$ that define a triangle block partition of $\A$
        \ForAll{$k \in \{0, \dots, K-1\}$}
            \State Load $TB(R_{k})$ of matrix $\A$ in the fast memory \label{line:seqSYMM:TBLoad}
            \ForAll{$j \in \{0, \dots, n_2-1\}$}
                \State Load $(\B_{i, j})_{i \in R_{k}}$ and $(\CC_{i, j})_{i \in R_{k}}$ in the fast memory \label{line:seqSYMM:colLoad}
                \ForAll{$i_1 \in R_{k}$}
                    \State \begin{displaymath}
                      \CC_{i_1, j} \pluseq \sum_{\underset{i_2 < i_1 \, \textrm{\textbf{or}} \, i_2 = i_1, \, i_2 \in D_{k}}{i_2 \in R_{k}}} \A_{i_1, i_2}\B_{i_2, j} + \sum_{\underset{i_2 > i_1}{i_2 \in R_{k}}} \A_{i_2, i_1}\B_{i_2, j}
                    \end{displaymath}
                \EndFor
                \State Write all elements of $(\CC_{i, j})_{i \in R_{k}}$ to the slow memory
                \State Flush $(\B_{i, j})_{i \in R_{k}}$ from the fast memory
            \EndFor
            \State Flush $TB(R_{k})$ from the fast memory
        \EndFor
        \EndFunction
    \end{algorithmic}
\end{algorithm}

\subsection{Analysis of the Algorithms under the Divisibility Conditions}
\label{subsec:seqAlgoAnalysis}

In this subsection we provide an analysis of all three sequential algorithms in terms of their required memory and the number of words they communicate. To simplify the analysis, we assume that the parameter $r$ derived from the memory requirement according to \cref{eq:cFitLB} is such that the divisibility conditions are satisfied and therefore there exists a triangle block partition of the symmetric matrix. In particular $\frac{n_1(n_1-1)}{r(r-1)}$, denoted by $K$, is an integer corresponding to the number of triangle blocks in the partition.
Furthermore, in the following we only focus on the case where the memory cannot hold the entire symmetric matrix plus one column of each non symmetric matrices, $M < \frac{n_{1}(n_{1}+1)}{2} + m n_{2}$. Otherwise, in the case where the memory size is so large that this inequality no longer holds, the proposed algorithms are still feasible and match the sequential lower bound (including the constant): the whole symmetric matrix is kept in memory as a single triangle block and updated by streaming through entire columns of the non-symmetric matrices.

\subsubsection{Memory Requirement}
\label{subsubsec:seqAlgoMem}

Let us consider a triangle block partition of the symmetric matrix according to $K$ sets $R_k \subset \{1, \dots, n_{1}\}$, each of size $r$.
At any point during the execution, each of the three sequential algorithms requires that a full triangle block of the symmetric matrix is present in the fast memory along with $m$ subsets of rows of the non-symmetric matrices. The number of elements present in the fast memory belonging to each subset $TB(R_{k}), \; {k \in \{1, \dots, K\}}$ is then $\frac{r(r-1)}{2} +|D_{k}|$ with $|D_{k}| \in \{0, 1\}$, which is smaller that $\frac{r^{2}}{2}$.
Now the memory required by each algorithm at any point should not be larger than the fast memory size, that is

\begin{displaymath}
  m|R_{k}| + |TB(R_{k})| \leq M
  \Leftarrow mr + \frac{r^{2}}{2} \leq M
  \Leftarrow r \leq \sqrt{2M+m^2}-m\text.
\end{displaymath}

Therefore a sufficient condition for our sequential algorithms to be feasible regarding the memory requirement is to use a triangle block partition such that

\begin{equation}
\label{eq:cFitLB}
    r = \left\lfloor \sqrt{2M+m^2}-m \right\rfloor\text.
\end{equation}

Note that this still assumes that the divisibility conditions will be satisfied with such $r$ and $n_1$.

\subsubsection{Bandwidth Cost}
\label{subsubsec:seqAlgoComm}

We now estimate the total number of words read by our algorithms. For each algorithm, there are as many iterations of the outer loop as triangle blocks, that is $K$. At the beginning of each iteration, a full triangle block of size $|TB(R_{k})|$ is read in the fast memory. For each triangle block, there are $n_2$ iterations of the inner loop. During each of them $|R_{k}|$ entries from each of the $m$ non-symmetric matrices are read in the fast memory. Thus, the number of words read by each algorithm is

\begin{align*}
  (m n_2 |R_{k}| + |TB(R_{k})|) K & = \left( m n_2 r + \frac{r(r-1)}{2} +|D_{k}| \right) \frac{n_1(n_1-1)}{r(r-1)}\\
                                  & \leq m \left( \frac{n_1(n_1-1) n_2}{r-1} \right) + \frac{n_1(n_1-1)}{2} + \frac{n_{1}(n_{1}-1)}{r(r-1)}\text.
\end{align*}

From \cref{eq:cFitLB}, we have $r-1 \geq \sqrt{2M+m^{2}}-(m+2) \geq \sqrt{2M}-(m+2)$. Therefore, the total number of words read by each algorithm is

\begin{align*}
(m n_2 |R_{k}| + |TB(R_{k})|) K & \leq m \left( \frac{n_1(n_1-1) n_2}{\sqrt{2M}-(m+2)} \right) + \frac{n_1(n_1-1)}{2} + \frac{n_{1}(n_{1}-1)}{r(r-1)}\\
& \leq m \left( \frac{n_1(n_1-1) n_2}{\sqrt{2M}} \right) + \frac{n_1(n_1-1)}{2} + \mathcal{O} \left( \frac{n_{1}^{2}n_{2}}{M} \right)\text.
\end{align*}

We can observe that the first term in the expression of the number of words read matches the sequential lower bound presented in \cref{subsec:seqLowerBound} when the matrix dimensions and the fast memory size grow to infinity: it is the leading order term in the cases when the dimension of the non-symmetric matrix $n_{2}$ is large compared to the memory size $M$. Otherwise, if $n_{2}$ is small compared to $M$, the cost incurred by reading the entire symmetric matrix, that is $\frac{n_{1}(n_{1}-1)}{2}$ words, dominates the overall bandwidth cost. In such cases, the number of words read by our algorithms still matches the sequential lower bound including the constant: the second term in the expression becomes the leading order term.

\subsection{Analysis of the Algorithms in the General Case}
\label{subsec:seqAlgoGeneralCase}

We now consider the execution of our sequential algorithms in the general case, that is when the value of parameter $r$ selected according to the memory requirement of \cref{eq:cFitLB} and the matrix size $n_{1}$ do not satisfy the divisibility conditions. In such case, there exists no triangle block partition of the symmetric matrix with each triangle block associated to an index subset $R_{k}$ of size $r$. Then we search for a value $\hat{n}_{1} \geq n_{1}$ close from $n_{1}$ such that $r$ and $\hat{n}_{1}$ satisfy the divisibility conditions; according to \cref{thm:divConditions} there exists a triangle block partition such that $|R_{k}| = r$ for each triangle block of any symmetric matrix of size $\hat{n}_{1}$. Considering the non-symmetric matrices of dimensions $\hat{n}_{1} \times n_{2}$ and the symmetric matrix of dimensions $\hat{n}_{1} \times \hat{n}_{1}$, extensions of the original matrices padded with zeros, we can apply our sequential algorithms, ignoring zeros entries.

\paragraph{Dimensions of the Padded Matrices}

In this paragraph, we define a feasible value for the padded matrix size $\hat{n}_{1}$ and provide a bound for the gap between $n_{1}$ and $\hat{n}_{1}$. Let $r$ the value derived from \cref{eq:cFitLB} given the fast memory size $M$ and algorithm parameter $m$. Recall that the divisibility conditions for $r$ and $\hat{n}_{1}$ are:

\begin{displaymath}
  \left\lbrace\begin{array}{l}
    r-1 \; | \; \hat{n}_{1}-1\text,\\
    r(r-1) \; | \; \hat{n}_{1}(\hat{n}_{1}-1)\text.
  \end{array}\right.
\end{displaymath}

First, let us define $n'$, the smallest integer larger or equal to $n_{1}$ divisible by $r-1$. Then: $n' \leq n_{1}+r-1$. Then we define $n''$, the smallest integer larger or equal to $\frac{n'}{r-1}$ divisible by $r$. Then: $n'' \leq \frac{n'}{r-1}+r$. Finally, let us define $\hat{n}_{1} = n''(r-1)$. Then $\hat{n}_{1}$ is divisible by $r$, by definition of $n''$, and is obviously divisible by $r-1$. Therefore $\hat{n}_{1}$ satisfy the divisibility conditions with $r$.

Furthermore, we can upper bound $\hat{n}_{1}$ according to $n_{1}$ and $r$:

\begin{align*}
  \hat{n}_{1} & = n''(r-1)\\
              & \leq \left( \frac{n'}{r-1}+r \right) (r-1) = n'+r(r-1)\\
              & \leq (n_{1}+r-1)+r(r-1) = n_{1}+r^{2}-1\text.
\end{align*}

Hence, given $r$ we can extend the original matrices by padding up to size $\hat{n}_{1}$ such that there exists a triangle block partition of the padded symmetric matrix. The extended size is such that

\begin{equation}
  \label{eq:n1PaddedSize}
  n_{1} \leq \hat{n}_{1} < n_{1}+r^2\text.
\end{equation}

\paragraph{Bandwidth Cost}

Let us now evaluate the total number of words read by our algorithms in the general case. Given the parameter $r$ selected according to \cref{eq:cFitLB} and a feasible value $\hat{n}_{1}$ derived using the procedure described above, there exists a triangle block partition of the padded symmetric matrix usable to apply each of our algorithms. The bandwidth cost analysis is very similar to the special case presented in \cref{subsubsec:seqAlgoComm} where the number of triangle blocks necessary to partition the padded matrix is $\hat{K} = \frac{\hat{n}_{1}(\hat{n}_{1}-1)}{r(r-1)}$.
Using \cref{eq:n1PaddedSize}, it follows that:

\begin{displaymath}
  \hat{K} = \frac{\hat{n}_{1}(\hat{n}_{1}-1)}{r(r-1)} \leq \frac{(n_{1}+r^{2})(n_{1}+r^{2}-1)}{r(r-1)} = \frac{n_{1}(n_{1}-1)}{r(r-1)} + \frac{2 r n_{1}}{r-1} + r(r+1)\text.
\end{displaymath}

When $M=o(n_1)$, $\hat{n}_{1}$ is close to $n_{1}$. Under such assumption the number of words read by each algorithm is

\begin{equation*}
(m n_{2} |R_{k}| + |TB(R_{k})|)\hat{K} = m \left( \frac{n_1(n_1-1) n_2}{\sqrt{2M}} \right) + \frac{n_1(n_1-1)}{2} + o \left( \frac{n_{1}^{2}n_{2}}{\sqrt M} + n_1^2 \right)\text.
\end{equation*}

Thus, when the matrix dimension $n_{1}$ is large compared to the memory size $M$, the first terms of the expression are the leading order terms.
Hence the number of words read by our algorithms matches the sequential lower bounds, including the constants.

\subsection{Number of Words Written}
\label{subsec:seqAlgoLimitExtend}
In \cref{alg:seqSYRK,alg:seqSYR2K} for SYRK and SYR2K computations, the symmetric matrix is the output of the algorithms. It is written to the slow memory exactly once because the triangle blocks define a partition of the output matrix. Assuming that the fast memory does not contain any output element in the beginning or end of the computation, the total number of words written is $n_1(n_1+1)/2$ which is a lower order term compared to the total number of words read when $n_{2}$ is large compared to $M$. On the contrary, the output is a non-symmetric matrix for our SYMM algorithm (\cref{alg:seqSYMM}). During the computation, each element in $\CC$ is written $(n_1-1)/(r-1)$ times in the slow memory, inducing $n_{1}n_{2}\frac{n_1-1}{r-1} \approx \frac{n_{1}^2 n_{2}}{\sqrt{2M}}$ words written, which is half the number of words read. Whether there exists a SYMM algorithm that simultaneously minimizes the total number of words communicated from and to the fast memory is an open problem that we plan to explore in future work.

\section{Communication Optimal Memory Independent Parallel Algorithms}
\label{sec:memIndepAlgorithms}
In this section, we present three families of parallel algorithms where the communication costs exactly match the leading order terms of the lower bounds of \cref{sec:memindeplb}.
Each family extends one of the algorithms for SYRK given by Al~Daas et al.~\cite{ABGKR23}.
The families are defined by how they partition the 3-dimensional iteration space of the full computations, and each family contains an algorithm for every computation.
Following \cite{ABGKR23}, we call the families of algorithms 1D, 2D, or 3D algorithms based upon the number of dimensions of the iteration space that are partitioned.
Extending the 1D algorithm for SYRK in \cite{ABGKR23}, the 1D family of algorithms all partition the computations only in the dimension of size $n_2$ (\cref{sec:memIndepAlgorithms:1Dalgorithms}) with 1D processor grids.
Similarly, the 2D family of algorithms all partition the computations in both dimensions of size $n_1$ (\cref{sec:memIndepAlgorithms:2Dalgorithms}). We also employ 1D processor grids for our 2D algorithms, but they are indexed differently. A processor index is associated with a block from the affine triangle block construction (see \Cref{sec:TBD}, \Cref{fig:TBD-SYMM-PAR}).
The 3D family of algorithms partition all 3 dimensions (\cref{sec:memIndepAlgorithms:3Dalgorithms}) using a 2D processor grid, and employ the corresponding 2D algorithms multiple times for different subsets of the $n_2$ dimension.
To achieve the optimal communication cost for each computation, we select the algorithm and processor grid based on the relative sizes of $n_1, n_2$, and $P$.
In all algorithms presented, each processor stores roughly 1/Pth of each input in the beginning of the computation and 1/Pth of the output at the end of the computation.
As mentioned in \cref{sec:TBD}, we will work with a chunk of contiguous elements instead of an individual element in our parallel algorithms.

\subsection{1D Parallel Algorithms}
\label{sec:memIndepAlgorithms:1Dalgorithms}
1D algorithms are optimal when only one dimension is large, and $P$ is relatively small.
In 1D algorithms, matrices that are indexed by the large dimension are not communicated, while all other matrices are communicated.
In the case of symmetric three nested loop computations the conditions for a 1D algorithm to be optimal are that $n_1 \leq mn_2$ and $P \leq mn_2/\sqrt{n_1(n_1-1)}$.
As $n_2$ is the large dimension, any non-symmetric matrix is short and wide, and the symmetric matrix is smaller than the non-symmetric matrices.
Only the large dimension, $n_2$, is partitioned, so only the symmetric matrix is communicated.

\subsubsection{Data Distribution and Parallel Algorithms}
We generalize the 1D data distribution presented in \cite[5.1]{ABGKR23} to apply it to all three computations.
We partition the $n_2$ dimension to get a 1D block column distribution of the non-symmetric matrices where a processor owns the same 1/Pth of the columns from every non-symmetric matrix.
The distribution of the lower triangle of the symmetric matrix is arbitrary, but must be even across the processors.
We identify a processor rank with an index $\ell \in \Pi$, where $0 \leq \ell < P$.
In defining our notation, we will use the matrix names from SYRK, so $\CC$ is the symmetric matrix, and $\A$ is a non-symmetric matrix.
We will follow this notation in the other computations using the appropriate matrix names to the computation.
If $\A$ is a non-symmetric matrix, as in SYRK and SYR2K, we denote the column block of $\A$ associated with processor $\ell$ by $\A_{*\ell}$, and use the same notation for the column block of any other non-symmetric matrix.
When $P$ divides $n_2$ evenly, each column block has dimensions $n_1 \times (n_2 /P)$.
If $\CC$ is the symmetric matrix, $\CC^{(\ell)}$ denotes the portion of the lower triangle owned by processor $\ell$ at the start or end of the computation.

\begin{algorithm}
	\caption{1D SYRK {\cite[Algorithm 1]{ABGKR23}}}
	\label{alg:1Dsyrk}
	\begin{algorithmic}[1]
		\Require $\Pi$ is a set of $P$ processors.
		\Require $\A$ is evenly column distributed across $\Pi$ with $\A_{*\ell}$ is column block owned by processor $\ell$.
		\Require $\CC$ is evenly distributed across $\Pi$ with $\CC^{(\ell)}$ owned by processor $\ell$.
		\Ensure $\CC \pluseq \A\A^T$ with $\CC^{(\ell)}$ owned by processor $\ell$.
		\Function{$\CC^{(\ell)}=$ 1D-SYRK}{$\CC^{(\ell)}, \A_{*\ell},\Pi$}
		\State $\ell = \Call{MyRank}{\Pi}$
		\State $\bar{\CC} = \A_{*\ell} \A_{*\ell}^T $ \label{line:1D:local-syrk} \Comment{Local SYRK computation}
		\State $\CC^{(\ell)} \pluseq \Call{Reduce-Scatter}{\bar{\CC},\Pi}$
		\EndFunction
	\end{algorithmic}
\end{algorithm}

\begin{algorithm}
	\caption{1D SYR2K}
	\label{alg:1Dsyr2k}
	\begin{algorithmic}[1]
		\Require $\Pi$ is a set of $P$ processors.
		\Require $\A$ and $\B$ are evenly column distributed across $\Pi$ with $\A_{*\ell}$ and $\B_{*\ell}$ are column blocks owned by processor $\ell$.
		\Require $\CC$ is evenly distributed across $\Pi$ with $\CC^{(\ell)}$ owned by processor $\ell$.
		\Ensure $\CC \pluseq \A\B^T + \B\A^T$ with $\CC^{(\ell)}$ owned by processor $\ell$.
		\Function{$\CC^{(\ell)}=$ 1D-SYR2K}{$\CC^{(\ell)}, \A_{*\ell}, \B_{*\ell},\Pi$}
		\State $\ell = \Call{MyRank}{\Pi}$
		\State $\bar{\CC} = \A_{*\ell}\B_{*\ell}^T + \B_{*\ell}\A_{*\ell}^T $ \label{line:1D:local-syr2k} \Comment{Local SYR2K computation}
		\State $\CC^{(\ell)} \pluseq \Call{Reduce-Scatter}{\bar{\CC},\Pi}$ \label{line:1D:reduce-scatter}
		\EndFunction
	\end{algorithmic}
\end{algorithm}

\begin{algorithm}
	\caption{1D SYMM}
	\label{alg:1Dsymm}
	\begin{algorithmic}[1]
		\Require $\Pi$ is a set of $P$ processors.
		\Require $\A$ is evenly distributed across $\Pi$ with $\A^{(\ell)}$ owned by processor $\ell$.
		\Require $\B$ and $\CC$ are evenly column distributed across $\Pi$ with $\B_{*\ell}$ and $\CC_{*\ell}$ are column blocks owned by processor $\ell$.
		\Ensure $\CC \pluseq \A\B$ with $\CC_{*\ell}$ owned by processor $\ell$.
		\Function{$\CC_{*\ell}=$1D-SYMM}{$\CC_{*\ell}, \A^{(\ell)}, \B_{*\ell}, \Pi$}
		\State $\ell = \Call{MyRank}{\Pi}$			
		\State $\A = \Call{All-Gather}{\A^{(\ell)}, \Pi}$ \label{line:1Dsymm:all-gather}
		\State $\CC_{*\ell} \pluseq \A\B_{*\ell}$ \label{line:1D:local-symm} \Comment{Local SYMM computation}
		\EndFunction
	\end{algorithmic}
\end{algorithm}

\subsubsection{Cost Analysis}
Following the analysis in \cite{ABGKR23} we consider both the computational and communication costs for each algorithm.
Recall that in our $\alpha-\beta-\gamma$ parallel computation model, $\alpha$, and $\beta$ represent the per message latency cost and per word bandwidth cost while $\gamma$ represents the cost of each arithmetic operation.
After deriving the computation and communication costs, we determine how much memory is required to perform the algorithms.

\paragraph{Computation Cost} All algorithms perform the local computations, \cref{line:1D:local-syr2k} in \cref{alg:1Dsyrk,alg:1Dsyr2k} and \cref{line:1D:local-symm} in \cref{alg:1Dsymm}, with blocks of dimension $n_1\times(n_2/P)$ of each non-symmetric matrix.
Thus the cost of the local computations are dominated by $m\gamma n_1^2n_2/P$ for all three algorithms. Note that we count both addition and multiplication operations of the computations.
As the additional computation performed by the \Call{Reduce-Scatter}{} in \cref{alg:1Dsyrk,alg:1Dsyr2k} is a lower order term, we do not consider it.

\paragraph{Communication Cost} Both SYRK and SYR2K perform communication in the \Call{Reduce-Scatter}{}, \cref{line:1D:reduce-scatter} of \cref{alg:1Dsyrk,alg:1Dsyr2k}, while SYMM performs communication in the \Call{All-Gather}{}, \cref{line:1Dsymm:all-gather} of \cref{alg:1Dsymm}.
In all three algorithms, the number of elements from the symmetric matrix being communicated is $n_1(n_1+1)/2$.
As mentioned in \cref{sec:prelim:collectiveCommunicationCosts}, we assume pairwise exchange algorithms are used for \Call{Reduce-Scatter}{} and \Call{All-Gather}{} collectives.
Thus, the total communication cost for each of the three algorithms is
\begin{equation}
  \label{eq:1D:commcost}
    \alpha(P-1)+\beta \frac{n_1(n_1+1)}{2}\frac{P-1}{P}\text.
\end{equation}

\paragraph{Memory Requirement} At the start of the computation, each processor owns $n_1n_2/P$ elements of each non-symmetric matrix and $n_1(n_1+1)/(2P)$ elements of the symmetric matrix.
Each processor requires access to all $n_1(n_1+1)/2$ elements of the symmetric matrix, Thus, the total number of elements accessed on each processor is at most $m\frac{n_1n_2}{P}+\frac{n_1(n_1+1)}{2}+\frac{n_1(n_1+1)}{2P}$, which is $n_1(n_1+1)/2$ elements more than the owned elements of the processor.
As $n_1 \leq mn_2$ and $P \leq mn_2/\sqrt{n_1(n_1-1)}$, $n_1(n_1+1)/2$ is roughly bounded by $m\frac{n_1n_2}{P}$.
Hence, the memory use of our 1D algorithms is less than twice the memory needed by a processor to store its owned data.

\subsection{2D Algorithms}
\label{sec:memIndepAlgorithms:2Dalgorithms}

2D algorithms are optimal when one dimension is small, and $P$ is also relatively small.
In 2D algorithms, matrices that are indexed by the small dimensions are communicated.
In the case of symmetric three nested loop computations the conditions for a 2D algorithm to be optimal are that $mn_2 < n_1$ and $P \leq n_1(n_1-1)/(m^2n_2^2)$.
As $n_2$ is the small dimension, any non-symmetric matrix is tall and skinny, and the symmetric matrix is larger than the non-symmetric matrices.
The symmetric matrix is not communicated, as both the dimensions of size $n_1$ are partitioned, but all non-symmetric matrices are communicated.

\subsubsection{Data Distribution and Algorithms}
As with our 1D algorithms, our 2D algorithms extend the data distribution presented in \cite[5.2]{ABGKR23} to apply to all three computations.
We index the processors with $k$, with $0\leq k < P$, and distribute the symmetric matrix using a triangle block partition specified in \cref{sec:TBD}.
Note that we use a different index than in the 1D case to indicate a different partition of the data and computation.
We assume throughout this section that $P=c(c+1)$ for a prime power $c$ and that the affine triangle block construction is utilized.
For the following discussion we will assume that $c^2$ evenly divides $n_1$.
As before, we will define our notation using the matrix names for SYRK and the notation can be extended to the other computations.
When $\CC$ is the symmetric matrix, given a pair $i,j$ with $i,j\in R_k$, $\CC_{ij}$ is a block of dimension $n_1/c^2 \times n_1/c^2$ which is assigned to processor $k$.
When $\A$ is a non-symmetric matrix, $A_{i*}$ denotes a $n_1/c^2 \times n_2$ block of rows of $\A$ that is evenly distributed across the $c+1$ processors $k$ for which $i\in R_k$.
We denote the portion of $\A_{i*}$ assigned to processor $k$ with $\A_{i*}^{(k)}$.

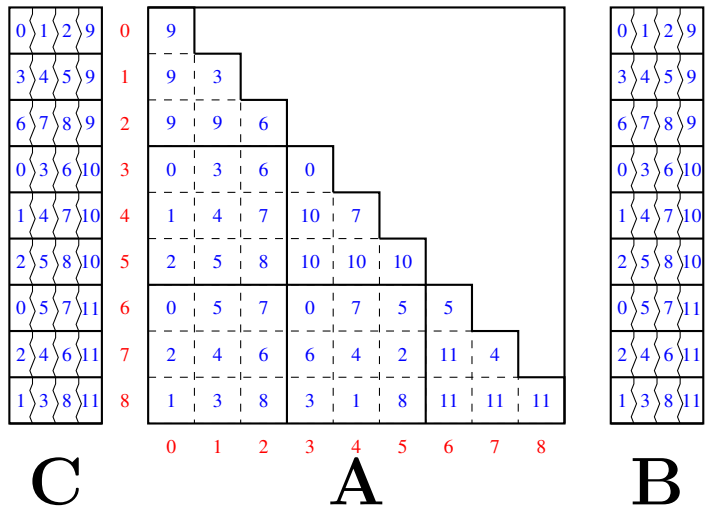
\begin{figure}[ht]
	\centering
	%!TEX root = ../paper.tex

% set offset
\pgfmathsetmacro{\offset}{.33}
% set styles
\tikzset{proclabel/.style={yscale=-1,blue}}
\tikzset{rowlabel/.style={yscale=-1,red}}

% for drawing help	
\newcommand{\gridhelp}{
	\coordinate (ll) at (-1,-1);
	\coordinate (ur) at (10,10);
	\draw[help lines,dashed] (ll) grid (ur);
	\node at (0,0) {0};
}	

%\begin{tikzpicture}[every node/.append style={transform shape},yscale=-1,scale=.675]
\begin{tikzpicture}[every node/.append style={transform shape},yscale=-1,scale=.615]
%\gridhelp
	% draw little boxes
	\foreach \x in {1,...,9} {\draw[dashed] (0,\x) -- (\x,\x) -- (\x,9);}
	% draw big boxes
	\foreach \x in {3,6} {\draw[thick] (0,\x) -- (\x,\x) -- (\x,9);}
	% draw diagonal
	\foreach \x in {1,...,9} {\draw[thick] (\x-1,\x-1) -- (\x,\x-1) -- (\x,\x);	}
	% draw outer box
	\draw[thick] (0,0) rectangle +(9,9);
	% draw processor labels
	\foreach \x/\y in {0/3,0/6,3/6,3/3} {\node[proclabel] at (\x+.5,\y+.5) {0};}
	\foreach \x/\y in {0/4,0/8,4/8} {\node[proclabel] at (\x+.5,\y+.5) {1};}
	\foreach \x/\y in {0/5,0/7,5/7} {\node[proclabel] at (\x+.5,\y+.5) {2};}
	\foreach \x/\y in {1/3,1/8,3/8,1/1} {\node[proclabel] at (\x+.5,\y+.5) {3};}
	\foreach \x/\y in {1/4,1/7,4/7,7/7} {\node[proclabel] at (\x+.5,\y+.5) {4};}
	\foreach \x/\y in {1/5,1/6,5/6,6/6} {\node[proclabel] at (\x+.5,\y+.5) {5};}
	\foreach \x/\y in {2/3,2/7,3/7,2/2} {\node[proclabel] at (\x+.5,\y+.5) {6};}
	\foreach \x/\y in {2/4,2/6,4/6,4/4} {\node[proclabel] at (\x+.5,\y+.5) {7};}
	\foreach \x/\y in {2/5,2/8,5/8} {\node[proclabel] at (\x+.5,\y+.5) {8};}
	\foreach \x/\y in {0/1,0/2,1/2,0/0} {\node[proclabel] at (\x+.5,\y+.5) {9};}
	\foreach \x/\y in {3/4,3/5,4/5,5/5} {\node[proclabel] at (\x+.5,\y+.5) {10};}
	\foreach \x/\y in {6/7,6/8,7/8,8/8} {\node[proclabel] at (\x+.5,\y+.5) {11};}
	% draw row/col labels
	\foreach \x in {0,...,8} {
		\node[rowlabel] at (-.5,\x+.5) {\x};
		\node[rowlabel] at (\x+.5,9+.5) {\x};
	}
	% label C
	\node[yscale=-1,scale=2] at (4.5,10.25) {\LARGE $\A$};
	
	\begin{scope}[shift={(-3,0)}]
		% draw row blocks
		\foreach \x in {1,...,9} {
			\foreach \y in {1,...,3} { \draw[decorate,decoration={zigzag,amplitude=1pt}] (\y/2,\x) -- (\y/2,\x-1);}
			\draw[thick] (0,\x) -- (2,\x);
		}
		% draw outer box
		\draw[thick] (0,0) rectangle +(2,9);
		% draw processor labels
		\foreach \x/\y in {1/0,2/1,3/2,4/9} {\node[proclabel] at (\x*.5-.25,.5) {\y};}
		\foreach \x/\y in {1/3,2/4,3/5,4/9} {\node[proclabel] at (\x*.5-.25,1.5) {\y};}
		\foreach \x/\y in {1/6,2/7,3/8,4/9} {\node[proclabel] at (\x*.5-.25,2.5) {\y};}
		\foreach \x/\y in {1/0,2/3,3/6,4/10} {\node[proclabel] at (\x*.5-.25,3.5) {\y};}
		\foreach \x/\y in {1/1,2/4,3/7,4/10} {\node[proclabel] at (\x*.5-.25,4.5) {\y};}
		\foreach \x/\y in {1/2,2/5,3/8,4/10} {\node[proclabel] at (\x*.5-.25,5.5) {\y};}
		\foreach \x/\y in {1/0,2/5,3/7,4/11} {\node[proclabel] at (\x*.5-.25,6.5) {\y};}
		\foreach \x/\y in {1/2,2/4,3/6,4/11} {\node[proclabel] at (\x*.5-.25,7.5) {\y};}
		\foreach \x/\y in {1/1,2/3,3/8,4/11} {\node[proclabel] at (\x*.5-.25,8.5) {\y};}
		% label A
		\node[yscale=-1,scale=2] at (1,10.25) {\LARGE $\CC$};
	\end{scope}
	\begin{scope}[shift={(+10,0)}]
		% draw row blocks
		\foreach \x in {1,...,9} {
			\foreach \y in {1,...,3} { \draw[decorate,decoration={zigzag,amplitude=1pt}] (\y/2,\x) -- (\y/2,\x-1);}
			\draw[thick] (0,\x) -- (2,\x);
		}
		% draw outer box
		\draw[thick] (0,0) rectangle +(2,9);
		% draw processor labels
		\foreach \x/\y in {1/0,2/1,3/2,4/9} {\node[proclabel] at (\x*.5-.25,.5) {\y};}
		\foreach \x/\y in {1/3,2/4,3/5,4/9} {\node[proclabel] at (\x*.5-.25,1.5) {\y};}
		\foreach \x/\y in {1/6,2/7,3/8,4/9} {\node[proclabel] at (\x*.5-.25,2.5) {\y};}
		\foreach \x/\y in {1/0,2/3,3/6,4/10} {\node[proclabel] at (\x*.5-.25,3.5) {\y};}
		\foreach \x/\y in {1/1,2/4,3/7,4/10} {\node[proclabel] at (\x*.5-.25,4.5) {\y};}
		\foreach \x/\y in {1/2,2/5,3/8,4/10} {\node[proclabel] at (\x*.5-.25,5.5) {\y};}
		\foreach \x/\y in {1/0,2/5,3/7,4/11} {\node[proclabel] at (\x*.5-.25,6.5) {\y};}
		\foreach \x/\y in {1/2,2/4,3/6,4/11} {\node[proclabel] at (\x*.5-.25,7.5) {\y};}
		\foreach \x/\y in {1/1,2/3,3/8,4/11} {\node[proclabel] at (\x*.5-.25,8.5) {\y};}
		% label A
		\node[yscale=-1,scale=2] at (1,10.25) {\LARGE $\B$};
	\end{scope}

\end{tikzpicture}
	\caption{Triangle block distribution using the affine construction for SYMM of $\A$, $\B$ and $\CC$ with $c=3$, $P=12$. Processor ranks $0\leq k < P$ are shown in blue to indicate ownership of a block and block indices $0\leq i < c^2$ are shown in red. Distribution of each row block of $\B$  and each row block of $\CC$ among their $c+1$ processors is arbitrary as long as it is even.}\label{fig:TBD-SYMM-PAR}
\end{figure}

Our 2D algorithms (\cref{alg:2Dsyrk,alg:2Dsyr2k,alg:2Dsymm}) partition the computation based on the distribution of the symmetric matrix.
In this way, no communication of the symmetric matrix is required, and communication involves only the non-symmetric matrices.
As the triangle block distribution of the symmetric matrix partitions both rows and columns, the algorithms are classified as 2D even though they are indexed by a single value $k$ (see \Cref{fig:TBD-SYMM-PAR}).
Each processor computes all computations associated with any block of the symmetric matrix it owns. Two types of local computations are performed. The first is a \Call{GEMM}{} computation involving off-diagonal blocks. The second is a symmetric computation involving a diagonal block of the symmetric matrix.

\begin{algorithm}
	\caption{2D SYRK {\cite[Algorithm 2]{ABGKR23}}}
	\label{alg:2Dsyrk}
\begin{algorithmic}[1]
	\Require $\Pi$ is a set of processors with $|\Pi|=P=c(c+1)$ for some prime power $c$.
	\Require $\A$ is evenly subdivided into $c^2$ row blocks, and each row block $\A_{i*}$ is evenly divided across a set of $c+1$ processors $Q_i$ with $\A_{i*}^{(k)}$ owned by processor $k$ so that $\A_{R_k*}^{(k)} = \{\A_{i*}^{(k)}: i \in R_k\}$ is a set of data blocks of $\A$ owned by processor $k$.
	\Require $\CC_{{T}_{k}} = \{ \CC_{ij}: i>j \in R_k \} \cup \{\CC_{ii}: i\in D_k\}$ is extended triangle block owned by processor $k$.
	\Ensure $\CC \pluseq \A\A^T$ with $\CC_{{T}_{k}}$ owned by processor $k$.
	\Function{$\CC_{{T}_{k}} =$ 2D-SYRK}{$\CC_{{T}_{k}}, \A_{R_k*}^{(k)}, \Pi$}
	\State $k = \Call{MyRank}{\Pi}$
	\Statex\Comment{Store $c$ row blocks in row block set}
	\State Allocate array $\bar{\A}$ of $P$ blocks, each of size $\frac{n_1n_2}{c^2(c+1)}$ \label{line:2D:startcomm}
	\ForAll{$i \in R_k$}
	\ForAll{$k' \in Q_i \backslash \{k\}$}
	\State $\bar{\A}_{k'} = \A_{i*}^{(k)}$
	\EndFor
	\EndFor
	\State $\bar{\A}$ = \Call{All-to-All}{$\bar{\A},\Pi$}  \label{line:2D:AlltoAll}
	\ForAll{$i \in R_k$}
	\ForAll{$k' \in Q_i \backslash \{k\}$}
	\State Store $\bar{\A}_{k'}$ into $\A_{i*}$
	\EndFor
	\EndFor \label{line:2D:stopcomm}
	\Statex\Comment{Perform local GEMM and SYRK computations}
	\ForAll{$\CC_{ij} \in \CC_{{T}_{k}}$} \label{line:2D:startcomp}
	\State $\CC_{ij} \pluseq \A_{i*}{\A_{j*}}^T$ \label{line:2D:local-gemm-syrk}
	\EndFor\label{line:2D:stopcomp}
	\EndFunction
\end{algorithmic}
\end{algorithm}

\begin{algorithm}
	\caption{2D SYR2K}
	\label{alg:2Dsyr2k}
	\begin{algorithmic}[1]
		\Require $\Pi$ is a set of processors with $|\Pi|=P=c(c+1)$ for some prime power $c$.
		\Require $\A$ is evenly subdivided into $c^2$ row blocks, and each row block $\A_{i*}$ is evenly divided across a set of $c+1$ processors $Q_i$ with $\A_{i*}^{(k)}$ owned by processor $k$ so that $\A_{R_k*}^{(k)} = \{\A_{i*}^{(k)}: i \in R_k\}$ is a set of data blocks of $\A$ owned by processor $k$. Similarly, $\B_{R_k*}^{(k)}$ is defined and it is owned by processor $k$.
		\Require $\CC_{{T}_{k}} = \{ \CC_{ij}: i>j \in R_k \} \cup \{\CC_{ii}: i\in D_k\}$ is extended triangle block owned by processor $k$.
		\Ensure $\CC \pluseq \A\A^T$ with $\CC_{{T}_{k}}$ owned by processor $k$.
		\Function{$\CC_{{T}_{k}} =$ 2D-SYR2K}{$\CC_{{T}_{k}}, \A_{R_k*}^{(k)}, \B_{R_k*}^{(k)}, \Pi$}
		\State $k = \Call{MyRank}{\Pi}$			
		\Statex\Comment{Store $c$ row blocks in row block set}
		\State Allocate array $\bar{\A}, \bar{\B}$ of $P$ blocks, each of size $\frac{n_1n_2}{c^2(c+1)}$ \label{line:2Dsyr2k:startcomm}
		\ForAll{$i \in R_k$}
		\ForAll{$k' \in Q_i \backslash \{k\}$}
		\State $\bar{\A}_{k'} = \A_{i*}^{(k)}$
		\State $\bar{\B}_{k'} = \B_{i*}^{(k)}$
		\EndFor
		\EndFor
		\State $\bar{\A}$ = \Call{All-to-All}{$\bar{\A},\Pi$}\label{line:2Dsyr2k:AlltoAllA}
		\State $\bar{\B}$ = \Call{All-to-All}{$\bar{\B},\Pi$}\label{line:2Dsyr2k:AlltoAllB}
		\ForAll{$i \in R_k$}
		\ForAll{$k' \in Q_i \backslash \{k\}$}
		\State Store $\bar{\A}_{k'}$ into $\A_{i*}$
		\State Store $\bar{\B}_{k'}$ into $\B_{i*}$
		\EndFor
		\EndFor \label{line:2Dsyr2k:stopcomm}
		\Statex\Comment{Perform local GEMM and SYR2K computations}
		\ForAll{$\CC_{ij} \in \CC_{{T}_{k}}$} \label{line:2Dsyr2k:startcomp}
		\State $\CC_{ij} \pluseq  \A_{i*}{\B_{j*}}^T + \B_{i*}{\A_{j*}}^T$\label{line:2Dsyr2k:local-gemm-syr2k}
		\EndFor\label{line:2Dsyr2k:stopcomp}
		\EndFunction
	\end{algorithmic}
\end{algorithm}

\begin{algorithm}
	\caption{2D SYMM}
	\label{alg:2Dsymm}
	\begin{algorithmic}[1]	
		\Require $\Pi$ is a set of processors with $|\Pi|=P=c(c+1)$ for some prime power $c$.
		\Require $\A_{{T}_{k}} = \{ \A_{ij}: i>j \in R_k \} \cup \{\A_{ii}: i\in D_k\}$ is extended triangle block owned by processor $k$.
		\Require $\B$ is evenly subdivided into $c^2$ row blocks, and each row block $\B_{i*}$ is evenly divided across a set of $c+1$ processors $Q_i$ with $\B_{i*}^{(k)}$ owned by processor $k$ so that $\B_{R_k*}^{(k)} = \{\B_{i*}^{(k)}: i \in R_k\}$ is a set of data blocks of $\B$ owned by processor $k$. Similarly, $\CC_{R_k*}^{(k)}$ is defined and it is owned by processor $k$.
		\Ensure $\CC \pluseq \A\B$ with $\CC_{R_k*}^{(k)}$ owned by processor $k$.
		\Function{$\CC_{R_k*}^{(k)} =$ 2D-SYMM}{$\CC_{R_k*}^{(k)}, \A_{{T}_{k}}, \B_{R_k*}^{(k)}, \Pi$}
		\State $k = \Call{MyRank}{\Pi}$			
		\Statex\Comment{Store $c$ blocks of a row in a row block set}
		\State Allocate array $\bar{\B}$ of $P$ blocks, each of size $\frac{n_1n_2}{c^2(c+1)}$ \label{line:2Dsymm:startcomm}
		\ForAll{$i \in R_k$}
		\ForAll{$k' \in Q_i \backslash \{k\}$}
		\State $\bar{\B}_{k'} = \B_{i*}^{(k)}$
		\EndFor
		\EndFor
		\State $\bar{\B}$ = \Call{All-to-All}{$\bar{\B},\Pi$}\label{line:2Dsymm:AlltoAllB}
		\ForAll{$i \in R_k$}
		\ForAll{$k' \in Q_i \backslash \{k\}$}
		\State Store $\bar{\B}_{k'}$ into $\B_{i*}$
		\EndFor
		\EndFor \label{line:2Dsymm:stopcomm}
		\State Store initial $\CC_{R_k*}^{(k)}$ in $\tilde{\CC}_{R_k*}^{(k)}$
		\State {Initialize all $c$ row blocks of $\CC$ to $\bm{0}$ that are partially owned by processor $k$}
		\Statex\Comment{Perform local GEMM and SYMM computations}
		\ForAll{$\A_{ij} \in \A_{{T}_{k}}$} \label{line:2Dsymm:startcomp}
		\State $\CC_{i*} \pluseq  \A_{ij}\B_{j*}$ \label{line:2Dsymm:local-gemm1}
		\State If ($i\ne j$) $\CC_{j*} \pluseq  \A_{ij}^T\B_{i*}$ \label{line:2Dsymm:local-gemm2-symm}
		\EndFor \label{line:2Dsymm:stopcomp}
		\Statex\Comment{Distribute and reduce results for each row block set}
		\State Allocate array $\bar{\CC}$ of $P$ blocks, each of size $\frac{n_1n_2}{c^2(c+1)}$
		\ForAll{$i \in R_k$}
		\ForAll{$k' \in Q_i \backslash \{k\}$}
		\State $\bar{\CC}_{k'} = \CC_{i*}^{(k')}$
		\EndFor
		\EndFor
		\State $\bar{\CC}$ = \Call{All-to-All}{$\bar{\CC},\Pi$}\label{line:2Dsymm:AlltoAllC}
		\ForAll{$i \in R_k$}
		\State $\CC_{i*}^{(k)} \pluseq \tilde{\CC}_{i*}^{(k)}$
		\ForAll{$k' \in Q_i \backslash \{k\}$}
		\State $\CC_{i*}^{(k)} \pluseq \bar{\CC}_{k'}$
		\EndFor
		\EndFor
		\EndFunction
	\end{algorithmic}
\end{algorithm}

We now illustrate how a row block of $\CC$ is updated by \cref{alg:2Dsymm}.
Let us focus on the sixth row block of $\CC$, $\CC_{6*}$ of \cref{fig:TBD-SYMM-PAR}.
Processors $\{0,5,7,11\}$ are responsible for the following updates:
processor $0$ performs $\A_{60} \B_{0*} + \A_{63} \B_{3*}$, processor $5$ performs $\A_{61} \B_{1*} +\A_{65} \B_{5*} + \text{SYMM}(\A_{66}, \B_{6*})$, processor $7$ performs $\A_{62} \B_{2*} + \A_{64} \B_{4*}$, and processor $11$ performs $\A_{76}^T \B_{7*} + \A_{86}^T \B_{8*}$.
After these updates are complete, a reduce-scatter operation is performed on these processors to update $\CC_{6*}$.
Each of these processors owns $\frac{1}{c+1}$th portion of updated $\CC_{6*}$.
Each processor is involved in partial update of $c$ blocks.
For example, processor $0$ is partially responsible for the update of $\CC_{0*}, \CC_{3*}$ and $\CC_{6*}$ row blocks.

\subsubsection{Cost Analysis}

As with our 1D algorithms, we derive the computational and communication costs by following the analysis in \cite{ABGKR23}, then derive how much memory is required for the computations.

\paragraph{Computation Cost}
All algorithms perform their local computations, \cref{line:2D:local-gemm-syrk} of \cref{alg:2Dsyrk}, \cref{line:2Dsyr2k:local-gemm-syr2k} of \cref{alg:2Dsyr2k}, and \cref{line:2Dsymm:local-gemm1,line:2Dsymm:local-gemm2-symm} of \cref{alg:2Dsymm}, as matrix multiplications where the block of the symmetric matrix involved has dimensions $(n_1/c^2)\times(n_1/c^2)$ and the row blocks of each non-symmetric matrix involved has dimensions $(n_1/c^2) \times n_2$.
Thus the local computations, which involve a matrix multiplication for each non-symmetric matrix, have computational cost $m \gamma \cdot 2\left(\frac{n_1}{c^2}\right)^2 n_2$.
As  $|R_k|=c$ and $|D_k| \in \{0,1\}$ for each $k$, most processors perform $c(c-1)/2$ local GEMM computations and one local symmetric computation, giving a total cost of
$$m\gamma \cdot \left[ \frac{c(c-1)}{2} \cdot 2\left(\frac{n_1}{c^2}\right)^2 n_2 + \left(\frac{n_1}{c^2}\right)^2 n_2 \right] = m\gamma \cdot \left[\frac{n_1^2n_2}{c^2} + O\left(\frac{n_1^2n_2}{c^3}\right)\right]\text.$$
Noting that $P=c^2+c$ implies $c=\sqrt{P+1/4}-1/2$, the leading order cost becomes
\begin{equation}
\label{eq:2D:compcost}
m\gamma \cdot \frac{n_1^2n_2}{P-c} = m\gamma \cdot \left[ \frac{n_1^2n_2}{P} +O\left(\frac{n_1^2n_2}{P^{3/2}}\right) \right]\text.
\end{equation}
As $c$ processors do not perform the computations with a diagonal block of the symmetric matrix, with $c\approx \sqrt P$, the computation is not perfectly load balanced.
However, the imbalance does not affect the leading order term.

\paragraph{Communication Cost}
All algorithms perform communication in the \Call{All-To-All}{} collectives, \cref{line:2D:AlltoAll} of \cref{alg:2Dsyrk}, \cref{line:2Dsyr2k:AlltoAllA,line:2Dsyr2k:AlltoAllB} of \cref{alg:2Dsyr2k}, and \cref{line:2Dsymm:AlltoAllB,line:2Dsymm:AlltoAllC} of \cref{alg:2Dsymm}.
We perform the \Call{All-to-All}{} collectives using a pairwise exchange algorithm over $|\Pi|=P$ processors, and accumulate the elements into arrays of size $\frac{n_1n_2}{c^2(c+1)}P = \frac{n_1n_2}{c}$.
Thus, the total communication cost for each of the three algorithms is $\alpha \cdot (P-1) + \beta \cdot m\frac{n_1n_2}{c}\left(1-1/P\right).$
As $c=\sqrt{P+1/4}-1/2$, this can be rewritten as
\begin{equation}
\label{eq:2D:commcost}
  \alpha\cdot(P-1)+m\beta\cdot\frac{n_1n_2}{\sqrt{P}}\left(\sqrt{1+\frac{1}{4P}} + \frac{1}{2\sqrt{P}}\right)\left(1-\frac{1}{P}\right)\text.
\end{equation}

\paragraph{Memory Requirement}
At the start of the computation, each processor owns $\frac{n_1n_2/c^2}{c+1} \cdot c = \frac{n_1n_2}{P}$ elements of each non-symmetric matrix.
Thus each \Call{All-to-All}{} requires $\frac{n_1n_2}{c^2(c+1)} \cdot P = \frac{n_1n_2}{c}$ additional words of memory per non-symmetric matrix.
While we use accumulate blocks in the statement of the algorithms to simplify the presentations, we assume that a pairwise exchange algorithm is used to perform \Call{All-to-All}{}.
This directly enables one to accumulate data from each row block into contiguous locations in memory of the receiving processor so no extra memory is required to rearrange the received data.
In addition to the memory required for the non-symmetric matrices, each processor computes $\frac{c(c-1)}{2} \cdot \frac{n_1^2}{c^4}$ elements of the symmetric matrix, and an additional $\frac{n_1}{2c^2}\left(\frac{n_1}{c^2}+1\right)$ elements for a processor that also computes a diagonal block.
Thus the total memory required for the algorithms is  $m\frac{n_1n_2}{P} + m\frac{n_1n_2}{c} + \frac{c(c-1)}{2} \cdot \frac{n_1^2}{c^4} + \frac{n_1}{2c^2}\left(\frac{n_1}{c^2}+1\right)$, whose leading order terms are $m \frac{n_1n_2}{\sqrt{P}} + \frac{n_1^2}{2P}$.
As $mn_2<n_1$ and $P \leq \frac{n_1(n_1-1)}{m^2n_2^2}$, $\frac{mn_1n_2}{\sqrt{P}}$ is bounded by $\frac{n_1^2}{P}$ and the memory use of our 2D algorithms is only a constant time more than the memory needed by a processor to store its owned data.

\subsection{3D Algorithms}
\label{sec:memIndepAlgorithms:3Dalgorithms}

3D algorithms are optimal when $P$ is large enough that it exceeds its bound for either the 1D or the 2D algorithm.
In the case of symmetric three nested loop computations, when the symmetric matrix is small relative to the non-symmetric matrices, $n_1 \leq mn_2$, a 3D algorithm becomes optimal when $P > mn_2/\sqrt{n_1(n_1-1)}$.
When the symmetric matrix is large relative to the non-symmetric matrices, $mn_2 < n_1$, a 3D algorithm becomes optimal when $P > n_1(n_1-1)/(m^2n_2^2)$.
In 3D algorithms, all matrices are communicated.

\subsubsection{Data Distribution and Algorithms}
The 3D algorithms combine the 1D and 2D algorithms by performing the 2D algorithms across each slice of processors in the 1D distribution.
In order for this to work, the data distributions can not be arbitrary for any array.
The distribution of the symmetric matrix must be compatible with the 2D triangle block distribution for the symmetric matrix, and the distribution of the non-symmetric matrices must be in row blocks which are compatible with the 2D distribution of the symmetric matrix.
SYRK and SYR2K, perform the 2D symmetric computation using $n_2/p_2$ columns of the non-symmetric matrices in each slice.
Then the outputs from each 2D computation are summed across the slice to compute $\CC$.
For SYMM, triangle blocks of the symmetric input matrix $\A$ are all gathered.
Then the 2D SYMM is performed using $n_2/p_2$ columns of $\B$ to compute the final result $\CC$ in the slices.

\begin{algorithm}
	\caption{3D SYRK {\cite[Algorithm 3]{ABGKR23}}}
	\label{alg:3D}
	\begin{algorithmic}[1]
		\Require $\Pi$ is a $p_1\times p_2$ grid of processors where $p_1=c(c+1)$ for some prime power $c$.
		\Require $\A$ is evenly divided into a $c^2\times p_2$ grid of blocks, and each block $\A_{i\ell}$ is evenly divided across a set of $c+1$ processors $Q_{i}\times \{\ell\}$ with $\A_{i\ell}^{(k)}$ owned by processor rank $(k,\ell)$ so that $\A_{R_k\ell}^{(k)} = \{\A_{i\ell}^{(k)}: i \in R_k\}$ is a set of data blocks of $\A$ owned by processor rank $(k,\ell)$.
		\Require $\CC_{{T}_{k}} = \{ \CC_{ij}: i>j \in R_k\} \cup \{\CC_{ii}: i\in D_k\}$ is extended triangle block distributed across processor ranks $(k,*)$ with $\CC_{{T}_{k}}^{(\ell)}$ owned by processor rank $(k,\ell)$.
		\Ensure $\CC \pluseq \A\A^T$ with $\CC_{{T}_{k}}^{(\ell)}$ owned by processor rank $(k,\ell)$.		
		\Function{$\CC_{{T}_{k}}^{(\ell)} =$ 3D-SYRK}{$\CC_{{T}_{k}}^{(\ell)}, \A_{R_k\ell}^{(k)}, \Pi$}	
		\State $(k,\ell) = \Call{MyRank}{\Pi}$			
		\State $\overline{\CC}_{{T}_{k}} = \Call{2D-SYRK}{\mathbf{0}, \A_{R_k\ell}^{(k)},\Pi_{*\ell}}$ \label{line:3D:2Dsyrk}
		\State $\CC_{{T}_{k}}^{(\ell)} \pluseq \Call{Reduce-Scatter}{\overline{\CC}_{{T}_{k}}, \Pi_{k*}}$ \label{line:3D:reduce-scatter}
		\EndFunction
	\end{algorithmic}
\end{algorithm}

\begin{algorithm}
\caption{3D SYR2K}
\label{alg:3Dsyr2k}
\begin{algorithmic}[1]
	\Require $\Pi$ is a $p_1\times p_2$ grid of processors where $p_1=c(c+1)$ for some prime power $c$.
	\Require $\A$ is evenly divided into a $c^2\times p_2$ grid of blocks, and each block $\A_{i\ell}$ is evenly divided across a set of $c+1$ processors $Q_{i}\times \{\ell\}$ with $\A_{i\ell}^{(k)}$ owned by processor rank $(k,\ell)$ so that $\A_{R_k\ell}^{(k)} = \{\A_{i\ell}^{(k)}: i \in R_k\}$ is a set of data blocks of $\A$ owned by processor rank $(k,\ell)$. Similarly, $\B_{R_k\ell}^{(k)}$ is defined and it is owned by processor rank $(k,\ell)$.
	\Require $\CC_{{T}_{k}} = \{ \CC_{ij}: i>j \in R_k\} \cup \{\CC_{ii}: i\in D_k\}$ is extended triangle block distributed across processor ranks $(k,*)$ with $\CC_{{T}_{k}}^{(\ell)}$ owned by processor rank $(k,\ell)$.
	\Ensure $\CC \pluseq \A\B^T + \B\A^T$ with $\CC_{{T}_{k}}^{(\ell)}$ owned by processor rank $(k,\ell)$.	
	\Function{$\CC_{{T}_{k}}^{(\ell)} =$ 3D-SYR2K}{$\CC_{{T}_{k}}^{(\ell)}, \A_{R_k\ell}^{(k)}, \B_{R_k\ell}^{(k)}, \Pi$}	
	\State $\overline{\CC}_{{T}_{k}} = \Call{2D-SYR2K}{\mathbf{0}, \A_{R_k\ell}^{(k)}, \B_{R_k\ell}^{(k)}, \Pi_{*\ell}}$ \label{line:3Dsyr2k:2Dsyr2k}
	\State $\CC_{{T}_{k}}^{(\ell)} \pluseq \Call{Reduce-Scatter}{\overline{\CC}_{{T}_{k}}, \Pi_{k*}}$ \label{line:3Dsyr2k:reduce-scatter}
	\EndFunction
\end{algorithmic}
\end{algorithm}

\begin{algorithm}
	\caption{3D SYMM}
	\label{alg:3Dsymm}
	\begin{algorithmic}[1]
		\Require $\Pi$ is a $p_1\times p_2$ grid of processors where $p_1=c(c+1)$ for some prime power $c$.
		\Require $\A_{{T}_{k}} = \{ \A_{ij}: i>j \in R_k\} \cup \{\A_{ii}: i\in D_k\}$ is extended triangle block distributed across processor ranks $(k,*)$ with $\A_{{T}_{k}}^{(\ell)}$ owned by processor rank $(k,\ell)$.
		\Require $\B$ is evenly divided into a $c^2\times p_2$ grid of blocks, and each block $\B_{i\ell}$ is evenly divided across a set of $c+1$ processors $Q_{i}\times \{\ell\}$ with $\B_{i\ell}^{(k)}$ owned by processor rank $(k,\ell)$ so that $\B_{R_k\ell}^{(k)} = \{\B_{i\ell}^{(k)}: i \in R_k\}$ is a set of data blocks of $\B$ owned by processor rank $(k,\ell)$. Similarly, $\CC_{R_k\ell}^{(k)}$ is defined and it is owned by processor rank $(k,\ell)$.
		\Ensure $\CC \pluseq \A\B$ with $\CC_{R_k\ell}^{(k)}$ owned by processor $(k,\ell)$.		
		\Function{$\CC_{R_k\ell}^{(k)} =$ 3D-SYMM}{$\CC_{R_k\ell}^{(k)}, \A_{{T}_{k}}^{(\ell)}, \B_{R_k\ell}^{(k)}, \Pi$}
		\State $(k,\ell) = \Call{MyRank}{\Pi}$			
		\State $\A_{{T}_{k}} = \Call{All-Gather}{\A_{{T}_{k}}^{(\ell)}, \Pi_{k*}}$ \label{line:3Dsymm:all-gather}
		\State $\CC_{R_k\ell}^{(k)} = \Call{2D-SYMM}{\CC_{R_k\ell}^{(k)}, \A_{{T}_{k}}, \B_{R_k\ell}^{(k)}, \Pi_{*\ell}}$ \label{line:3Dsymm:2Dsymm}
		\EndFunction
	\end{algorithmic}
\end{algorithm}

\subsubsection{Cost Analysis}

\paragraph{Computation Cost}
All algorithms perform their 2D computations, \cref{line:3D:2Dsyrk} of \cref{alg:3D,alg:3Dsyr2k} and \cref{line:3Dsymm:2Dsymm} of \cref{alg:3Dsymm}, with $p_1$ processors and non-symmetric input matrices of dimension $n_1\times n_2/p_2$.
Hence, we can determine the computation cost from the cost for the 2D algorithms, \cref{eq:2D:compcost}, by replacing $P$ with $p_1$ and using $n_2/p_2$ instead of $n_2$  to get $m\gamma \cdot \left[\frac{n_1^2n_2}{P} + O\left(\frac{n_1^2n_2}{Pp_1^{1/2}}\right)\right].$
Additional computation is performed in the \Call{Reduce-Scatter}{}, \cref{line:3D:reduce-scatter} of \cref{alg:3D,alg:3Dsyr2k}, but the cost is a lower order term.

\paragraph{Communication Cost}
To begin, we consider the communication cost of the call to the relevant 2D algorithm, \cref{line:3D:2Dsyrk} of \cref{alg:3D,alg:3Dsyr2k} and \cref{line:3Dsymm:2Dsymm} of \cref{alg:3Dsymm}.
We derive the relevant communication cost from \cref{eq:2D:commcost} by noting that the 2D algorithm is performed on non-symmetric matrices with dimensions $n_1\times n_2/p_2$ using $p_1$ processors so the relevant cost is $\alpha \cdot (p_1-1) + m\beta\cdot\frac{n_1n_2}{cp_2}\left(1-\frac{1}{p_1}\right)$.
Additional communication occurs in the \Call{Reduce-Scatter}{}, \cref{line:3D:reduce-scatter} of \cref{alg:3D,alg:3Dsyr2k}, or \Call{All-Gather}{}, \cref{line:3Dsymm:all-gather} of \cref{alg:3Dsymm}, collectives on the symmetric matrix.
The communication collective uses $p_2$ processors and is called on a triangle block of blocks, possibly containing a diagonal block, which has a maximum size of $\frac{c(c{-}1)}{2}\cdot\frac{n_1^2}{c^4} + \frac12 \cdot \frac{n_1}{c^2}\left(\frac{n_1}{c^2}+1\right).$
Thus its cost is dominated by $\alpha \cdot \left(p_2-1\right) + \beta \cdot \frac12 \frac{n_1^2}{c^2} \left(1-\frac{1}{p_2}\right).$

Combining the communication cost of the 2D computations and communicating the symmetric matrix across slices, we obtain a total communication cost, in leading order terms, of
$$\alpha\cdot (p_1 + p_2) +\beta\cdot\left( m\frac{n_1n_2}{cp_2}\left(1-\frac{1}{p_1}\right) + \frac12\frac{n_1^2}{c^2} \left(1-\frac{1}{p_2}\right)\right)\text.$$
Given that $c=\sqrt{p_1+1/4} -\frac{1}{2}$, the communication cost is dominated by
\begin{equation}
\label{eq:3D:commcost}
\alpha\cdot \left(p_1+ p_2\right) + \beta \cdot\left(m\frac{n_1n_2}{\sqrt{p_1}p_2}+ \frac{n_1^2}{2p_1}\right)\text.
\end{equation}

\paragraph{Memory Requirement}
To determine the memory use, we note that the algorithms perform the relevant 2D algorithm on a problem of size $n_1\times n_2/p_2$ using $p_1$ processors.
Additionally, \cref{alg:3Dsymm} performs the initial \Call{All-Gather}{} on $p_2$ processors and \cref{alg:3D,alg:3Dsyr2k} perform the final \Call{Reduce-Scatter}{} on $p_2$ processors require storage for $\frac{n_1(n_1+1)}{2P}p_2 = \frac{n_1(n_1+1)}{2p_1}$ elements.
However this memory requirement is also included in the relevant 2D algorithm.
Thus the total memory required (in leading order terms) for our 3D algorithms is $m\frac{n_1n_2}{\sqrt{p_1}p_2}+\frac{n_1^2}{2p_1}$.

\subsection{Optimal Processor Grid Selection}
\label{sec:opt}
The selection of the optimal processor grid follows that given for SYRK in \cite{ABGKR23}.
We derive the communication minimizing processor grid dimensions from the communication lower bound, \cref{thm:memindeplb}.
Communication minimizing values of $p_1$ and $p_2$ are specific to each of the three cases of the communication lower bound, so we will address each case separately.
For simplicity, in each case we will assume that $p_1$ and $p_2$ are integers, that each expression involving division has an integer solution, and, in case 2 and 3, that there exists a prime power $c$ with the property that $c(c+1)=p_1$.

Case 1:  $n_1\leq mn_2\text{ and }P \leq \frac{mn_2}{\sqrt{n_1(n_1-1)}}$.
In this scenario, the 1D algorithms, \cref{alg:1Dsyrk,alg:1Dsyr2k,alg:1Dsymm}, are optimal.
To use the 1D algorithms, the non-symmetric matrices are split into column blocks so $p_1=1$ and $p_2=P$.
Then the bandwidth cost, $\beta\cdot\frac{n_1(n_1+1)}{2}\left(1-\frac{1}{P}\right)$, given in \cref{eq:1D:commcost} exactly matches the leading order term of the lower bound.

Case 2: $n_1> mn_2\text{ and }P \leq \frac{n_1(n_1-1)}{m^2n_2^2}$.
In this scenario, the 2D algorithms, \cref{alg:2Dsyrk,alg:2Dsyr2k,alg:2Dsymm}, are optimal.
To use the 2D algorithms, the symmetric matrix is divided into triangle blocks of blocks so $p_1=P$ and $p_2=1$.
Then the bandwidth cost, $m\beta\cdot \frac{n_1n_2}{\sqrt{P}} \left((1+\frac{1}{4P})^{1/2} + \frac{1}{2P^{1/2}}\right)\left(1-\frac{1}{P}\right)$, given in \cref{eq:2D:commcost} exactly matches the leading order term of the lower bound.

Case 3: $n_1\leq mn_2\text{ and }P > \frac{mn_2}{\sqrt{n_1(n_1-1)}} \text{ or } n_1{>}mn_2\text{ and } P > \frac{n_1(n_1-1)}{m^2n_2^2}$.
In this scenario, the 3D algorithms, \cref{alg:3D,alg:3Dsyr2k,alg:3Dsymm}, are optimal.
To determine the optimal settings of $p_1$ and $p_2$ we note that the communication cost for each matrix should match the communication lower bound for this scenario.
Before we do this, recall that $p_2 = P/p_1$ so the communication cost associated with communicating each non-symmetric matrix, $(n_1n_2)/(\sqrt{p_1}p_2)$, from the first term of \cref{eq:3D:commcost} can instead be written as $\frac{n_1n_2\sqrt{p_1}}{P}$.
We set this equal to  $\left(\frac{n_1^2n_2}{\sqrt{m}P}\right)^{2/3}$
and solve for $p_1$ to get $p_1 = \left(\frac{n_1P}{mn_2}\right)^{2/3}$.
Again using $p_1=P/p_1$ we must have $p_2=\left(\frac{mn_2}{n_1}\right)^{2/3}P^{1/3}$.
We use these values in the bandwidth cost for the 3D algorithms, $m\frac{n_1n_2}{\sqrt{p_1}p_2}+ \frac{n_1^2}{2p_1}$, given in \cref{eq:3D:commcost} and simplify to get the communication cost $\frac{3m}{2}\left(\frac{n_1^2n_2}{\sqrt{m}P}\right)^{2/3}$ which exactly matches the leading order term of the lower bound.

\section{Limited Memory Parallel Algorithms}
\label{sec:limitedMemory}
In this section, we focus on scenarios where processors do not have enough memory to run the 3-dimensional algorithms.
We modify the 3-dimensional algorithms for all the three computations to limit the memory needed to store intermediate data and call them 3-dimensional algorithms for limited memory.
Note that we do not focus on limited memory scenarios for 1D and 2D algorithms as their memory requirements are on the order of the data owned by a processor.

We assume that each processor has sufficient memory to store the elements of the symmetric matrix that it will use during the computation along with some additional memory.
Specifically we assume there exists $x > 1$ such that the processor has at least  $x\frac{n_1^2}{2P}$ words of memory.
Our algorithms keep the required portion of the symmetric matrix in the processor's local memory, and gather or compute the required portions of non-symmetric matrices in small chunks.
This requires that each small chunk of a non-symmetric matrix is evenly distributed on a set processors from which data is gathered for input matrices and on which reduce-scatter is performed for the output matrix.
Note that this is a more precise distribution for the non-symmetric matrices than the distribution needed to run the algorithms of \cref{sec:memIndepAlgorithms:3Dalgorithms}.

We present our SYRK, SYR2K and SYMM algorithms for limited memory in \cref{alg:3Dsyrk-limitedmemory,alg:3Dsyr2k-limitedmemory,alg:3Dsymm-limitedmemory}.
These algorithms work with $b$ columns of non-symmetric matrices at once.
They call their corresponding 2D algorithms $\frac{n_2/p_2}{b}$ times.
For $b=n_2/p_2$, the algorithms are equivalent to the 3D algorithms of \cref{sec:memIndepAlgorithms:3Dalgorithms}.
We use $\ell[t]$ to denote the $b$ columns accessed at the $t$th step from all columns of index $\ell$.
All columns of $\ell$ are processed in $\frac{n_2/p_2}{b}$ steps.

\begin{algorithm}
	\caption{3D SYRK for limited memory}
	\label{alg:3Dsyrk-limitedmemory}
	\begin{algorithmic}[1]		
		\Require $\Pi$ is a $p_1\times p_2$ grid of processors where $p_1=c(c+1)$ for some prime power $c$.
		\Require $\A$ is evenly divided into a $c^2\times p_2$ grid of blocks, and each block $\A_{i\ell}$ is evenly divided across a set of $c+1$ processors $Q_{i}\times \{\ell\}$ with $\A_{i\ell}^{(k)}$ owned by processor rank $(k,\ell)$ so that $\A_{R_k\ell}^{(k)} = \{\A_{i\ell}^{(k)}: i \in R_k\}$ is a set of data blocks of $\A$ owned by processor rank $(k,\ell)$.
		\Require $b$ is the number of columns of $\A$ processed at each step.
		\Require $\CC_{{T}_{k}} = \{ \CC_{ij}: i>j \in R_k\} \cup \{\CC_{ii}: i\in D_k\}$ is extended triangle block distributed across processor ranks $(k,*)$ with $\CC_{{T}_{k}}^{(\ell)}$ owned by processor rank $(k,\ell)$.
		\Ensure $\CC \pluseq \A\A^T$ with $\CC_{{T}_{k}}^{(\ell)}$ owned by processor rank $(k,\ell)$.		
		\Function{$\CC_{{T}_{k}}^{(\ell)} =$ 3D-SYRK}{$\CC_{{T}_{k}}^{(\ell)}, \A_{R_k\ell}^{(k)}, \Pi, b$}
		\State $(k,\ell) = \Call{MyRank}{\Pi}$			
		\State $\overline{\CC}_{{T}_{k}} = 	\mathbf{0}$
		\Statex\Comment{Work with $b$ columns of $\A$ at each step}
		\For{$t=1 \text{ to } \frac{n_2}{p_2}$ in step size of $b$}
		\State $\overline{\CC}_{{T}_{k}} = \Call{2D-SYRK}{\overline{\CC}_{{T}_{k}}, \A_{R_k\ell[t]}^{(k)},\Pi_{*\ell}}$ \label{line:3Dsyrk-eachstep:2Dsyrk}
		\Comment{$\A_{R_k\ell[t]}^{(k)}$ is set of data blocks of $b$ columns}
		\EndFor
		\State $\CC_{{T}_{k}}^{(\ell)} \pluseq \Call{Reduce-Scatter}{\overline{\CC}_{{T}_{k}}, \Pi_{k*}}$ \label{line:3Dsyrk-limitedMemory:reduce-scatter}
		\EndFunction
	\end{algorithmic}
\end{algorithm}

\begin{algorithm}
	\caption{3D SYR2K for limited memory}
	\label{alg:3Dsyr2k-limitedmemory}
	\begin{algorithmic}[1]
		\Require $\Pi$ is a $p_1\times p_2$ grid of processors where $p_1=c(c+1)$ for some prime power $c$.
		\Require $\A$ is evenly divided into a $c^2\times p_2$ grid of blocks, and each block $\A_{i\ell}$ is evenly divided across a set of $c+1$ processors $Q_{i}\times \{\ell\}$ with $\A_{i\ell}^{(k)}$ owned by processor rank $(k,\ell)$ so that $\A_{R_k\ell}^{(k)} = \{\A_{i\ell}^{(k)}: i \in R_k\}$ is a set of data blocks of $\A$ owned by processor rank $(k,\ell)$.  Similarly, $\B_{R_k\ell}^{(k)}$ is defined and it is owned by processor rank $(k,\ell)$.
		\Require $b$ is the number of columns of $\A$ processed at each step.
		\Require $\CC_{{T}_{k}} = \{ \CC_{ij}: i>j \in R_k\} \cup \{\CC_{ii}: i\in D_k\}$ is extended triangle block distributed across processor ranks $(k,*)$ with $\CC_{{T}_{k}}^{(\ell)}$ owned by processor rank $(k,\ell)$.
		\Ensure $\CC \pluseq \A\B^T + \B\A^T$ with $\CC_{{T}_{k}}^{(\ell)}$ owned by processor rank $(k,\ell)$.
		\Function{$\CC_{{T}_{k}}^{(\ell)} =$ 3D-SYR2K}{$\CC_{{T}_{k}}^{(\ell)}, \A_{R_k\ell}^{(k)}, \B_{R_k\ell}^{(k)}, (k,\ell), \Pi, b$}
		\State $(k,\ell) = \Call{MyRank}{\Pi}$	
		\State $\overline{\CC}_{{T}_{k}} = 	\mathbf{0}$
		\Statex\Comment{Work with $b$ columns of $\A$ and $\B$ at each step}
		\For{$t=1 \text{ to } \frac{n_2}{p_2}$ in step size of $b$}
		\Statex\Comment{$\A_{R_k\ell[t]}^{(k)}$ and $\B_{R_k\ell[t]}^{(k)}$ are sets of data blocks of $b$ columns}
		\State $\overline{\CC}_{{T}_{k}} = \Call{2D-SYR2K}{\overline{\CC}_{{T}_{k}}, \A_{R_k\ell[t]}^{(k)}, \B_{R_k\ell[t]}^{(k)},\Pi_{*\ell}}$  \label{line:3Dsyr2k-eachstep:2Dsyr2k}
		\EndFor
		\State $\CC_{{T}_{k}}^{(\ell)} \pluseq \Call{Reduce-Scatter}{\overline{\CC}_{{T}_{k}}, \Pi_{k*}}$  \label{line:3Dsyr2k-limitedMemory:reduce-scatter}
		\EndFunction
	\end{algorithmic}
\end{algorithm}

\begin{algorithm}
	\caption{3D SYMM for limited memory}
	\label{alg:3Dsymm-limitedmemory}
	\begin{algorithmic}[1]
		\Require $\Pi$ is a $p_1\times p_2$ grid of processors where $p_1=c(c+1)$ for some prime power $c$.
		\Require $\A_{{T}_{k}} = \{ \A_{ij}: i>j \in R_k\} \cup \{\A_{ii}: i\in D_k\}$ is extended triangle block distributed across processor ranks $(k,*)$ with $\A_{{T}_{k}}^{(\ell)}$ owned by processor rank $(k,\ell)$.
		\Require $\B$ is evenly divided into a $c^2\times p_2$ grid of blocks, and each block $\B_{i\ell}$ is evenly divided across a set of $c+1$ processors $Q_{i}\times \{\ell\}$ with $\B_{i\ell}^{(k)}$ owned by processor rank $(k,\ell)$ so that $\B_{R_k\ell}^{(k)} = \{\B_{i\ell}^{(k)}: i \in R_k\}$ is a set of data blocks of $\B$ owned by processor rank $(k,\ell)$.  Similarly, $\CC_{R_k\ell}^{(k)}$ is defined and it is owned by processor rank $(k,\ell)$.
		\Require $b$ is the number of columns of $\B$ processed at each step.
		\Ensure $\CC \pluseq \A\B$ with $\CC_{R_k\ell}^{(k)}$ owned by processor rank $(k,\ell)$.				
		\Function{$\CC_{R_k\ell}^{(k)} =$ 3D-SYMM}{$\CC_{R_k\ell}^{(k)}, \A_{{T}_{k}}^{(\ell)}, \B_{R_k\ell}^{(k)}, \Pi, b$}
		\State $(k,\ell) = \Call{MyRank}{\Pi}$
		\State $\A_{{T}_{k}} = \Call{All-Gather}{\A_{{T}_{k}}^{(\ell)}, \Pi_{k*}}$ \label{line:3Dsymm-limitedmemory:all-gather}
		\Statex\Comment{Work with $b$ columns of $\B$ at each step}
		\For{$t=1 \text{ to } \frac{n_2}{p_2}$ in step size of $b$}
		\Statex\Comment{$\B_{R_k\ell[t]}^{(k)}$ and $\CC_{R_k\ell[t]}^{(k)}$ are sets of data blocks of $b$ columns}
		\State $\CC_{R_k\ell[t]}^{(k)} = \Call{2D-SYMM}{\CC_{R_k\ell[t]}^{(k)}, \A_{{T}_{k}}, \B_{R_k\ell[t]}^{(k)}, \Pi_{*\ell}}$ \label{line:3Dsymm-eachstep:2Dsymm}
		\EndFor
		\EndFunction
	\end{algorithmic}
\end{algorithm}

\subsection{Cost Analysis}
\label{sec:limitedMemory:cost-analysis}
We remind the reader that $m$ denotes the number of non-symmetric matrices involved in a computation.
The values of $m$ for SYRK, SYR2K and SYMM are 1, 2 and 2, respectively. We also have $p_1=c(c+1)$ for some prime power $c$ and $p_1p_2=P$.

\paragraph{Computation Cost}
As with the 3D-algorithms of \cref{sec:memIndepAlgorithms:3Dalgorithms}, almost all the computations occur in the calls to the 2D-algorithms, \cref{line:3Dsyrk-eachstep:2Dsyrk} of \cref{alg:3Dsyrk-limitedmemory,alg:3Dsyr2k-limitedmemory,alg:3Dsymm-limitedmemory}, with $p_1$ processors and non-symmetric matrices of dimensions $n_1\times b$.
Thus we can derive the computational cost of each of the calls from the cost of the 2D algorithms, \cref{eq:2D:compcost}, to be $m\gamma\cdot\left[\frac{n_1^2b}{p_1}+O\left(\frac{n_1^2b}{p_1^{3/2}}\right)\right]$.
As each algorithm calls the 2D algorithm $n_2/(p_2b)$ times, the leading order term of the cost for all calls to the 2D algorithms is $m\gamma\left(\frac{n_1^2n_2}{P}\right)$.
Additional computations occur in the \Call{Reduce-Scatter}{} in \cref{line:3Dsyrk-limitedMemory:reduce-scatter} of \cref{alg:3Dsyrk-limitedmemory,alg:3Dsyr2k-limitedmemory}, but, as in our analysis of the original 3D-algorithms, the cost is a lower order term so we ignore it in our analysis.

\paragraph{Communication Cost}
As we did when determining the computation cost, we begin determining the communication cost by considering the calls to the relevant 2D algorithms, \cref{line:3Dsyrk-eachstep:2Dsyrk} of \cref{alg:3Dsyrk-limitedmemory,alg:3Dsyr2k-limitedmemory,alg:3Dsymm-limitedmemory}.
Again, we note that these 2D algorithms are being called over $p_1$ processors with non-symmetric matrices of dimensions $n_1\times b$, and thus the communication cost, derived from the 2D communication cost given in \cref{eq:2D:commcost}, is $\alpha(p_1-1)+m\beta\frac{n_1b}{\sqrt{p_1}}\left(\sqrt{1-\frac{1}{4p_1}}+\frac{1}{2\sqrt{p_1}}\right)\left(1-\frac{1}{p_1}\right)$.
As before the 2D algorithms are called $n_2/(p_2b)$ times for each algorithm.
Additional communication occurs in the \Call{Reduce-Scatter}{} collective, \cref{line:3Dsyrk-limitedMemory:reduce-scatter} of \cref{alg:3Dsyrk-limitedmemory,alg:3Dsyr2k-limitedmemory}, or the \Call{All-Gather}{} collective, \cref{line:3Dsymm-limitedmemory:all-gather} of \cref{alg:3Dsymm-limitedmemory}.
As was the case in the 3D algorithms, the relevant communication collective uses $p_2$ processors and is called on a triangle block of blocks, possibly containing a diagonal block.
Thus its communication cost is dominated by $\alpha p_2 +\beta\frac{n_1^2}{2p_1}$.
Thus the overall communication cost is dominated by
\begin{equation}
  \label{eq:3D-limitedMemory:commcost}
  \alpha\left(p_1\frac{n_2}{bp_2}+p_2\right)+\beta\left(m\frac{n_1n_2}{\sqrt{p_1}p_2}+\frac{n_1^2}{2p_1}\right)\text.
\end{equation}

\paragraph{Memory Cost}
Each algorithm requires $mb\frac{n_1}{c} + \frac{c(c-1)}{2} \cdot \left(\frac{n_1}{c^2}\right)^2 + \frac12 \frac{n_1}{c^2} \left(\frac{n_1}{c^2}+1\right)$ intermediate memory for one instance of 2D call.
As different chunks of non-symmetric matrices are gathered or computed in each iteration, we can use the same memory for all the iterations.
As the same elements of the symmetric matrix are updated or computed in all the iterations, we require only one copy of the symmetric matrix in memory.
Therefore the total memory required by each algorithm is $mb\frac{n_1}{c} + \frac{c(c-1)}{2} \cdot \left(\frac{n_1}{c^2}\right)^2 + \frac12 \frac{n_1}{c^2} \left(\frac{n_1}{c^2}+1\right) + O_{d}\text,$
where $O_{d}$ denotes the number of elements owned by each processor at the beginning and end of the algorithm.
This equals $m\frac{n_1n_2}{P} + \frac{n_1^2(c-1)}{2c^3p_2}$ with an additional $\frac{n_1}{2c^2p_2} \left(\frac{n_1}{c^2}+1\right)$ if the processor is assigned a diagonal block.

\subsection{Optimality of \cref{alg:3Dsyrk-limitedmemory,alg:3Dsyr2k-limitedmemory,alg:3Dsymm-limitedmemory}}
\label{sec:limitedMemory:optimality}
As each processor has memory for at least $x\cdot \frac{n_1^2}{2P}$ elements for $x > 1$, we set $p_2=x$ and $p_1=P/x$.
With this setting, each processor keeps $x\cdot \frac{n_1^2}{2P}$ elements of the symmetric matrix in the memory and gathers (or computes and writes) non-symmetric matrices in small chunks parameterized by $b$.
When $b=1$, the total memory requirement of each algorithm is $m\frac{n_1}{c} + \frac{c(c-1)}{2} \cdot \left(\frac{n_1}{c^2}\right)^2 + \frac12 \frac{n_1}{c^2} \left(\frac{n_1}{c^2}+1\right) + O_{d} \approx \frac{n_1^2}{2c^2} +  O_{d} \approx x \cdot \frac{n_1^2}{2P} + O_{d}$.

The bandwidth cost of the algorithms with the above values of $p_1, p_2$, and $b$ is
$$\beta\left(m\frac{n_1n_2}{\sqrt{p_1}p_2}+\frac{n_1^2}{2p_1}\right) = \beta\left(m\frac{n_1n_2}{\sqrt{Px}} + x\cdot \frac{n_1^2}{2P}\right) \text.$$

As mentioned at the start of this section, we assume that each processor has much less than the required memory to run the 3d-algorithm, i.e, $x\cdot\frac{n_1^2}{P} = o \left( \left(\frac{n_1^2n_2}{P}\right)^{2/3} \right)$.
Therefore, we have $x\cdot\frac{n_1^2}{P} = o \left( \frac{n_1n_2}{\sqrt{Px}} \right)$.
We also have $M \approx x \cdot \frac{n_1^2}{2P} + O_{d}$.
Thus the number of words communicated is dominated by $m\frac{n_1n_2}{\sqrt{Px}} \approx \frac{m}{\sqrt{2}} \frac{n_1^2n_2}{P\sqrt{M-O_{d}}}$ which matches the leading term in the memory dependent parallel lower bound exactly when $O_d = o(M)$.

Note that with $b=1$, the latency cost of the algorithm is $\Omega(n_2/p_2)$ which may be large.
In order to minimize the latency cost we propose that $b$ is chosen to be as large as possible given the constraint that the size of the gathered (or partially computed) non-symmetric matrix or matrices is still the dominating term.
For example, setting $b=\log_2(n_1/c)$ or $b=\sqrt{n_1/c}$ significantly minimizes the latency cost while maintaining the same leading order term of the memory requirement and with the same bandwidth cost.

\section{Conclusion}
\label{sec:conclusion}

In this work, we establish communication lower bounds of SYRK, SYR2K and SYMM computations for sequential and distributed-memory parallel computational models. Our lower bound proofs rely on the symmetric Loomis-Whitney inequality~\cite{ABGKR22} that relates a subset of 3D loop iterations to the required data access within 2D matrices. We use this inequality along with other constraints to express communication lower bounds as solutions to constrained optimization problems. We solve these problems analytically to derive bounds. We also show that our bounds are tight by presenting communication optimal algorithms for each setting. Our algorithms use triangle block partitioning schemes to define triangular blocks that reduce data accesses for a subset of symmetric computations. We obtain the triangle block partitions from the solutions of the balanced clique partitions problem. In our algorithms, we construct triangle blocks based on affine and projective planes and express more a more general set of partitions using Steiner systems.

We believe that the symmetric Loomis-Whitney inequality and lower bound approach apply to other matrix computations involving symmetry and that the results can be generalized to computations with iteration spaces involving more than three dimensions, such as those involving symmetric tensors.
While triangle block partitions in the sequential case will likely lead to poor spatial locality (sometimes referred to as latency in the sequential model), we believe the parallel algorithms have promise to deliver high performance with reasonable software cost.
We leave implementations of the triangle block partition-based algorithms and verifications of the theoretical improvements leading to practical speedups to future work.

\section*{Acknowledgements}
    The authors thank Jeremy Rouse for showing us the relationship between triangle blocks and the clique partition problem, pointing us towards the geometric solutions of the clique partition problem, and creating the Magma scripts that generate the sets $\{R_k\}$ for a given prime power.

    This project received funding from the European Research Council (ERC) under European Union's Horizon 2020 research and innovation program (grant agreement 810367).  This work is supported by the National Science Foundation under grant CCF-1942892 and OAC-2106920. This material is based upon work supported by the US Department of Energy, Office of Science, Advanced Scientific Computing Research program under award DE-SC-0023296.

\bibliographystyle{IEEEtran}
\bibliography{refs}

\newpage

\appendix
\label{sec:appendix}

\section{Triangle Block Partitions using Projective Plane Construction}
\label{sec:appendix:projexample}
\begin{figure}[htb]
	\pgfmathsetmacro{\c}{4}
	\def\R{%
		0/{1,5,9,13,17}/17,%
		1/{1,6,11,16,18}/18,%
		2/{1,7,12,14,19}/19,%
		3/{1,8,10,15,20}/20,%
		4/{2,5,12,15,18}/2,%
		5/{2,6,10,14,17}/6,%
		6/{2,7,9,16,20}/9,%
		7/{2,8,11,13,19}/11,%
		8/{3,5,10,16,19}/3,%
		9/{3,6,12,13,20}/13,%
		10/{3,7,11,15,17}/7,%
		11/{3,8,9,14,18}/14,%
		12/{4,5,11,14,20}/5,%
		13/{4,6,9,15,19}/15,%
		14/{4,7,10,13,18}/10,%
		15/{4,8,12,16,17}/4,%
		16/{1,2,3,4,21}/1,%
		17/{5,6,7,8,21}/8,%
		18/{9,10,11,12,21}/12,%
		19/{13,14,15,16,21}/16,%
		20/{17,18,19,20,21}/21%
	}
	\centering
	%!TEX root = ../paper.tex

% set offset and matrix dim
\pgfmathsetmacro{\offset}{.33}
\pgfmathsetmacro{\m}{\c*\c+\c+1}
\pgfmathsetmacro{\n}{\m-1}

% set styles
\tikzset{proclabel/.style={yscale=-1,blue}}
\tikzset{rowlabel/.style={yscale=-1,red}}

% for drawing help	
\newcommand{\gridhelp}{
	\coordinate (ll) at (-1,-1);
	\coordinate (ur) at (10,10);
	\draw[help lines,dashed] (ll) grid (ur);
	\node at (0,0) {0};
}	

%\begin{tikzpicture}[every node/.append style={transform shape},yscale=-1,scale=.675]
\begin{tikzpicture}[every node/.append style={transform shape},yscale=-1,scale=.425]
%\gridhelp
	% draw little boxes
	\foreach \x in {1,...,\m} {\draw[dashed] (0,\x) -- (\x,\x) -- (\x,\m);}
	% draw big boxes
	\foreach \x in {0,...,\c} {\draw[thick] (0,\c*\x+\c) -- (\c*\x+\c,\c*\x+\c) -- (\c*\x+\c,\m);}
	% draw diagonal
	\foreach \x in {1,...,\m} {
		\draw[thick] (\x-1,\x-1) -- (\x,\x-1) -- (\x,\x);
		\draw[thick] (\x-1,\x-1) -- (\x-1,\x) -- (\x,\x);
	}
	% draw outer box
	\draw[thick] (0,0) rectangle +(\m,\m);
	% draw processor labels
	\foreach \p/\list/\d in \R {
		\foreach \x in \list {
			\foreach \y in \list {
				\ifnum \x<\y 
					\node[proclabel] at (\x-.5,\y-.5) {\p};
				\fi
			}
		}
		% diagonal block
		\ifnum \d>0
			\node[proclabel] at (\d-.5,\d-.5) {\p};
		\fi
	}
	% draw row/col labels
	\foreach \x in {0,...,\n} {
		\node[rowlabel] at (-.5,\x+.5) {\x};
		\node[rowlabel] at (\x+.5,\m+.5) {\x};
	}
\end{tikzpicture}
	\vspace*{-0.275cm}\caption{Triangle block distribution for $c=4$, $c^2+c+1=21$ segments using the projective construction. Diagonal elements are assigned using a greedy procedure.}\label{fig:TBD-proj}
\end{figure}
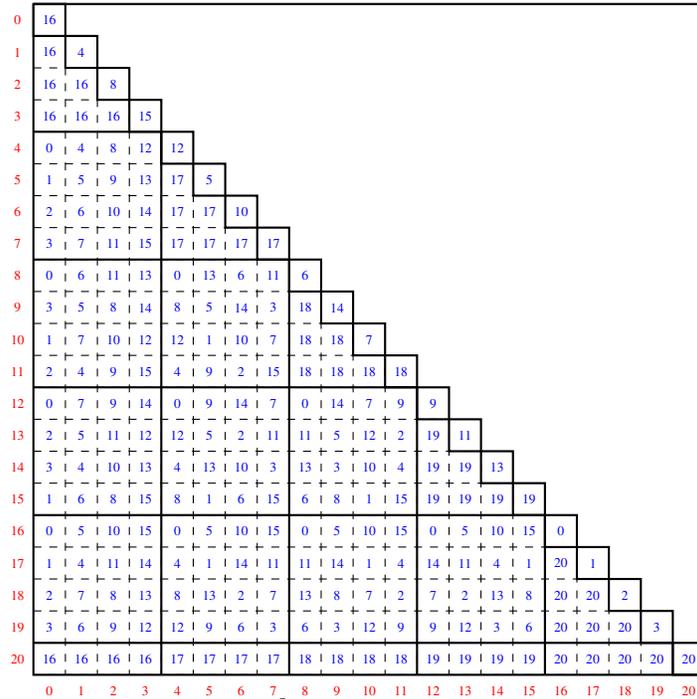

We provide an example of triangle block partition for $c=4$ using the projective plane construction in \cref{fig:TBD-proj}.
The sets used to define the partition are presented in \cref{tab:TBD-proj}.

\vspace*{-0.25cm}
\begin{table}[htb]
    \begin{center}
        {\scriptsize
        \begin{tabular}{|c|c|c||c|c|}
        \hline
        $k$ & $R_k$ & $D_k$ & $i$ & $Q_i$ \\
        \hline
        0 & \{0,4,8,12,16\} & \{16\} & 0 & \{0,1,2,3,16\}\\
        1 & \{0,5,10,15,17\} & \{17\} & 1 & \{4,5,6,7,16\} \\
        2 & \{0,6,11,13,18\} & \{18\} & 2 & \{8,9,10,11,16\} \\
        3 & \{0,7,9,14,19\} & \{19\} & 3 & \{12,13,14,15,16\} \\
        4 & \{1,4,11,14,17\} & \{1\} & 4 & \{0,4,8,12,17\} \\
        5 & \{1,5,9,13,16\} & \{5\} & 5 & \{1,5,9,13,17\} \\
        6 & \{1,6,8,15,19\} & \{8\} & 6 & \{2,6,10,14,17\} \\
        7 & \{1,7,10,12,18\} & \{10\} & 7 & \{3,7,11,15,17\}\\
        8 & \{2,4,9,15,18\} & \{2\} & 8 & \{0,6,11,13,18\} \\
        9 & \{2,5,11,12,19\} & \{12\} & 9 & \{3,5,8,14,18\} \\
        10 & \{2,6,10,14,16\} & \{6\} & 10 & \{1,7,10,12,18\} \\
        11 & \{2,7,8,13,17\} & \{13\} & 11 & \{2,4,9,15,18\} \\
        12 & \{3,4,10,13,19\} & \{4\} & 12 & \{0,7,9,14,19\} \\
        13 & \{3,5,8,14,18\} & \{14\} & 13 & \{2,5,11,12,19\} \\
        14 & \{3,6,9,12,17\} & \{9\} & 14 & \{3,4,10,13,19\} \\
        15 & \{3,7,11,15,16\} & \{3\} & 15 & \{1,6,8,15,19\} \\
        16 & \{0,1,2,3,20\} & \{0\} & 16 & \{0,5,10,15,20\} \\
        17 & \{4,5,6,7,20\} & \{7\} & 17 & \{1,4,11,14,20\} \\
        18 & \{8,9,10,11,20\} & \{11\} & 18 & \{2,7,8,13,20\} \\
        19 & \{12,13,14,15,20\} & \{15\} & 19 & \{3,6,9,12,20\} \\
        20 & \{16,17,18,19,20\} & \{20\} & 20 & \{16,17,18,19,20\} \\
        \hline
        \end{tabular}}
        \caption{Row block sets/processor sets of triangle block distribution for $c=4$, $c^2+c+1=21$ segments using the projective construction.\vspace*{-0.35cm}}
        \label{tab:TBD-proj}
    \end{center}
\end{table}

\section{Triangle Block Partitions using Steiner System}
\label{sec:appendix:Steiner-example}
We provide an example of triangle block partition using the Steiner $(15, 3, 2)$ system in \cref{fig:steiner_15-3-2}.

\begin{figure}[htb]
  \pgfmathsetmacro{\c}{4}
  \pgfmathsetmacro{\m}{15}
  \def\R{%
    0/{1,2,3}/1,%
    1/{1,4,5}/5,%
    2/{1,6,7}/7,%
    3/{1,8,9}/9,%
    4/{1,10,11}/11,%
    5/{1,12,13}/13,%
    6/{1,14,15}/15,%
    7/{2,4,6}/2,%
    8/{2,5,7}/0,%
    9/{2,8,10}/0,%
    10/{2,9,11}/0,%
    11/{2,12,14}/0,%
    12/{2,13,15}/0,%
    13/{3,4,7}/4,%
    14/{3,5,6}/6,%
    15/{3,8,11}/3,%
    16/{3,9,10}/10,%
    17/{3,12,15}/0,%
    18/{3,13,14}/14,%
    19/{4,8,12}/0,%
    20/{4,9,13}/0,%
    21/{4,10,14}/0,%
    22/{4,11,15}/0,%
    23/{5,8,13}/0,%
    24/{5,9,12}/0,%
    25/{5,10,15}/0,%
    26/{5,11,14}/0,%
    27/{6,8,14}/0,%
    28/{6,9,15}/0,%
    29/{6,10,12}/0,%
    30/{6,11,13}/0,%
    31/{7,8,15}/8,%
    32/{7,9,14}/0,%
    33/{7,10,13}/0,%
    34/{7,11,12}/12%
  }
  \centering
  %!TEX root = ../sec7.tex

% set offset and matrix dim
\pgfmathsetmacro{\offset}{.33}
\pgfmathsetmacro{\n}{\m-1}

% set styles
\tikzset{proclabel/.style={yscale=-1,blue}}
\tikzset{rowlabel/.style={yscale=-1,red}}

% for drawing help	
\newcommand{\gridhelp}{
	\coordinate (ll) at (-1,-1);
	\coordinate (ur) at (10,10);
	\draw[help lines,dashed] (ll) grid (ur);
	\node at (0,0) {0};
}	

%\begin{tikzpicture}[every node/.append style={transform shape},yscale=-1,scale=.675]
\begin{tikzpicture}[every node/.append style={transform shape},yscale=-1,scale=.425]
%\gridhelp
	% draw little boxes
	\foreach \x in {1,...,\m} {\draw[dashed] (0,\x) -- (\x,\x) -- (\x,\m);}
	% draw diagonal
	\foreach \x in {1,...,\m} {
		\draw[thick] (\x-1,\x-1) -- (\x,\x-1) -- (\x,\x);
		\draw[thick] (\x-1,\x-1) -- (\x-1,\x) -- (\x,\x);
	}
	% draw outer box
	\draw[thick] (0,0) rectangle +(\m,\m);
	% draw processor labels
	\foreach \p/\list/\d in \R {
		% off-diagonal blocks
		\foreach \x in \list {
			\foreach \y in \list {
				\ifnum \x<\y 
					\node[proclabel] at (\x-.5,\y-.5) {\p};
				\fi
			}
		}
		% diagonal block
		\ifnum \d>0
			\node[proclabel] at (\d-.5,\d-.5) {\p};
		\fi
	}
	% draw row/col labels
	\foreach \x in {0,...,\n} {
		\node[rowlabel] at (-.5,\x+.5) {\x};
		\node[rowlabel] at (\x+.5,\m+.5) {\x};
	}
\end{tikzpicture}
  \caption{Example of triangle block partition of the lower triangle of the matrix based on the Steiner $(15, 3, 2)$ system. Each set $R_k$ is of size $3$ and the corresponding triangle block $TB(R_k)$ contains $3$ elements in the strict lower triangle. Diagonal elements are assigned using a greedy procedure.}
  \label{fig:steiner_15-3-2}
\end{figure}

\end{document}